\documentclass[aps,prx,showpacs,notitlepage,floatfix, twocolumn,superscriptaddress,nofootinbib]{revtex4-1}
\usepackage{bbm}
\usepackage{mathrsfs}
\usepackage{epsfig}
\usepackage{graphicx}
\usepackage{amsfonts}
\usepackage[figuresright]{rotating}
\usepackage{amssymb}
\usepackage{amsmath}
\usepackage{dcolumn}
\usepackage{bm}
\usepackage{braket}
\usepackage[colorlinks,linkcolor=blue,anchorcolor=blue,citecolor=blue,urlcolor=blue]{hyperref}
\usepackage[dvipsnames]{xcolor}
\usepackage[titletoc]{appendix}

\def\be{\begin{equation}}
\def\ee{\end{equation}}
\def\bea{\begin{equation}\begin{aligned}}
\def\eea{\end{aligned}\end{equation}}

\definecolor{navy}{RGB}{81, 114, 233}

\def\Re{\rm{Re}}

\def\w{\mathbf{w}}
\def\e{\mathbf{e}}
\def \erf {\text{erf}}
\def \tr {\text{tr}}

\begin{document}
\title{Locality, Quantum Fluctuations, and Scrambling}
\author{Shenglong Xu}
\affiliation{Condensed Matter Theory Center and Department of Physics, University of Maryland, College Park, MD 20742, USA}
\author{Brian Swingle}
\affiliation{Condensed Matter Theory Center, Maryland Center for Fundamental Physics, Joint Center for Quantum Information and Computer Science, and Department of Physics, University of Maryland, College Park, MD 20742, USA}

\begin{abstract}
Thermalization of chaotic quantum many-body systems under unitary time evolution is related to the growth in complexity of initially simple Heisenberg operators. Operator growth is a manifestation of information scrambling and can be diagnosed by out-of-time-order correlators (OTOCs). However, the behavior of OTOCs of local operators in generic chaotic local Hamiltonians remains poorly understood, with some semiclassical and large N models exhibiting exponential growth of OTOCs and a sharp chaos wavefront and other random circuit models showing a diffusively broadened wavefront. In this paper we propose a unified physical picture for scrambling in chaotic local Hamiltonians. We construct a random time-dependent Hamiltonian model featuring a large N limit where the OTOC obeys a Fisher-Kolmogorov-Petrovsky-Piskunov (FKPP) type equation and exhibits exponential growth and a sharp wavefront. We show that quantum fluctuations manifest as noise (distinct from the randomness of the couplings in the underlying Hamiltonian) in the FKPP equation and that the noise-averaged OTOC exhibits a cross-over to a diffusively broadened wavefront. At small N we demonstrate that operator growth dynamics, averaged over the random couplings, can be efficiently simulated for all time using matrix product state techniques. To show that time-dependent randomness is not essential to our conclusions, we push our previous matrix product operator methods to very large size and show that data for a time-independent Hamiltonian model are also consistent with a diffusively-broadened wavefront.
\end{abstract}
\maketitle
\section{Introduction}
\label{sec:introduction}

Information scrambling describes a process whereby information about the initial condition of a unitarily evolving system spreads over the entire system, becoming inaccessible to any local measurement ~\cite{ Hayden2007, Sekino2008, Shenker2014, Hosur2016}. Because it describes an effective loss of memory, scrambling is relevant for understanding quantum thermalization  (e,g.,~\cite{Deutsch1991, Srednicki1994,Tasaki1998a,Rigol2008}), i.e., the emergence of irreversibly from unitary time evolution, and is also tied to the black hole information problem. Scrambling is also closely related to the dynamics of initially simple Heisenberg operators, with the growth in size and complexity of these operators probing the spreading of quantum information~\cite{Nahum2017a, VonKeyserlingk2017,Xu2018,Roberts2018,Jonay2018, Mezei2018, You2018, Chen2018}.

Given two local operators $W_0$ and $V_r$ at positions $0$ and $r$, the out-of-time ordered correlator (OTOC),
\be
F(r,t)=\braket{ W_{0}^\dagger(t) V_{r}^\dagger W_{0}(t) V_{r}},
\label{eq:otoc}
\ee
provides one way to quantify scrambling by probing how the Heisenberg operator $W_0(t)$ grows with time. Although scrambling can also be usefully characterized in entropic terms, OTOCs are more directly measurable, with early experiments having already been carried out in a variety of platforms~\cite{swingle2016,Zhu2016,Yao2016a,Halpern2016,Halpern2017,Campisi2017,Yoshida2017,Garttner2016, Wei2016, Li2017a,  Meier2017}. A closely related quantity is the squared commutator between $V$ and $W$, defined as,
\be
C(r,t)=\braket{[W_{0}(t),V_r]^\dagger [W_{0}(t),V_r]} = 2[1  -  \Re(F)].
\ee

The physical picture is that under Heisenberg dynamics, the operator $W_0$ expands and eventually fails to commute with $V_{r}$, as manifested by the growth of $C(r,t)$ from zero. For chaotic local Hamiltonians, $W_0(t)$ is expected to expand ballistically, with speed called the butterfly velocity, so that the OTOC exhibits a causal light-cone-like structure in space-time. The squared commutator remains small outside the light-cone and grows rapidly the boundary of the light-cone is crossed. Inside the light-cone, $C(r,t)$ saturates for chaotic systems regardless of the specific form of operator $W$ and $V$.

A particularly interesting question concerns the specific growth form of $C(r,t)$ near the wavefront of the light-cone. In some models, $C(r,t)$ grows exponentially with time, a phenomenon proposed as a quantum analog of classical butterfly effect, the exponential divergence of initially nearby trajectories. This observation has led to an emphasis on probing the footprint of quantum chaos at an intermediate time-scale, especially with a view towards defining a notion of quantum Lyapunov exponent. A well-defined Lyapunov exponent $\lambda_L$, i.e., purely exponential growth of $C(r,t)$, plus the ballistic growth of the OTOC, implies that the wavefront is sharp, and sharp wavefronts have been identified in a broad class of holographic/large $N$ models, including the O($N$) model ~\cite{Chowdhury2017}, the diffusive metal ~\cite{Patel2017} and the coupled SYK model~\cite{Sachdev1993, kitaev2015, Gu2017}. On the other hand, although significant efforts have been made ~\cite{Luitz2017, Bohrdt2017a, Heyl2018, Lin2018}, a clear signature of purely exponential growth of the OTOC in more physical systems with finite on-site degrees of freedom is absent, and there are some counterexamples in random circuit models~\citep{Nahum2017a,	Nahum2017, VonKeyserlingk2017, Rakovszky2017, Khemani2017}.

To reconcile the many different scenarios, in a recent paper \cite{Xu2018} we proposed a universal form for the early growth region of the squared commutator,
\be \label{eq:broadexp}
C(r,t)\sim\exp \left(-\lambda_p (x/v_B-t)^{1+p}/t^p \right),
\ee
assuming there is a well-defined butterfly velocity $v_B$ (a different ansatz is needed for localized systems~\cite{Swingle2016a, Chen2016a, Fan2017, Huang2017}), furthered studied by \cite{Khemani2018}.  The shape of the wavefront is controlled by a single parameter $p$, denoted  as the broadening exponent, associated with the growth rate $\lambda_p$. For large $N$/holographic models, $p=0$ and the corresponding $\lambda_p$ is the Lyapunov exponent. However, an exact calculation in a Haar random brickwork circuit model gave $p=1$ in one dimension, indicating a diffusive broadening of the wavefront. Saddle point analysis shows $p=\frac{1}{2}$ for general non-interacting systems with translational invariance~\cite{Lin2018, Xu2018, Khemani2018}. Large-scale matrix-product state simulations using the time-dependent variational principle~\cite{Leviatan2017} and  matrix product operator simulations~\cite{Xu2018} also gave strong evidence of wavefront broadening for chaotic local Hamiltonian systems.

In this work, we make two contributions to understanding the early growth region behavior of the OTOC. First, to understand the intriguing differences between the large $N$ models and the Haar random brickwork circuit models, we design and analyze a new random circuit model, denoted as the Brownian coupled cluster model (BCC). BCC, as an extension of the single cluster version~\cite{Lashkari2012, Shenker2014a}, describes the dynamics of clusters of $N$ spins connected in a one-dimensional array (or more generally, connected according to any graph), similar to coupled SYK cluster models but with the couplings random in both space and time. We show that in the large $N$ limit, BCC is similar to other large $N$ models and has a well defined Lyapunov exponent, but the finite $N$ correction qualitatively changes the broadening exponent from $p=0$ to $p=1$ in one dimension. We find that finite $N$ corrections are actually quite dramatic, with the broadening of the wavefront characterized by a diffusion constant which scales as $1/\log ^3N$ at large $N$. We also find that there is a finite region in space-time where the wavefront remains sharp, indicating strong finite-size effects on the broadening exponent.

With this new point of view, our second contribution is to push our numerical matrix product operator simulations of operator growth in a local Hamiltonian Ising system to include $200$ spins in the wavefront and up to time $250$ in units of the nearest neighbor Ising coupling. By directly analyzing the way contours of constant $C$ deviate in space-time, we find that the broadening exponent indeed converges to $p=1$ in the large space-time limit. Therefore we conclude that diffusive broadening of the wavefront is generic for one-dimensional chaotic systems.

In more detail, our analysis of the BCC proceeds by focusing on operator dynamics, suitably averaged over the random couplings in the Hamiltonian. Any operator may be expanded in a complete basis of operators, with the expansion coefficients called operator amplitudes and with the square of the amplitudes forming a probability distribution, the operator probability distribution. This procedure is completely analogous to expanding a time evolving wavefunction in a complete basis of states. The starting point of the analysis is the derivation of an equation of motion for the circuit-averaged operator probability distribution of a Heisenberg operator. The effect of averaging over the couplings in the quantum Hamiltonian is to give a closed stochastic equation for the operator probability distribution; physically, the operator amplitudes evolve via unitary time evolution for each choice of couplings, and the averaging dephases this dynamics to yield a master equation for the operator probability distribution. One point should be emphasized: The randomness of the couplings in the Hamiltonian, which we sometimes call ``disorder'', is physically distinct from the quantum randomness manifested in the operator probability distribution. The latter will, in a certain limit, be instantiated as a random process which we call ``noise''. One of the key assertions of this paper is that the disorder average is a technical convenience while the noise average contains essential physics of quantum fluctuations.

Starting from the master equation for the operator probability distribution, the analysis proceeds from two limits. In a large $N$ limit, a mean-field-like treatment of the operator distribution becomes exact, and the operator dynamics can be translated into a closed non-linear partial differential equation for the operator weight, a measure of the size of the operator which is linearly related to the OTOC. The resulting dynamical equation is similar to the Fisher-Kolmogorov-Petrovsky-Piskunov (FKPP) equation \cite{Fisher1937, Kolmogorov1937}  which occurs, for example, in studies of combustion waves, invasive species, and quantum chromodynamics, among others, and was recently introduced in the context of scrambling to describe the growth of OTOCs~\cite{Aleiner2016, Chowdhury2017, Grozdanov2018}. The key physical effects embodied by FKPP-type equations are unstable exponential growth, diffusion, and eventual saturation; together these lead to traveling wave solutions with a sharp wavefront which describe the spreading of local Heisenberg operators. The leading finite $N$ correction results in a stochastic partial differential equation, a noisy FKPP-like equation, in which the noise is multiplicative and $1/N$ suppressed. Drawing from the noisy FKPP literature~\cite{Brunet1997, Brunet2006}, we argue that the noisy FKPP-like in the BCC has a diffusively broadened wavefront after averaging over noise. These analytical arguments are verified by direct numerical integration of the large-but-finite-$N$ BCC stochastic equation. It should be emphasized again that the noise in the noisy FKPP-like equation represents quantum fluctuations, not different instances of the microscopic couplings.

In the small $N$ limit, a different analytical treatment shows that the OTOC exhibits the same broadening diffusive broadening as in the Haar random brickwork circuit model. Moreover, we show that by representing the operator probability distribution as a ``stochastic'' matrix product state, it is possible to numerically solve the master equation for the time dynamics. Thanks to the dephasing provided by the disorder average over couplings, one can show that the late time operator probability distribution has low ``correlation/entanglement'' when viewed as a matrix product state. We further find that the operator probability distribution never has high correlation/entanglement, so that matrix product state techniques can accurately capture the operator dynamics for all times. A modest bond dimension of $\chi=32$ is already sufficient to converge the dynamics for $200$ sites for all time.

Finally, taking these lessons from the BCC, especially the crucial role of noise, meaning quantum fluctuations, we argue that the diffusive broadening of the operator growth wavefront is generic in one-dimension. This idea has been previously conjectured based on work with random circuit models \cite{Nahum2017a,Nahum2017, VonKeyserlingk2017, Rakovszky2017, Khemani2017}. One piece of evidence is direct numerical simulation of the time-independent Hamiltonian dynamics of Heisenberg operators in a system of $200$ spins for very long time. A new analysis of the space-time contours of constant squared commutator conclusively demonstrates diffusive broadening of the wavefront at the largest sizes. Another piece of evidence is the prevalence of noiseless FKPP-like equations describing OTOC dynamics in large $N$ or weakly coupled models, including linearized FKPP-like equations obtained in resummed perturbation theory \cite{Chowdhury2017} and fully non-linear FKPP-like equations obtained from self-consistent Keldysh treatments \cite{Aleiner2016}. We argue that, starting from these known results, quantum fluctuations should invariably be described by adding noise, specifically multiplicative noise of the type found in the BCC. Hence these models will also suffer similar dramatic finite $N$ effects resulting in diffusively broadened wavefronts.

The remainder of the paper is organized as follows. Section \ref{sec:general_formalism} describes in detail the notions of operator dynamics used throughout the paper and their relations to OTOCs. Section \ref{sec:BCC} introduces and analyzes the Brownian coupled cluster model (BCC), both at small and large $N$. Section \ref{sec:local_ham} discusses implications of the results for generic Hamiltonian systems. We conclude with some outlook, including the effects of conserved quantities and going beyond one dimension, as well as open questions.

\section{General formalism of operator dynamics and its relation to the out-of-time-ordered correlator}
\label{sec:general_formalism}
Consider a generic quantum system consisting of $L$ sites with $N$ spin-1/2s per site. The dynamics of the system is governed by a local unitary circuit $U(t)$. In the Pauli basis, a Heisenberg operator $W(t)$ takes the form \be
W(t)=\sum c(\mathcal{S}) \mathcal{S}
\ee
where $\mathcal{S}$ is a product of Pauli operators with length $NL$ and the time dependence is encoded in the coefficients $c(\mathcal{S},t)$. The normalization is $\tr (W(t)^\dagger W(t))=2^{NL}$, so that $\sum\limits_\mathcal{S} |c(\mathcal{S},t)|^2=1$. Under Heisenberg time evolution in a chaotic quantum system, an initially-localized operator grows and eventually equilibrates as far as local probes are concerned. The initial configuration cannot be recovered from local data, a manifestation of scrambling. The OTOC is designed precisely to quantify this process since the deviation of the correlator from its initial value for certain position indicates that the Heisenberg operator has developed a nontrivial component at that site. To understand the relationship between the OTOC and the operator string picture, we consider the following averaged OTOC,
\be
F(r,t)=\frac{1}{4N}\sum_{a,\alpha} \frac{1}{2^{NL}}\text{Tr}(W(t)^\dagger \sigma_{r,a}^\alpha W(t) \sigma_{r,a}^\alpha)
\ee
where $a$ runs from $1$ to $N$ and $\alpha$ from 0 to 3, representing the identity and the three Pauli matrices.

Using the decoupling channel identity, we obtain that
\be
\phi(r,t)=1-F(r,t)=\sum\limits |c(\mathcal{S})|^2 \frac{1}{N}\left ( \sum\limits_a w(\mathcal{S}_{r,a}) \right )
\ee
where $\mathcal{S}_{r,a}$ is the $r,a$ Pauli operator in the string $\mathcal{S}$ and the weight is $w(\sigma^\alpha)=1-\delta_{\alpha 0}$. Therefore $\phi(r,t)$ measures the average number of nontrivial local operators within a single spin cluster located at $r$, growing from $0$ and saturating to $3/4$ in a thermalizing system. Similarly, the averaged squared commutator $C(r,t)$ equals $\frac{8}{3}\phi(r,t)$. There are two implications. First, $C(r,t)$ does not depend on the phase information of the coefficients $c(\mathcal{S})$ (which are in fact real for Hermitian $W(t)$), since it is completely determined by the probability distribution $|c(\mathcal{S})|^2$. Second, $C(r,t)$ does not care about the specific operator configuration $\mathcal{S}$, since only the number of non-trivial operators matters. Both features significantly simplify the calculations.

In general, dealing with the Schr\"odinger equation is overcomplicated for the purposes of studying operator dynamics due to the unimportant phase information and associated high operator entanglement entropy. However, deriving a closed set of dynamical equations for the operator probability $|c(\mathcal{S})|^2$ from the Schr\"odinger equation is difficult. Random circuit models are useful to overcome this difficulty by introducing disorder as a dephasing mechanism, making the probability distribution dynamics tractable both numerically and analytically, for example, in recent studies of the Haar random brickwork circuit. This paper introduces another class of random circuit models, the Brownian coupled cluster (BCC), describing the dynamics of a system of coupled spin clusters with interaction random both in space and time. BCC can be regarded as a smoother version of the random brickwork circuit, in which the interactions are only between pairs of spins (or more generally, few-body) even in the large $N$ limit, and thus is naturally more tied to holographic and SYK models. While in the random brickwork circuit, the dimension of the on-site Hilbert space does not qualitatively affect the operator dynamics, in the BCC, we expect a smooth crossover from the random brickwork circuit result for small $N$ to holographic and SYK physics physics in the large $N$ limit.

\section{Brownian coupled cluster model}
\label{sec:BCC}

\begin{figure}
\includegraphics[height=0.5\columnwidth, width=0.9\columnwidth]
{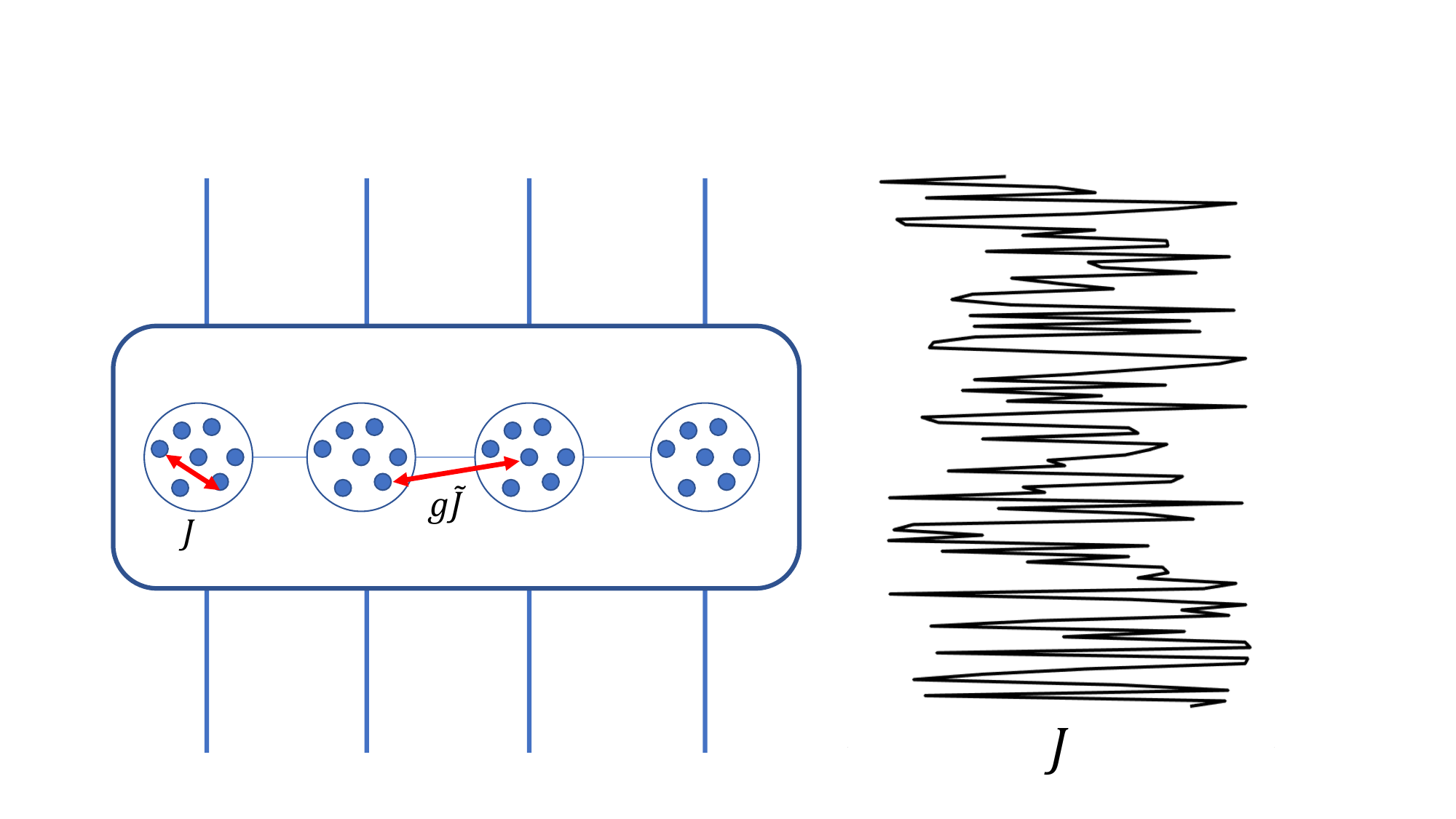}
\caption{The Brownian coupled cluster model. Spins within a same cluster interact with each other and also interact with the spins in the neighboring clusters. The intra-cluster coupling $J$ and the inter-cluster coupling $\tilde{J}$ are random in both space and time. }
\label{fig:BCC}
\end{figure}

Fig.~\ref{fig:BCC} shows a schematic of the BCC model. It is best described using discrete time steps $dt$, with the limit $dt\rightarrow 0$ taken later. For small $dt$, the whole time evolution unitary breaks into pieces,
\be
U(t)=\prod\limits_{m=1}^{t/dt}\exp\left(-i \sum\limits_r H^{(m)}_r -i\sum\limits_{rr'} H^{(m)}_{rr'}\right),
\ee
with $m$ a discrete time index. The on-site terms and the bond terms read
\bea
H^{(m)}_r&=\sum\limits_{a,b,\alpha,\beta }J_{m,r,a,b}^{\alpha\beta}\sigma_{r,a}^\alpha \sigma_{r,b}^\beta\\
H^{(m)}_{rr'}&=g  \sum\limits_{a,b,\alpha,\beta } \tilde{J}_{m,r,r', a,b}^{\alpha\beta}\sigma_{r,a}^\alpha \sigma_{r',b}^\beta
\eea
where $\alpha$ and $\beta$ are the Pauli matrix indices running over from 0 to 3 (including the identity for convenience), $a$ and $b$ from 1 to $N$ label the spins in the cluster, $r$ and $r'$ label clusters (sometimes called sites), and $\braket{rr'}$ means nearest neighbors. At each time step, the models contains two sets of uncorrelated random variables $J$ and $\tilde{J}$ with mean zero and variance $\frac{1}{8(N-1)}dt$ and $\frac{1}{16N}dt$, respectively.

With the help of the random couplings, one can derive a master equation for the averaged probability distribution $h=\overline{|c(\mathcal{S})|^2}$. To simplify the calculation, we assume that $h$ only depends on the operator weight $w_r=\sum \limits_a w(\mathcal{S}_{r,a})$ of each cluster instead of on the details of the operator configuration $\mathcal{S}$. This approximation is valid after a short relaxation time even though $W$ starts as a specific operator. To proceed further, introduce the operator weight probability
\be
\tilde{h}(\mathbf{w})=h(\mathbf{w})D(\mathbf{w})
\ee
with $D$ the number of operators with weight configuration $\mathbf{w}$,
\be
D(\mathbf{w}) = \prod \limits_r {N \choose w_r}3^{w_r}.
\ee
The operator weight probability is a properly normalized probability distribution over the $(N+1)^L$ possible weight strings.

The derivation of the master equation for $\tilde{h}$ is recorded in the appendix \ref{appsec:master_eq}, with the result being,
\bea
\partial_t \tilde{h}=\sum\limits_r &\left [-\gamma_r^+(w)\tilde{h}(\w)+\gamma^+_r(w+1)\tilde{h}(\w+\e_r)\right ]+\\
  &\left [-\gamma^-_r(w)\tilde{h}(\w)+\gamma^-_r(w-1)\tilde{h}(\w-\e_r)\right ] \\
+\sum\limits_{\braket{rr'}} &\left [-\gamma^+_b(w_r, w_{r'})\tilde{h}(\w)+\gamma^+_b(w_r+1,w_{r'})\tilde{h}(\w+\e_r) \right]\\
+&\left [-\gamma^-_b(w_r,w_{r'})\tilde{h}(\w)+\gamma^-_b(w_r-1,w_{r'})\tilde{h}(\w-\e_r) \right ] \\
+&[r\longleftrightarrow r'] \\
 \label{eq:master}
\eea
The evolution equation manifestly conserves the total probability, $\sum\limits_{\{\w\}} \tilde{h}=1$ for all time, independent of the specific form of the functions $\gamma^+$ and $\gamma^-$. For this particular problem, these functions are
\bea
&\gamma^+_r(w)=\frac{1}{N-1}w(w-1), \  \gamma^-_r(w)=\frac{1}{N-1}3(N-w)w \\
&\gamma^+_b(w_1,w_2)=\frac{g^2}{2N}w_1w_2, \  \gamma^-_b(w_1,w_2)=\frac{g^2}{2N}3 (N-w_1)w_2.
\eea

In the following subsections, we will analyze this master equation in the infinite-$N$ limit, study its large-$N$ expansion, and compare the result with small-$N$ results. Complementing these analytical results are numerical simulations of the master equation for $200$ spin clusters using tensor network methods.

\subsection{The infinite $N$ limit}
\label{sec:inf_N}
In the infinite-$N$ limit, the master equation Eq.~\eqref{eq:master} can be approximated by a Fokker-Planck equation,
\bea
\partial_t \tilde{h}&=\sum\limits_r  \partial_{\phi_r}\left [ \alpha(\phi_r)\tilde{h}(\phi) \right ] +\partial^2_{\phi_r} \left [\beta(\phi_r)\tilde{h}(\phi)\right ]+\mathcal{O}\left ( \frac{1}{N^2}\right )\\
\alpha(\phi_r)&=(4\phi_r-3)(\phi_r+\frac{g^2}{2}(\phi_{r-1}+\phi_{r+1}))+\frac{1}{N}4(\phi_r-1)\phi_r\\
\beta(\phi_r)&=\frac{1}{4N}(3-2\phi_r)(2\phi_r+g^2(\phi_{r-1}+\phi_{r+1}))
\label{eq:fokker-planck}
\eea
where $\phi_r=w_r/N$ is the scaled operator weight.

Using the Ito stochastic calculus, the Fokker-Planck equation can be mapped to a Langevin equation,
\be
\partial_t \phi_r=-\alpha(\phi_r)+\sqrt{2\beta(\phi_r)}\eta_r(t)
\label{eq:langevin}
\ee
with $\braket{\eta_{r}(t)\eta_{r'}(t')}=\delta_{rr'}\delta(t-t')$. This mapping explicitly demonstrates that the noise $\eta$ arises from the deterministic master equation for the operator weight probability as $1/N$ effect. It is important that this noise $\eta$ is conceptually different from the randomness of the Brownian circuit introduced to obtain the master equation; it originates purely from the quantum fluctuation in the BCC. Later we will show that the noise, although suppressed at large $N$, has a drastic effect on the operator dynamics.

First, we study the infinite-$N$ limit in which the noise is set to zero and the stochastic Langevin equation becomes deterministic. After taking the continuum limit of the Langevin equation, in which $\phi(r,t)$ is assumed to vary slowly with respect to $r$, we obtain a FKPP-type equation,
\bea
\partial_t \phi(r,t)=(3-4\phi(r,t))\left (\frac{g^2}{2}\partial_r^2 \phi(r,t)+(1+g^2)\phi(r,t)\right ),
\label{eq:fisher}
\eea
describing a growth-diffusion-saturation process. For simplicity of presentation, we hereafter refer to Eq.~\eqref{eq:fisher} as an FKPP equation. There are two fixed points of the dynamics, an unstable solution $\phi(r,t)=0$  and a stable solution $\phi(r,t)=\frac{3}{4}$. The stable solution describes the equilibrium state where every operator string is equally probable. An initially localized operator configuration translates to an initial condition for the FKPP-type equation which is the unstable solution everywhere away from the initial local operator.

Similar to the FKPP equation, Eq.~\eqref{eq:fisher} admits a traveling wave solution $\phi(r,t)=f(r-vt)$ when the velocity $v$ is larger than $v_c=\sqrt{18g^2(1+g^2)}$. Ahead of the wavefront, $r \gg vt$, the traveling wave decays exponentially with $r$.
For initial operator profile that is sufficiently localized, the wavefront travels with the minimal velocity $v_c$ and approaches the traveling-wave solution at late times. A detailed analysis can be found in the appendix \ref{appsec:noisy_FKPP}. Ahead of the wavefront, the traveling wave decays as $\exp(6(1+g^2)(t-r/v_c))$, consistent with a sharp wavefront. Therefore, the infinite-$N$ limit of the Brownian coupled cluster model exhibits a well-defined Lyapunov exponent. The butterfly velocity and the Lyapunov exponent are
\bea
v_B&=\sqrt{18g^2(1+g^2)}\\
\lambda_L &=6(1+g^2).
\label{eq:inf_N_v_l}
\eea
Within the framework described by Eq.~\eqref{eq:broadexp}, the infinite-$N$ limit has broadening exponent $p=0$.

The existence of a Lyapunov exponent in the infinite-$N$ limit is in sharp contrast with the random brickwork circuit model result, where the diffusive-spreading nature of the wavefront is independent of the dimension of the on-site Hilbert space. The reason is that the brickwork model has no notion of few-body interactions within an on-site cluster due to the use of Haar random unitary matrices in the circuit. In Fig.~\ref{fig:inf_N}(b), we explicitly verify the sharp wavefront by numerically solving Eq. \ref{eq:fisher}. 
\begin{figure}
\includegraphics[height=0.9\columnwidth, width=0.9\columnwidth]
{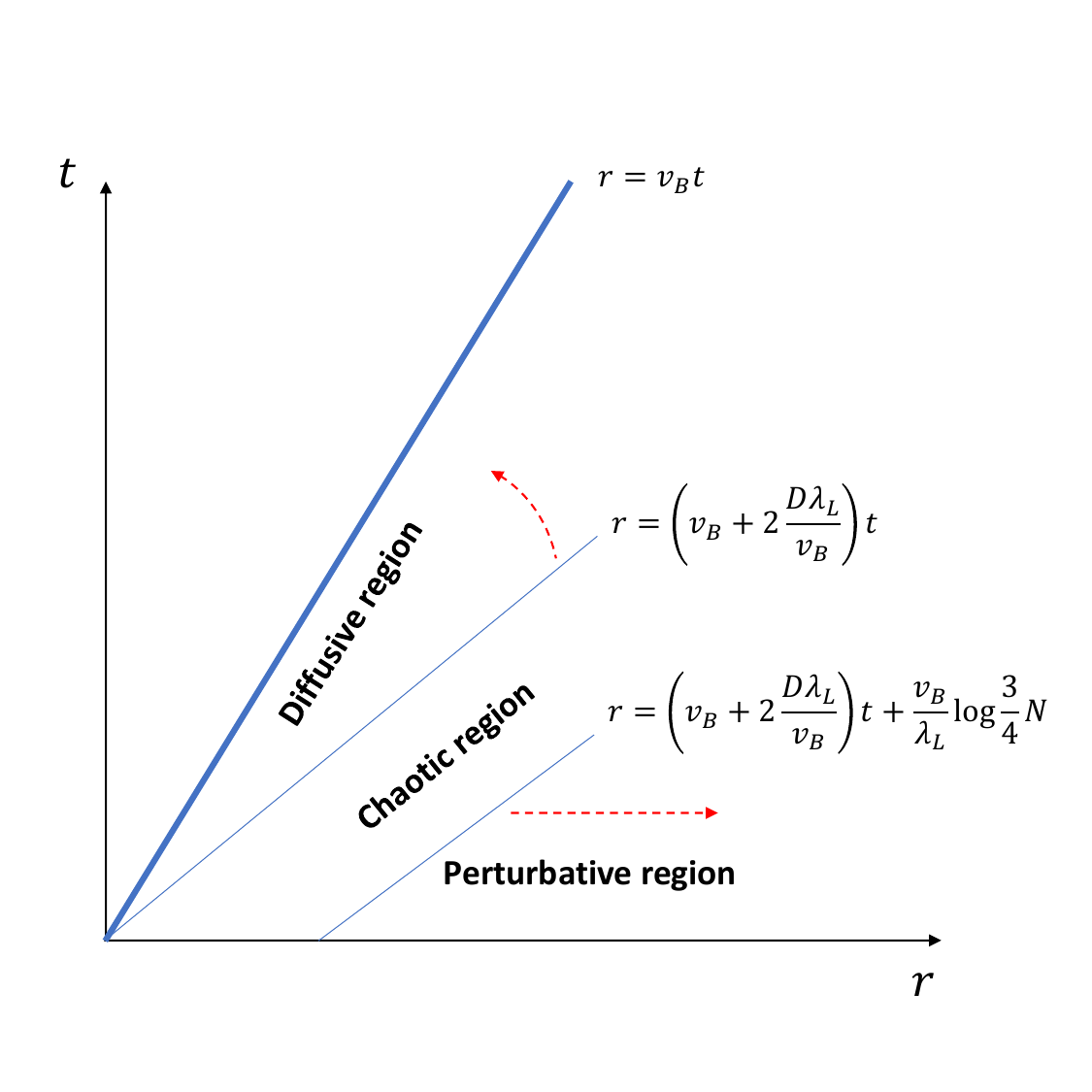}
\caption{The scrambling ``phase diagram'' in space-time. An overall unimportant spatial offset is omitted. The bold line marks the wavefront. The space-time is roughly divided into three regions. In the diffusive region, the squared commutator exhibits a diffusively broadening wavefront, quantitatively different from the exponential growth in the large $N$ limit. Away from the wavefront, there is a region where the squared commutator grows exponentially with a modified Lyapunov exponent. Further ahead the wavefront, the chaotic region gives way to the perturbative region. As $N$ increases, the chaotic region expands as indicated by dashed arrows and eventually dominates the wavefront behavior in the infinite $N$ limit. }
\label{fig:spacetime}
\end{figure}

\subsection{The large-$N$ expansion}
\label{sec:large_N}

Having established the purely exponential growth of the squared commutator in the infinite-$N$ limit, we now investigate the behavior away from this limit. Comparing with the infinite-$N$ limit, the large-$N$ expansion affects Eq.~\eqref{eq:fisher} in two significant ways. First, $\phi(r,t)$ in principle only takes discrete values $0, \frac{1}{N}, \frac{2}{N} ...$. Therefore, in Eq.~\ref{eq:fisher}, $\phi(r,t)$ is set to zero when it is below $1/N$. This hard cutoff allows the traveling wave to propagate with velocity smaller than $v_c$, and as such the cutoff is important for obtaining the correction to butterfly velocity. Second, the noise term in the Langevin equation $\eqref{eq:langevin}$ becomes important.  The deterministic differential equation Eq.~\eqref{eq:fisher} is augmented by a multiplicative noise term
\bea
&f_{noise}\\
&=\sqrt{\frac{1}{N}\left (3-2\phi(r,t)\right )\left (\frac{g^2}{2}\partial_r^2 \phi(r,t)+(1+g^2)\phi(r,t)\right )}\eta(r,t).
\label{eq:noise_fisher}
\eea
Due to its multiplicative nature, the noise term only affects the physics when $\phi(r,t)$ is non-zero, and therefore the noise does not violate the causal structure of the noiseless FKPP equation . The effect of the noise is most prominent near the forward edge of the wave, with its most important effect being to make the position of the wavefront a random variable described by a biased random walk. The resulting noise-averaged front spreads diffusively with the diffusion constant $D$ in addition to the drift $v_B(N) t$. Following an analysis of the original noisy FKPP equation \cite{Brunet1997, Brunet2006}, which we review in the appendix \ref{appsec:noisy_FKPP}, we are able to obtain the scaling of the $v_B(N)$ and $D$ in the large $N$ limit,
\bea
&\delta v_B=v_B(N)-v_B(\infty)\sim - \frac{1}{(\log N)^2}\\
&D\sim \frac{1}{(\log N)^3}
\label{eq:large_N_v_d}
\eea
This is a remarkable result, indicating that the system approaches the infinite-$N$ limit very slowly. It also shows that at large but finite $N$, the broadening exponent becomes $p=1$ instead of $0$.

In each realization of the noise, $\phi(r,t)$ still grows exponentially. But as $\sqrt{2Dt}$ grows larger than the width of the traveling wave, the exponential growth of $\phi(r,t)$ is smoothed out by the diffusive movement of the wavefront's position, leading to a diffusive broadening of the noise-averaged wavefront. To quantitatively understand the effect of the noise induced by finite $N$ on the wavefront, we approximate the traveling wave solution in a single realization of the noise by the following phenomenological model,
\be
\phi(r,t)=\begin{cases}
    \frac{3}{4}, & \text{if $r<v_B t+r_0+\frac{v_B}{\lambda}\log \frac{4}{3}$}\\
    0, & \text{if $r>v_B t+r_0+\frac{v_B}{\lambda}\log (N)$} \\
     e^{\lambda_L (t-(r-r_0)/v_B)}, & \text{otherwise}\\
  \end{cases}
\ee
which accounts for the saturation behind the wavefront and the growth ahead of the wavefront.

Using this simple model, the noise-averaged squared commutator, which is proportional to the noise-averaged $\phi(r,t)$, is the convolution of $\phi(r,t)$ with a Gaussian distribution describing the diffusive motion,
\bea
C&(r,t)=\frac{1}{\sqrt{4\pi D t}}\int d \Delta x \frac{8}{3}\phi(r+\Delta x,t)e^{-\frac{\Delta x^2}{4Dt}}.\\
\eea
To simplify the notation, we introduce the dimensionless units $\tau=\lambda_L t$, $u=\frac{\lambda_L}{v_B}r$ and $\xi=\frac{D \lambda_L}{v_B^2}$, with $\xi$ describing the strength of the noise. The result of the convolution is

\bea
C(u,\tau)=&\erf\left(\frac{\log(4/3)-z}{
   \sqrt{4\xi  \tau }}\right)+1\\
-&\frac{4}{3} e^{\xi  \tau -z} \erf\left(\frac{\log (4/3)+2\xi  \tau
   -z}{ \sqrt{4\xi  \tau }}\right)\\
  +&\frac{4}{3}e^{\xi  \tau -z} \erf\left(\frac{\log (N)+2\xi  \tau -z}{
   \sqrt{4\xi  \tau }}\right)\\
   \label{eq:largeN_c}
\eea
where $\erf(x)$ is the error function $\frac{2}{\sqrt{\pi}}\int\limits_0^x e^{-t^2}dt$, and $z=u-u_0-\tau$ is the position in the traveling frame with some unimportant offset $u_0$ determined by the initial condition.

The next step is to analyze the behavior of Eq.~\eqref{eq:largeN_c} in space-time. It exhibits a light cone structure with a butterfly velocity independent of $\xi$, since the butterfly velocity is entirely set by the cutoff approximation and does not depend on the diffusion constant explicitly. This can be seen from the fact that $C(\tau,\tau)$ asymptotically approaches $\frac{1}{2}$. The space-time of $t-r$ plane can be approximately divided into three regions based on the behavior of Eq.~\eqref{eq:largeN_c}, as illustrated in Fig.~\ref{fig:spacetime}. The region near the wave-front is the diffusive region, where the last two terms of Eq.~\eqref{eq:largeN_c} roughly cancel each other and $C(u,\tau)$ is dominated by the single error function,
\be
\lim \limits_{\tau\rightarrow\infty}C(\tau+z,\tau)=1+\erf\left(-\frac{z}{\sqrt{4\xi \tau}}\right).
\ee
In the limit that $\sqrt{4\xi\tau}\ll z\ll \tau$, $C(r,t)\sim \exp\left (-\frac{(r-r_0-v_bt)^2}{4D t} \right )$, consistent with the universal form with broadening exponent $p=1$. This clearly demonstrate that the wavefront spreads diffusively. The growth behavior near the wavefront is dominated by the noise, and the original Lyapunov exponents does not enter.

This should be contrasted with the chaotic region where the first two error functions are far away from saturation, but the last error function is already saturated. In this region, the squared commutator is $C(u,\tau)\sim \frac{4}{3}e^{\xi\tau-z}$, a pure growth form with a modified Lyapunov exponent $\lambda_L(1+D\lambda_L/v_B^2)$. This size of this region scales as $\log N$, and the value of $C$ in this region can be arbitrarily small in the long-time limit since it is enclosed by two lines with a bigger velocity $v=v_B(1+2\xi)$ than the butterfly velocity. Therefore, it is difficult to extract this region from numerical data of finite $N$ spin chains.

There is also a third region denoted the perturbative region, where $z$ is the largest scale in the system. In this case the squared commutator is infinitesimally small and behaves as $\frac{8}{3N}\sqrt{\frac{\xi \tau}{\pi z^2}}\exp(-z^2/4\xi \tau)$.

\subsection{The small $N$ limit}
\label{sec:small_N}
The large $N$ analysis presented in the last section cannot be naively generalized to the case of small $N$. In the small $N$ limit, the master equation Eq.~\eqref{eq:master} is still valid,  but the approach of approximating the master equation with the Fokker-Planck equation, Eq.~\eqref{eq:fokker-planck}, to derive the Langevin equation, Eq.~\eqref{eq:langevin}, is not.

Instead, we take a rather different approach by considering the probability of the operator string ending on a specific site at a given time, similar to what was studied in the random brickwork model. This probability $\rho(r,t)$ is defined as
\be
\rho(r,t)= \sum\limits_{\{\mathbf{w}\}} \left (\tilde{h}(\mathbf{w}, t)(1-\delta_{w_r,0})\prod\limits_{s>r} \delta_{w_s,0}\right )
\ee
Note that the sum of $\rho(r,t)$ is conserved.

From Eq.~\eqref{eq:master}, one can derive the rate equation for $\rho(r,t)$ as,
\bea
\partial_t \rho(r,t)=&-\xi \rho(r,t)+\sum\limits_l\gamma^-_b(0,l)\rho_l(r-1,t)\\
           &+\sum\limits_l\gamma^+_b(1,l)\rho_{l1}(r+1,t),
\label{eq:smallN_rho}
\eea
 where a subindex on $\rho$ indicates a restriction: the operator string must end with that particular configuration. For example, $\rho_{l1}(r+1,t)$ is the probability of the operator string with $w=l$ on site $r$, $w=1$ on site $r+1$, and $w=0$ for all sites beyond $r+1$.

Now we use the approximation of local equilibrium, which is crucially different from the large $N$ case, to relate $\rho_l(r,t)$ and $\rho_{l'l}(r,t)$ to $\rho(r)$. 
The approximation of local equilibrium states that all the local Pauli strings instantly have the same probability to appear once the operator front reaches there, as a result of small on-site degrees of freedom. This suggests that $\rho_l(r,t)=\frac{3^l}{4^N-1}{N \choose l} \rho(r,t)$ and $\rho_{l1}(r,t)=\frac{3^l}{4^N}{N \choose l}{\frac{3 N}{4^N-1}}\rho(r,t)$. With this approximation, we obtain a closed equation for $\rho(r,t)$,
\bea
\partial_t \rho(r,t)=&-\xi \rho(r,t)+\frac{9N}{8}\frac{4^N}{4^N-1}g^2 \rho(r-1,t)\\
 &+\frac{9N}{8}\frac{1}{4^N-1}g^2 \rho(r+1,t).
\eea
The conservation law of $\rho(r,t)$ determines that $\xi =\frac{9N}{8}\frac{4^N+1}{4^N-1}$. In the continuum limit, the equation reads
\be
\partial_t \rho(r,t)=-\frac{9N}{8}g^2 \partial_r \rho(r,t) +\frac{9N}{16}\frac{4^N+1}{4^N-1}g^2\partial_r^2 \rho(r,t)
\ee
This leads to
\be
\rho(r,t)=\frac{1}{2\sqrt{\pi Dt}}\exp\left (-\frac{(x-v_Bt)^2}{4Dt}\right )
\ee
with
\bea
v_B&=\frac{9N}{8}g^2, \\
D&=\frac{9N}{16}\frac{4^N+1}{4^N-1}g^2.
\label{eq:smallN_vd}
\eea
The average squared commutator is related to $\rho(r,t)$ as
\be
C(r,t)=2\int _r^\infty \rho(s,t)ds=1+\erf \left(\frac{v_Bt-x}{\sqrt{4Dt}}\right)
\label{eq:smallN_c}
\ee
The final result is consistent with the Haar random circuit result and has broadening exponent $p=1$.

The above analysis relies on the local equilibrium approximation, hold in the small N limit. However, as $N$ increases, the relaxation time $t_{loc}$ increases as well, scaling as $\log N$.  Consequently the Eq. \ref{eq:smallN_c} needs to be modified to incorporate the delay in the relaxation. 
For finite $t_{loc}$, the operator on the right most site never reaches full local equilibrium, and the conditional probability $\rho_l(r,t)/\rho(r,t)$ is always biased towards small $l$  compared with the equilibrium value $\frac{3^l}{4^N-1}{N \choose l} $, with the extreme case $\rho_l(r,t)/\rho(r,t)=\delta_{l,1} $. Therefore, to incorporate the effect of finite $t_{loc}$, we write $\rho_l(r,t)$ as
\bea
\frac{\rho_l(r,t)}{\rho(r,t)} = \left [ \left(\frac{3^l}{4^N-1}{N \choose l}-\delta_{l,1}\right ) (1 - e^{-\Delta t/t_{loc}})+\delta_{l,1}\right ]
\label{eq:finite_scrambling}
\eea
which approaches to the equilibrium distribution as $t_{loc}\rightarrow 0$ and approaches the extreme case as $t_{loc}\rightarrow \infty$. Here $\Delta t $ can be interpreted  as $1/v_B$, the time scale for operator expansion.

Based on Eq. \ref{eq:finite_scrambling}, the summation of $\rho_l$ in Eq. \ref{eq:smallN_rho} can be written as
\bea
&\sum\limits_l \gamma^-_b(0,l)\rho_l(r-1,t)= \\
&\left ( \left (\frac{9N}{8}\frac{4^N}{4^N-1}-\frac{3}{2} \right )g^2 (1-e^{-\Delta t/ t_{loc}}) + \frac{3}{2}g^2 \right )\rho(r-1,t)
\eea
Applying  the similar argument to $\rho_{l1}$, we obtain,
\bea
&\sum\limits_l \gamma^+_b (1, l) \rho_{l1} (r+1, t) \\
&\ \ \ \ \ = \left (\frac{9N}{8}\frac{1}{4^N-1}g^2(1-e^{-\Delta t/ t_{loc}})+\frac{g^2}{2N} \right ) \rho(r+1, t)
\eea
leading to modified $v_B$ and $D$,
\bea
&\tilde{v}_B = v_B (1- e^{-\Delta t/t_{loc }} ) + \left (\frac{3}{2}-\frac{1}{2N} \right )  g^2 e^{-\Delta t/t_{loc}} \\
&\tilde{D} =D (1-e^{-\Delta t/t_{loc}}) + \left (\frac{3}{4}+\frac{1}{4N} \right ) g^2 e^{-\Delta t/t_{loc}}
\eea
The above analysis suggests that Eq. \ref{eq:smallN_vd} overestimates both $v_B$ and $D$ for $N$ larger than 1, and the deviation increases as $N$. We also note  that Eq. \ref{eq:smallN_c} also relies on the assumption of local equilibrium and is in general not valid at finite $N$. To study the operator dynamics and scrambling at finite $N$, we directly simulate the master equation using matrix-product form, as discussed in the next section.  We will see that,  indeed, the results deviate from the prediction in Eq. \ref{eq:smallN_vd} as $N$ increases and gradually crossovers to the large $N$ results discussed in the last section.

\subsection{Finite $N$ results from tensor network simulation}
\label{sec:finite_N}
The large $N$ analysis together the small $N$ analysis convincingly demonstrates that away from infinite $N$, the behavior of the squared commutator is dominated by the error function near the wavefront, leading to a broadening exponent $p=1$. The effect of $N$ is mostly encoded in the butterfly velocity and the diffusion constant. But the $N$-dependence of $v_B$ and $D$ obtained from the two limits are not consistent. The small $N$ analysis suggests that both $v_B$ and $d$ increase with $N$, while the large-$N$ analysis indicates that $v_B$ saturates to a certain value and $D$ decreases with $N$. Therefore it is interesting to study how the two quantities interpolate between the two limits, especially the diffusion constant which is expected to be non-monotonic.

For this purpose, we directly simulate the master equation $\ref{eq:master}$ by representing the probability distribution as a matrix product state (MPS) with physical dimension $N+1$. Comparing with the usual TEBD method \cite{Vidal2003, Vidal2004}, simulating a stochastic process with MPS (S-MPS) \cite{Johnson2010} is quite different.  In the appendix \ref{appsec:SMPS}, we discuss the difference and introduce several useful techniques, including the canonical form for S-MPS along with the truncation schemes that are helpful for preserving the 1-norm of the S-MPS instead of the 2-norm.   Notably the truncation scheme developed by White et. al~\cite{White2017} can be directly applied to here to exactly preserve the 1-norm for all time.

Generally, S-MPS requires higher bond dimension to capture local observables accurately compared with unitary MPS with the same entanglement. This is because the $1$-norm normalization appropriate to S-MPS tends to amplify errors. Nevertheless, the entanglement entropy is still a good measure for determining whether the probability distribution can be represented efficiently as an MPS. Initially, the probability distribution is a product state with operator weight $1$ in the center cluster of system and operator weight $0$ elsewhere. In the early growth region ahead of the lightcone, the probability is not affected by the stochastic evolution and continues to enjoy low entanglement. Inside the lightcone, the probability distribution reaches the steady state where every operator string is equally probable and also admits a simple product state representation. Therefore, the entanglement entropy only accumulates around the wavefront. In practice, we find that the entanglement entropy never exceeds one bit, allowing us to obtain the whole scrambling curve up to $N=10$. The resulting scrambling curve can be fit with the error function with almost perfect quality to extract the butterfly velocity and the diffusion constant. The fitting result is shown in the appendix \ref{appsec:SMPS}.

\begin{figure}
\includegraphics[height=0.49\columnwidth, width=0.49\columnwidth]
{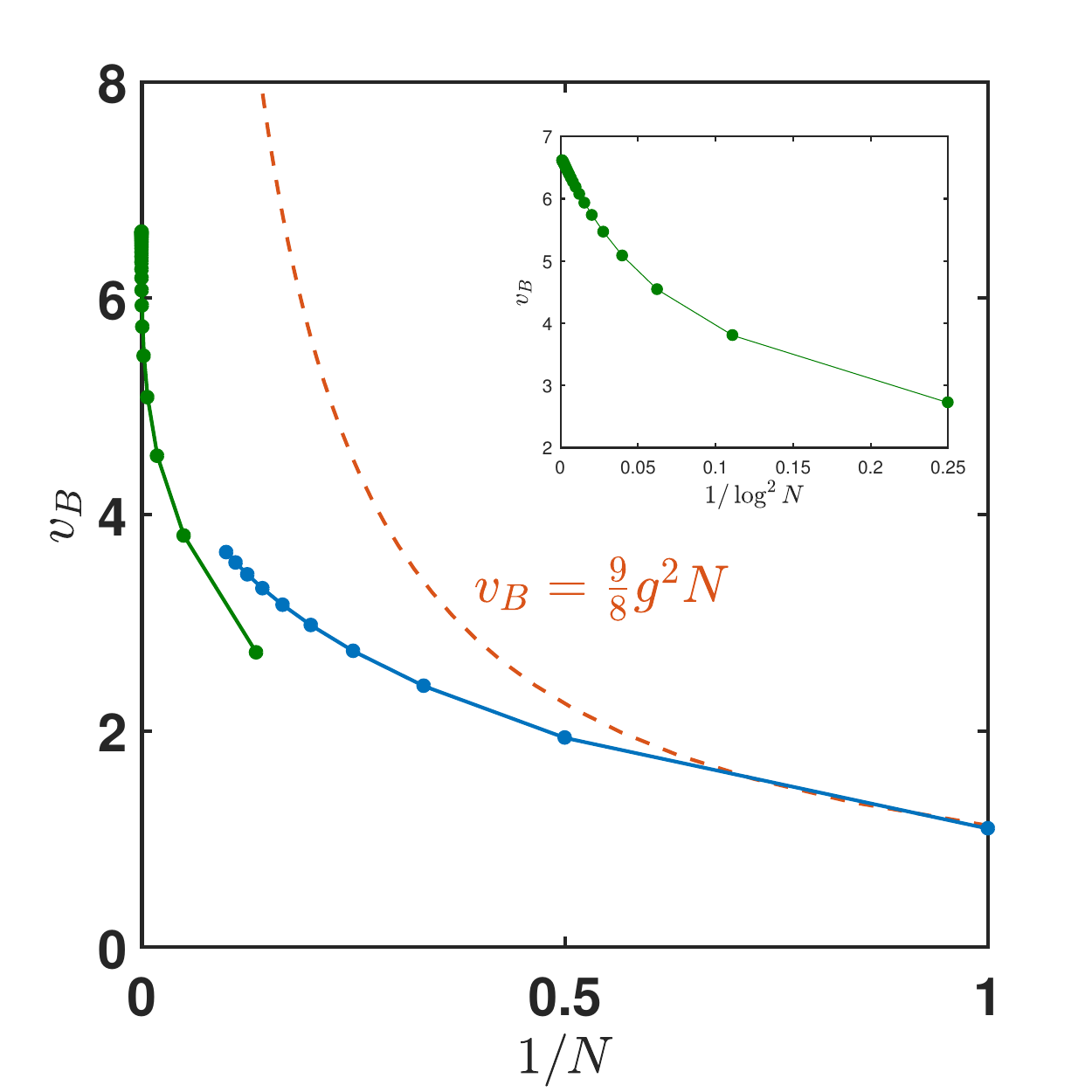}
\includegraphics[height=0.49\columnwidth, width=0.49\columnwidth]
{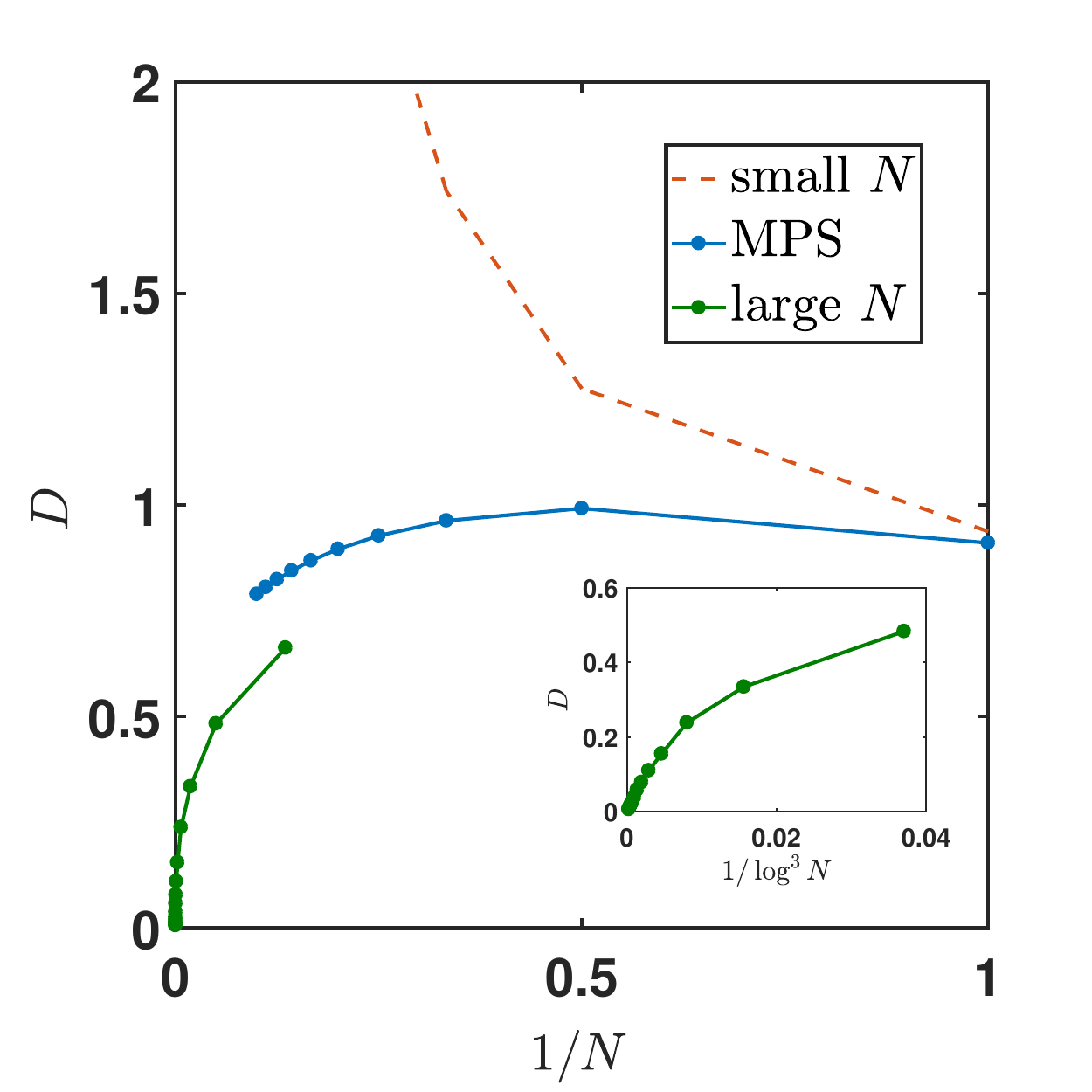}
\caption{Comparison between the butterfly velocity $v_B$ (a) and the wavefront diffusion constant $D$ (b)  obtained from MPS simulation of the probability distribution to the small $N$ and large $N$ analysis. When $N$ is small, $v_B$ and $D$ agree with the small $N$ result relying on the local equilibrium assumption. As $N$ increases, the approximation breaks down, and both $v_B$ and $D$ crossover to the large $N$ result.}
\label{fig:brownian_vd}
\end{figure}

The result of the butterfly velocity is shown in Fig.~\ref{fig:brownian_vd}(a) together with that obtained from small $N$ and large $N$ analysis for $g=1$. We find that at $N=1$ (single spin in the spin cluster), $v_B$ from S-MPS agrees with small $N$ analysis perfectly. As $N$ increases, $v_B$ deviates from the linear growth predicted at small $N$ and smoothly connects to result from the large-$N$ analysis. Fig.~\ref{fig:brownian_vd}(b) summarizes the $N$-dependence of the diffusion constant from the different methods of analysis. The result from the S-MPS indeed exhibits non-monotonic behavior. At $N$=1, it agrees with the small $N$ analysis. It peaks at $N=3$ and drops as $N$ increases, approaching the result from the large-$N$ analysis.
Therefore the numerical results agree with the analyses above from both limits. In principle, for large enough $N$, one should be able to identify the chaotic region.
\subsection{Comparing with Haar random brickwork circuit models}

It is instructive to compare the Brownian circuit model studied here with the previous studied random brickwork circuit models. By designation, in the brickwork circuit, the Haar random unitary matrices equilibrate the operator string on the two sites it connects to immediately if there is non-trivial weight. In this case, the analysis in section \ref{sec:small_N} becomes exact, the entire region ahead the wave-front is governed by the error function. On the other hand, in the BCC model studied here, the operator string takes a finite time to reach equilibrium even locally, the time scale being  $\sim \log N$. The direct consequence is that although near the wave-front, the behavior of the squared commutator is dominated by the error function, there is still a region in space-time ahead of the wavefront where $C(r,t)$ grows exponentially, as illustrated in Fig. \ref{fig:spacetime}. As $N$ increases, this chaotic region expands and finally dominates the wave-front in the infinite-$N$ limit.

\section{Implications for local Hamiltonian systems}
\label{sec:local_ham}

The analysis so far has shown two things. First, that there exists a random circuit model with a parameter $N$ such that, at infinite $N$, the model exhibits exponential growth of the squared commutator with $p=0$. Second, that for any non-infinite $N$, the dynamics of the model inevitably crosses over to a diffusively broadened wavefront with $p=1$. It is quite plausible that any sufficiently generic random circuit model with finite on-site Hilbert space will also exhibit a diffusively broadened wavefront with $p=1$ (with this already being established for the BCC and the random brickwork circuit). The key question is, what aspects of this analysis hold when the couplings are not random in time? We now argue that $p=1$ is generic for chaotic quantum many-body systems in $d=1$ with finite local Hilbert space dimension.

The argument has two thrusts. First, we directly numerically simulate a small $N$ Ising spin chain with conserved energy. Combining large scale numerical simulations with a new analysis technique, our previous result is improved to show that the system asymptotically approaches $p=1$. Second, based on previous work in energy conserving systems showing the existence of noiseless FKPP-type equations governing the spreading of chaos at large $N$ and/or weak coupling model, we argue that quantum fluctuations inevitably introduce multiplicative noise into these equations. The physics of the noisy FKPP equation then naturally leads to $p=1$. For the latter argument, recall that we were careful to distinguish the noise in the $1/N$ corrected FKPP equation, which was a manifestation of quantum fluctuations, from the space-time random couplings in the microscopic Hamiltonian.

Although we focus on energy conserving systems here, we conjecture that our analysis also applies to Floquet models where the couplings are not random in time, but energy is not conserved because the Hamiltonian is time-dependent. In the case of a conserved energy, it also makes sense to talk about non-infinite temperature states. We briefly discuss how the story might be modified in this case, with a focus on the physics of the chaos bound.

\subsection{Wavefront broadening for small $N$ energy conserving systems}
According to the analysis in section \ref{sec:large_N}, for finite on-site Hilbert space dimension, each contour of the squared commutator intersects with the chaotic region for a limited amount of time and eventually merges into the diffusive region. The chaotic regime would suppress wave front broadening, and the diffusive broadening is only clearly visible when the diffusive regime is much larger than the chaotic regime. This suggests a strong finite-size effect that will hinder extraction of the broadening exponent $p$ from fitting the squared commutator with the universal growth form.  This new insight obtained from analyzing the BCC model is consistent with what we observed in our earlier result \cite{Xu2018} from matrix product operator (MPO) calculation of the squared commutator, where the value of $p$ drifts upwards along the contour and the asymptotic value was not addressed.

To unambiguously analyze the broadening of wavefront, we improve our numerical result by pushing both the system size and the simulation time so that the wavefront travel through $\sim 200$ sites. We measure the spatial difference $\delta x$ between two contours of the squared commutator as a function of time. For the two contours we choose, we check that an MPO with bond dimension $\chi=8$ and bond dimension $\chi=16$ give identical results. Then assuming the general form in Eq.~\eqref{eq:broadexp} applies, the asymptotic value of the slope is related to the broadening exponent,
\be
\frac{d \log \delta x}{d \log t} = \frac{p}{p+1}.
\ee

\begin{figure}
\includegraphics[height=0.49\columnwidth, width=0.49\columnwidth]
{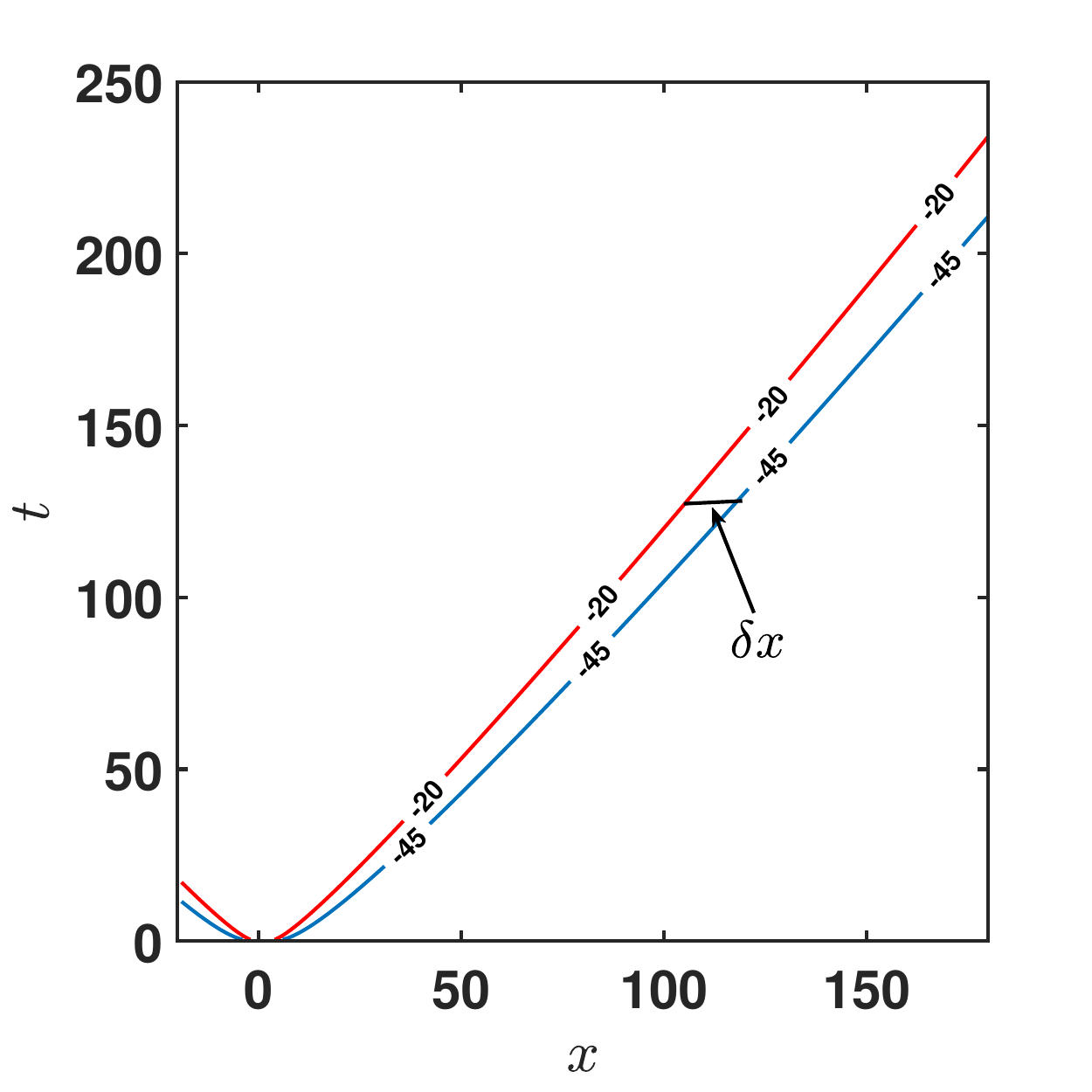}
\includegraphics[height=0.49\columnwidth, width=0.49\columnwidth]
{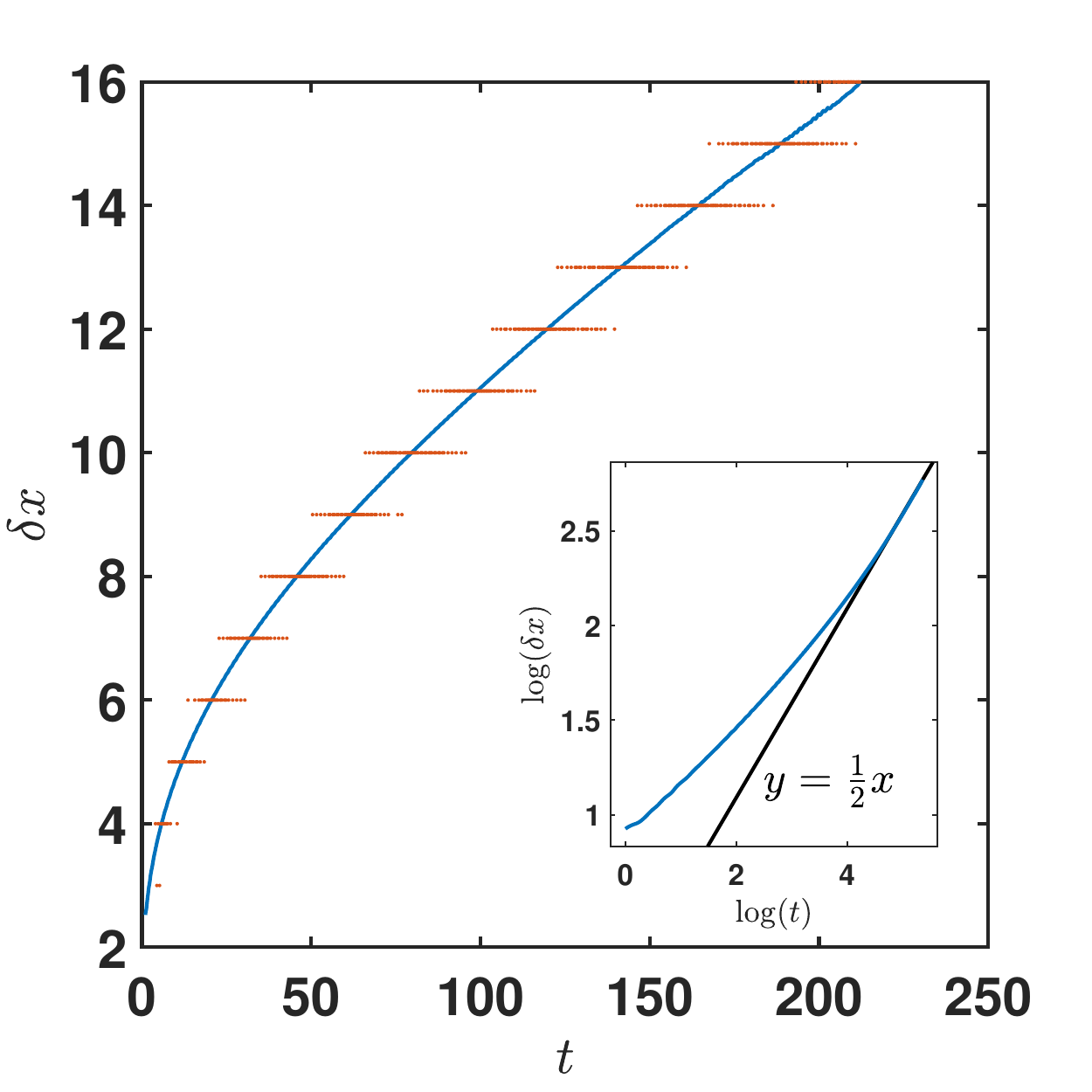}
\caption{The diffusive broadening wave front in the mixed field Ising chain. (a) The contour of $\log C(r,t)=-20$ and $\log C(r,t)=-45$. The system size and the time limit allow the front to travel through $\sim 200$ sites. The contours obtained from MPO simulation with bond dimension 8 and 16 are identical, showing excellent convergence of our method. (b) The spatial distance $\delta x$ between the two contours increases with time, showing definitive broadening of the wave front. In the inset, we plot $\delta x$ with $t$ on a log-log plot. The slope of the curve approaches to $1/2$, a strong evidence of diffusive broadening and the broadening exponent $p=1$.  }
\label{fig:chaotic_broadening}
\end{figure}

We perform this analysis on the mixed-field Ising chaotic spin chain described by the Hamiltonian,
\bea
H=-\frac{1}{E_0}\left ( J \sum\limits_{r=1}^{L} Z_r Z_{r+1} +h_x\sum\limits_{r=1}^L X_r+h_z\sum\limits_{r=1}^L Z_r \right)
\label{eq:ham}
\eea
where $X_r$ and $Z_r$ are local Pauli operators. The parameters are set to $J=1$, $h_x=1.05$ and $h_z=0.5$. The overall normalization factor $E_0=\sqrt{4J^2+2h_x^2+2h_z^2}$.
The result is shown in Fig.~\ref{fig:chaotic_broadening}. The inset of Fig.~\ref{fig:chaotic_broadening}(b) plots $\delta x$ with $t$ on a log-log plot. The slope gradually increases and approaches to $1/2$ in the large space-time limit, indicating that the wave-front broadens diffusively, $p=1$, at the largest sizes and times. The initial deviation may be due to the early-time microscopic physics where $C(r,t)$ behaves as $t^x/x!$. However, the fact that the deviation persists to an intermediate scale suggests that there may exist a chaotic region in the space-time causing a strong finite-size effect. Extracting this region for local Hamiltonian systems is an interesting future research direction.

\begin{figure}
\includegraphics[height=0.49\columnwidth, width=0.49\columnwidth]
{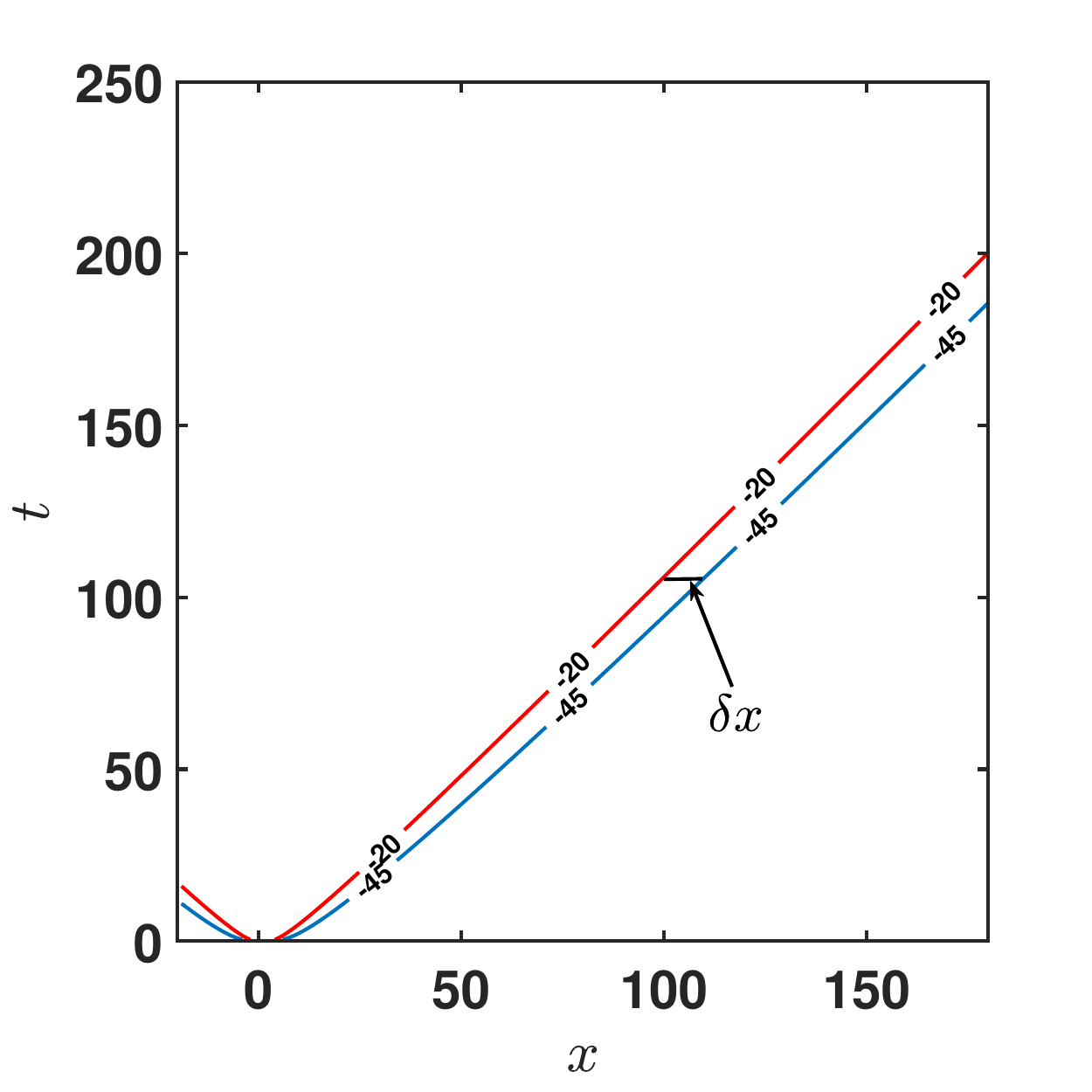}
\includegraphics[height=0.49\columnwidth, width=0.49\columnwidth]
{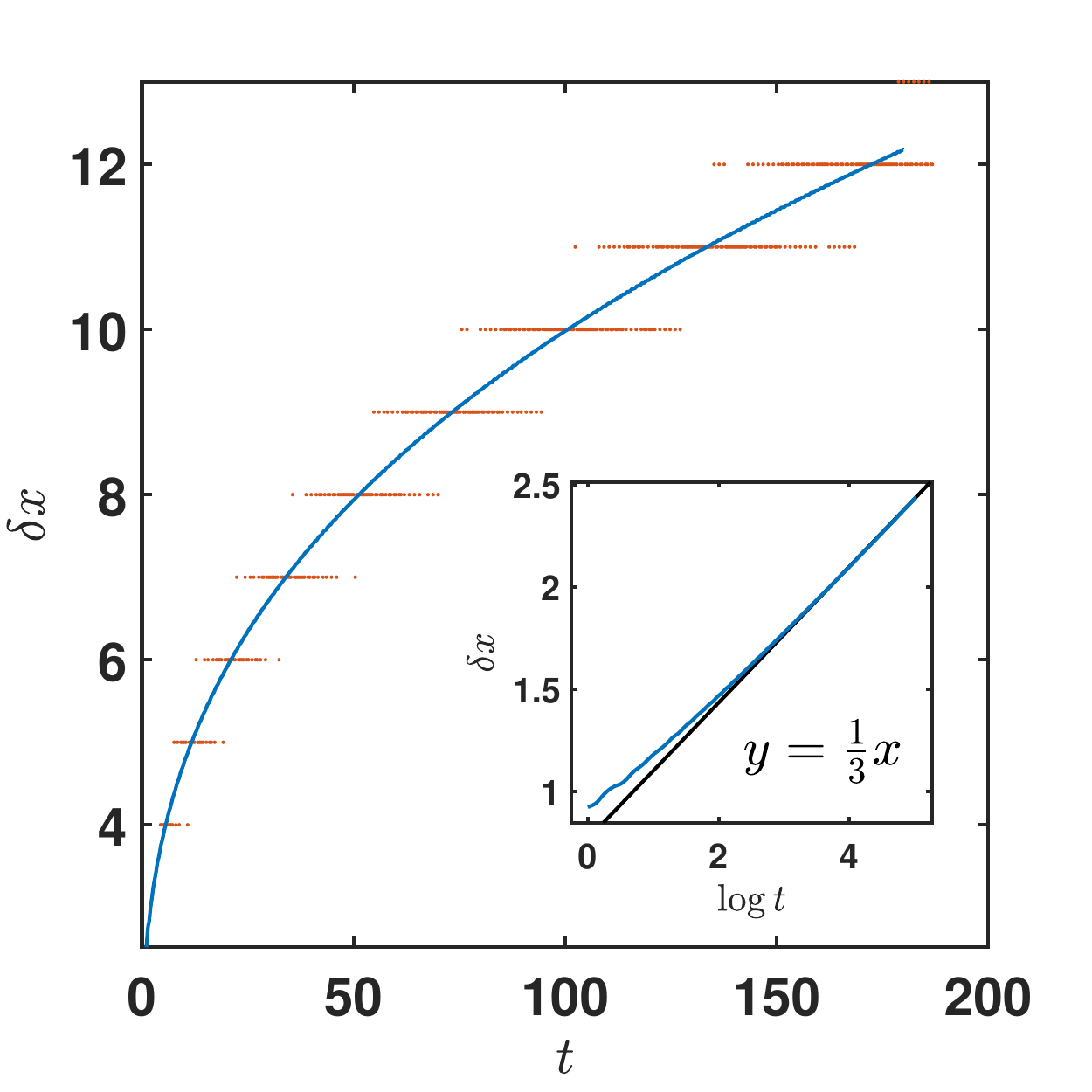}
\caption{The same figure as Fig. \ref{fig:chaotic_broadening} but for transverse-field Ising model describing non-interacting fermions. The slope of the curve on a log-log plot approaches to $1/3$, agreeing with exact results for this model.  }
\label{fig:free_broadening}
\end{figure}

To further validate this method, the same analysis is performed for the transverse-field Ising model (setting $h_z$ to zero) , which describes non-interacting fermions and has a broadening exponent $p=1/2$. The result is shown in Fig.~\ref{fig:free_broadening} where one can indeed see that the slope of curve on a log-log plot approaches to $1/3$.

\subsection{Conjectured wavefront broadening for large $N$ energy conserving systems}

Given that one generic energy conserving model exhibits $p=1$, it is plausible that this is a universal behavior among local chaotic Hamiltonians in one dimension. To bolster this conjecture and to give a physical picture for it in one limit, it is useful to return to the noiseless FKPP equation. Indeed, a number of different models have been shown to have operator growth described by a noiseless FKPP-type equation, in some cases linearized and in some cases fully non-linear, at large $N$ or weak coupling. What we argue is that such equations should inevitably be augmented by a noise term describing quantum fluctuations which has the form considered in this work. This would imply that this broad class of large $N$ models also have $p=1$ at the largest sizes and times. Assuming this is true, $p=1$ then occurs at large and small $N$ and weak and strong coupling, so it is reasonable to conjecture that it is universal property of one-dimensional chaotic systems.

The argument proceeds in three steps of increasing specificity. The background assumptions are that one has a closed dynamical equation governing $\phi \propto C$ and some parameter $N$ which measures the local degrees of freedom. First, quantum fluctuations are expected to add a noise term to any approximate set of closed equations governing the dynamics of the OTOCs of simple operators. This is because operator growth is not deterministic in a quantum system, since a single Heisenberg operator is a superposition of many different complex operator strings. Second, the specific form of the noise term must be multiplicative for local Hamiltonians, meaning proportional to some power of $\phi$. This is because operator growth arises from the failed cancellation between $U$ and $U^\dagger$ due to the insertion of the perturbation $W$. Operator growth never spontaneously occurs far away from the current support of $W(t)$ precisely because $U$ and $U^\dagger$ cancel in far away regions. Third, the form of the multiplicative noise term should be $\sqrt{\phi/N}$. This is necessary to ensure that the noise is most important when $\phi$ is small, which should be true since larger values of $\phi$ are more self-averaging. This also guarantees that the associated Fokker-Planck equation has a sensible $1/N$ expansion, i.e., with no unusual powers of $1/N$ appearing. Then assuming the dynamical equation for $C$ includes saturation effects, we have all the components necessary for the noisy FKPP analysis to apply.

One caveat here is that the analysis is framed in terms of large $N$ models. While FKPP type equations have also been derived in weak coupling approximations, it is less clear how to identify the precise role of the $N$ parameter in that case. One simple intuition is that $N$ should arise because the dynamics is effectively coarse-grained over a long length scale corresponding to the inelastic mean free path. Identifying $N \sim \ell$, with $\ell$ some kind of inelastic mean free path, would then predict butterfly velocity corrections and a diffusion constant going like $\delta v_B \sim 1/\log^2 \ell$ and $D \sim 1/\log^3 \ell$. It would be interesting to better understand the situation at weak coupling.

\subsection{Finite temperature}

Here we make some comments about the dynamics of operator growth at non-infinite $T$. The key point is that the FKPP equation and its noisy counterpart make no particular reference to temperature, so it is reasonable to suppose that they could hold at non-infinite $T$. The noiseless FKPP equation has already been derived at finite $T$ for a variety of models; temperature only enters in so far as the parameters in the FKPP equation are temperature dependent. Thus the results on the BCC from small to large $N$ provide some hints on the interplay between quantum fluctuation and local scrambling.

The manifestation of this interplay is the distinction between the diffusive region and the chaotic region in the space-time structure of the OTOC. In the infinite-$N$ limit, quantum fluctuations are completely suppressed and the local scrambling time $~\sim \log N$ is infinite. In this case, the chaotic region occupies the entire space-time. For large but finite $N$, as soon as quantum fluctuations are present, the wavefront broadens diffusively while the chaotic region occurs ahead of the front. In the small $N$ limit, with the local equilibrium assumption implying that the local scrambling time is short, the diffusive region extends to the entire space-time. This is also consistent with the analytical results on the random brickwork circuit.

Now consider a local Hamiltonian system, say a spin $1/2$ system. Assuming the mixed-field Ising behavior is generic, we showed that at infinite temperature the wavefront also broadens diffusively. But it is difficult to tell whether there exists a chaotic region in addition to the diffusive region ahead of the wavefront due to the microscopic details affecting the behavior of OTOC at early time. On the other hand, at finite temperature and assuming a separation of time-scales, there is a thermally regulated version of the OTOC whose growth rate is bounded by the temperature \cite{Maldacena2016}. If we assume the same bound applies to the non-thermally regulated object\footnote{We know this cannot be true in general, or at least the conditions for the bound to apply are different. The expectation for the thermally regulated object is that the bound applies after a time independent of distance $x$, but this cannot be true for the non-thermally regulated object. This is because the non-thermally regulated object always has a perturbative regime where $C \sim \frac{t^x}{x!}$ which has an unbounded derivative $\frac{d \log C}{dt}$ at large $x$.}, we would have
\bea
\frac{d \log C}{dt}\leq 2\pi T.
\eea

Given this chaos bound, the diffusive broadening form
\bea
C(r,t)=\exp\left(-\frac{(r-v_Bt)^2}{4Dt} \right)
\eea
can only be valid up to finite distance ahead the front, because its growth rate diverges in the large $r$ limit. Imposing the chaos bound, we can estimate maximum distance from the wavefront for which the diffusive behavior is valid. The growth rate of the diffusive growth from is
 \be
 \gamma(r,t)=\frac{v_B}{2D}\left(\frac{r}{t}-v_B\right)+\frac{1}{4D}\left(\frac{r}{t}-v_B\right)^2.
 \ee
Demanding that $\gamma(r,t)\leq 2\pi T$ predicts that the diffusive region can at most persist up to the space-time line given by
 \bea
 \frac{r}{v_B t} = \sqrt{1 + \frac{4 D (2\pi T)}{v_B^2}}.
 \eea

One might also imagine that even at infinite temperature, the logarithmic derivative cannot be too large, for example that it should be bounded by the size of the microscopic couplings. In the BCC, this was indeed true at large $N$ where the crossover from the diffusive to chaotic region was roughly where the rate of growth in the diffusive region was approaching the Lyapunov rate (which itself was of order the microscopic scale). A similar argument might also apply when the system has a Lyapunov exponent less than $2\pi T$. In any event, at finite $T$ there may generically be a region ahead of the wavefront, at least at large but finite $N$ or weak coupling, where the Lyapunov exponent is still visible. One issue is that one could run into the perturbative region before any exponent can be extracted. More generally, one should carefully study the thermally regulated commutators to see the precise consequences of the chaos bound, something we will report on in forthcoming work.

\section{Conclusion and outlook}
In this work, we studied a random time-dependent Hamiltonian model with a large $N$ limit in which we could study in detail the way the infinite $N$ Lyapunov exponent gave way to a diffusively broadened scrambling wavefront. Based on this model and an analysis of large scale MPO simulations in a time-independent Hamiltonian model, we conjectured that the local operator growth wavefront broadens diffusively in generic local chaotic Hamiltonians with finite local Hilbert space dimension. We also showed how a modified stochastic MPS formalism could be used to simulate the operator dynamics for all times after averaging over different Hamiltonian realizations in the random model. A unifying element was the emergence of a noiseless FKPP equation at infinite $N$ and a corresponding noisy FKPP equation at finite $N$. The noise was an effect of quantum fluctuations and ensured that both large $N$ and small $N$ exhibited $p=1$ dynamics.

It is straightforward to extend the BCC model to any dimension, or indeed, to any graph. In higher dimensions, there will still be a Lyapunov exponent and sharp wavefront at infinite $N$. Finite $N$ corrections will then introduce noise into the FKPP-type equation. The analog of the cutoff on $\phi$ is an extended cutoff front where $\phi=1/N$. This front will then experience some random dynamics with a constant drift (the butterfly velocity) and noisy local dynamics. Although the general long-distance structure may be complex, e.g., in high dimensions the noise may be relevant or irrelevant, one expects KPZ-like dynamics in low dimensions. It will be interesting to analyze the higher dimensional case in more detail, and possibly also study the model on more general graphs.

In terms of future directions, we have several works in progress. One is to consider the effect of a conserved $U(1)$ symmetry on the operator spreading. This has already been studied in the random brickwork circuit \cite{Rakovszky2017, Khemani2017}, but we anticipate new interesting physics associated with the interplay with the large $N$ effects. Another is a study of the entanglement dynamics in the model as a function of $N$. Another interesting direction is to directly obtaining the noise physics in a Hamiltonian large $N$ model.

\section{Acknowledgment}
S. Xu and B. Swingle thank S. Sahu, D. Huse, and E. Brunet for interesting discussions. This material is based upon work supported by the Simons Foundation via the It From Qubit
Collaboration, by the Air Force Office of Scientific Research under award number FA9550-17-1-0180, and by the NSF Physics Frontier Center at the Joint Quantum Institute (PHY-1430094).

\bibliography{scrambling,hydrodynamics,tensor}

\begin{thebibliography}{60}%
\makeatletter
\providecommand \@ifxundefined [1]{%
 \@ifx{#1\undefined}
}%
\providecommand \@ifnum [1]{%
 \ifnum #1\expandafter \@firstoftwo
 \else \expandafter \@secondoftwo
 \fi
}%
\providecommand \@ifx [1]{%
 \ifx #1\expandafter \@firstoftwo
 \else \expandafter \@secondoftwo
 \fi
}%
\providecommand \natexlab [1]{#1}%
\providecommand \enquote  [1]{``#1''}%
\providecommand \bibnamefont  [1]{#1}%
\providecommand \bibfnamefont [1]{#1}%
\providecommand \citenamefont [1]{#1}%
\providecommand \href@noop [0]{\@secondoftwo}%
\providecommand \href [0]{\begingroup \@sanitize@url \@href}%
\providecommand \@href[1]{\@@startlink{#1}\@@href}%
\providecommand \@@href[1]{\endgroup#1\@@endlink}%
\providecommand \@sanitize@url [0]{\catcode `\\12\catcode `\$12\catcode
  `\&12\catcode `\#12\catcode `\^12\catcode `\_12\catcode `\%12\relax}%
\providecommand \@@startlink[1]{}%
\providecommand \@@endlink[0]{}%
\providecommand \url  [0]{\begingroup\@sanitize@url \@url }%
\providecommand \@url [1]{\endgroup\@href {#1}{\urlprefix }}%
\providecommand \urlprefix  [0]{URL }%
\providecommand \Eprint [0]{\href }%
\providecommand \doibase [0]{https://doi.org/}%
\providecommand \selectlanguage [0]{\@gobble}%
\providecommand \bibinfo  [0]{\@secondoftwo}%
\providecommand \bibfield  [0]{\@secondoftwo}%
\providecommand \translation [1]{[#1]}%
\providecommand \BibitemOpen [0]{}%
\providecommand \bibitemStop [0]{}%
\providecommand \bibitemNoStop [0]{.\EOS\space}%
\providecommand \EOS [0]{\spacefactor3000\relax}%
\providecommand \BibitemShut  [1]{\csname bibitem#1\endcsname}%
\let\auto@bib@innerbib\@empty
\bibitem [{\citenamefont {Hayden}\ and\ \citenamefont
  {Preskill}(2007)}]{Hayden2007}%
  \BibitemOpen
  \bibfield  {author} {\bibinfo {author} {\bibfnamefont {P.}~\bibnamefont
  {Hayden}}\ and\ \bibinfo {author} {\bibfnamefont {J.}~\bibnamefont
  {Preskill}},\ }\bibfield  {title} {\bibinfo {title} {{Black holes as mirrors:
  quantum information in random subsystems}},\ }\href
  {http://arxiv.org/abs/0708.4025
  http://dx.doi.org/10.1088/1126-6708/2007/09/120
  http://stacks.iop.org/1126-6708/2007/i=09/a=120?key=crossref.eca9390b41dd1fa1b67fb6e27f9f17c0}
  {\bibfield  {journal} {\bibinfo  {journal} {JHEP}\ }\textbf {\bibinfo
  {volume} {2007}}\bibinfo  {number} { (09)},\ \bibinfo {pages}
  {120}}\BibitemShut {NoStop}%
\bibitem [{\citenamefont {Sekino}\ and\ \citenamefont
  {Susskind}(2008)}]{Sekino2008}%
  \BibitemOpen
\bibfield  {number} {  }\bibfield  {author} {\bibinfo {author} {\bibfnamefont
  {Y.}~\bibnamefont {Sekino}}\ and\ \bibinfo {author} {\bibfnamefont
  {L.}~\bibnamefont {Susskind}},\ }\bibfield  {title} {\bibinfo {title} {{Fast
  scramblers}},\ }\href {http://arxiv.org/abs/0808.2096
  http://dx.doi.org/10.1088/1126-6708/2008/10/065
  http://stacks.iop.org/1126-6708/2008/i=10/a=065?key=crossref.0eebed9ef6bfd908b2747d054e345a56}
  {\bibfield  {journal} {\bibinfo  {journal} {JHEP}\ }\textbf {\bibinfo
  {volume} {2008}}\bibinfo  {number} { (10)},\ \bibinfo {pages}
  {065}}\BibitemShut {NoStop}%
\bibitem [{\citenamefont {Shenker}\ and\ \citenamefont
  {Stanford}(2014)}]{Shenker2014}%
  \BibitemOpen
\bibfield  {number} {  }\bibfield  {author} {\bibinfo {author} {\bibfnamefont
  {S.~H.}\ \bibnamefont {Shenker}}\ and\ \bibinfo {author} {\bibfnamefont
  {D.}~\bibnamefont {Stanford}},\ }\bibfield  {title} {\bibinfo {title} {{Black
  holes and the butterfly effect}},\ }\href
  {http://link.springer.com/10.1007/JHEP03(2014)067} {\bibfield  {journal}
  {\bibinfo  {journal} {JHEP}\ }\textbf {\bibinfo {volume} {2014}}\bibinfo
  {number} { (3)},\ \bibinfo {pages} {67}}\BibitemShut {NoStop}%
\bibitem [{\citenamefont {Hosur}\ \emph {et~al.}(2016)\citenamefont {Hosur},
  \citenamefont {Qi}, \citenamefont {Roberts},\ and\ \citenamefont
  {Yoshida}}]{Hosur2016}%
  \BibitemOpen
\bibfield  {number} {  }\bibfield  {author} {\bibinfo {author} {\bibfnamefont
  {P.}~\bibnamefont {Hosur}}, \bibinfo {author} {\bibfnamefont {X.-L.}\
  \bibnamefont {Qi}}, \bibinfo {author} {\bibfnamefont {D.~A.}\ \bibnamefont
  {Roberts}},\ and\ \bibinfo {author} {\bibfnamefont {B.}~\bibnamefont
  {Yoshida}},\ }\bibfield  {title} {\bibinfo {title} {{Chaos in quantum
  channels}},\ }\href {http://arxiv.org/abs/1511.04021
  http://dx.doi.org/10.1007/JHEP02(2016)004
  http://link.springer.com/10.1007/JHEP02(2016)004} {\bibfield  {journal}
  {\bibinfo  {journal} {JHEP}\ }\textbf {\bibinfo {volume} {2016}}\bibinfo
  {number} { (2)},\ \bibinfo {pages} {4}}\BibitemShut {NoStop}%
\bibitem [{\citenamefont {Deutsch}(1991)}]{Deutsch1991}%
  \BibitemOpen
\bibfield  {number} {  }\bibfield  {author} {\bibinfo {author} {\bibfnamefont
  {J.~M.}\ \bibnamefont {Deutsch}},\ }\bibfield  {title} {\bibinfo {title}
  {{Quantum statistical mechanics in a closed system}},\ }\href
  {https://link.aps.org/doi/10.1103/PhysRevA.43.2046} {\bibfield  {journal}
  {\bibinfo  {journal} {Phys. Rev. A}\ }\textbf {\bibinfo {volume} {43}},\
  \bibinfo {pages} {2046} (\bibinfo {year} {1991})}\BibitemShut {NoStop}%
\bibitem [{\citenamefont {Srednicki}(1994)}]{Srednicki1994}%
  \BibitemOpen
  \bibfield  {author} {\bibinfo {author} {\bibfnamefont {M.}~\bibnamefont
  {Srednicki}},\ }\bibfield  {title} {\bibinfo {title} {{Chaos and quantum
  thermalization}},\ }\href {https://link.aps.org/doi/10.1103/PhysRevE.50.888}
  {\bibfield  {journal} {\bibinfo  {journal} {Phys. Rev. E}\ }\textbf {\bibinfo
  {volume} {50}},\ \bibinfo {pages} {888} (\bibinfo {year} {1994})}\BibitemShut
  {NoStop}%
\bibitem [{\citenamefont {Tasaki}(1998)}]{Tasaki1998a}%
  \BibitemOpen
  \bibfield  {author} {\bibinfo {author} {\bibfnamefont {H.}~\bibnamefont
  {Tasaki}},\ }\bibfield  {title} {\bibinfo {title} {{From Quantum Dynamics to
  the Canonical Distribution: General Picture and a Rigorous Example}},\ }\href
  {https://link.aps.org/doi/10.1103/PhysRevLett.80.1373
  http://arxiv.org/abs/cond-mat/9707253
  http://dx.doi.org/10.1103/PhysRevLett.80.1373} {\bibfield  {journal}
  {\bibinfo  {journal} {Phys. Rev. Lett.}\ }\textbf {\bibinfo {volume} {80}},\
  \bibinfo {pages} {1373} (\bibinfo {year} {1998})}\BibitemShut {NoStop}%
\bibitem [{\citenamefont {Rigol}\ \emph {et~al.}(2008)\citenamefont {Rigol},
  \citenamefont {Dunjko},\ and\ \citenamefont {Olshanii}}]{Rigol2008}%
  \BibitemOpen
  \bibfield  {author} {\bibinfo {author} {\bibfnamefont {M.}~\bibnamefont
  {Rigol}}, \bibinfo {author} {\bibfnamefont {V.}~\bibnamefont {Dunjko}},\ and\
  \bibinfo {author} {\bibfnamefont {M.}~\bibnamefont {Olshanii}},\ }\bibfield
  {title} {\bibinfo {title} {{Thermalization and its mechanism for generic
  isolated quantum systems}},\ }\href
  {http://www.nature.com/doifinder/10.1038/nature06838} {\bibfield  {journal}
  {\bibinfo  {journal} {Nature}\ }\textbf {\bibinfo {volume} {452}},\ \bibinfo
  {pages} {854} (\bibinfo {year} {2008})}\BibitemShut {NoStop}%
\bibitem [{\citenamefont {Nahum}\ \emph {et~al.}(2018)\citenamefont {Nahum},
  \citenamefont {Vijay},\ and\ \citenamefont {Haah}}]{Nahum2017a}%
  \BibitemOpen
  \bibfield  {author} {\bibinfo {author} {\bibfnamefont {A.}~\bibnamefont
  {Nahum}}, \bibinfo {author} {\bibfnamefont {S.}~\bibnamefont {Vijay}},\ and\
  \bibinfo {author} {\bibfnamefont {J.}~\bibnamefont {Haah}},\ }\bibfield
  {title} {\bibinfo {title} {{Operator Spreading in Random Unitary Circuits}},\
  }\href {http://arxiv.org/abs/1705.08975
  http://dx.doi.org/10.1103/PhysRevX.8.021014
  https://link.aps.org/doi/10.1103/PhysRevX.8.021014} {\bibfield  {journal}
  {\bibinfo  {journal} {Phys. Rev. X}\ }\textbf {\bibinfo {volume} {8}},\
  \bibinfo {pages} {021014} (\bibinfo {year} {2018})}\BibitemShut {NoStop}%
\bibitem [{\citenamefont {von Keyserlingk}\ \emph {et~al.}(2018)\citenamefont
  {von Keyserlingk}, \citenamefont {Rakovszky}, \citenamefont {Pollmann},\ and\
  \citenamefont {Sondhi}}]{VonKeyserlingk2017}%
  \BibitemOpen
  \bibfield  {author} {\bibinfo {author} {\bibfnamefont {C.~W.}\ \bibnamefont
  {von Keyserlingk}}, \bibinfo {author} {\bibfnamefont {T.}~\bibnamefont
  {Rakovszky}}, \bibinfo {author} {\bibfnamefont {F.}~\bibnamefont
  {Pollmann}},\ and\ \bibinfo {author} {\bibfnamefont {S.~L.}\ \bibnamefont
  {Sondhi}},\ }\bibfield  {title} {\bibinfo {title} {{Operator Hydrodynamics,
  OTOCs, and Entanglement Growth in Systems without Conservation Laws}},\
  }\href {http://arxiv.org/abs/1705.08910
  https://link.aps.org/doi/10.1103/PhysRevX.8.021013} {\bibfield  {journal}
  {\bibinfo  {journal} {Phys. Rev. X}\ }\textbf {\bibinfo {volume} {8}},\
  \bibinfo {pages} {021013} (\bibinfo {year} {2018})}\BibitemShut {NoStop}%
\bibitem [{\citenamefont {Xu}\ and\ \citenamefont {Swingle}(2018)}]{Xu2018}%
  \BibitemOpen
  \bibfield  {author} {\bibinfo {author} {\bibfnamefont {S.}~\bibnamefont
  {Xu}}\ and\ \bibinfo {author} {\bibfnamefont {B.}~\bibnamefont {Swingle}},\
  }\bibfield  {title} {\bibinfo {title} {{Accessing scrambling using matrix
  product operators}},\ }\href {http://arxiv.org/abs/1802.00801} {\bibfield
  {journal} {\bibinfo  {journal} {arXiv:1802.00801}\ } (\bibinfo {year}
  {2018})}\BibitemShut {NoStop}%
\bibitem [{\citenamefont {Roberts}\ \emph {et~al.}(2018)\citenamefont
  {Roberts}, \citenamefont {Stanford},\ and\ \citenamefont
  {Streicher}}]{Roberts2018}%
  \BibitemOpen
  \bibfield  {author} {\bibinfo {author} {\bibfnamefont {D.~A.}\ \bibnamefont
  {Roberts}}, \bibinfo {author} {\bibfnamefont {D.}~\bibnamefont {Stanford}},\
  and\ \bibinfo {author} {\bibfnamefont {A.}~\bibnamefont {Streicher}},\
  }\bibfield  {title} {\bibinfo {title} {{Operator growth in the SYK model}},\
  }\href {http://arxiv.org/abs/1802.02633
  http://link.springer.com/10.1007/JHEP06(2018)122} {\bibfield  {journal}
  {\bibinfo  {journal} {JHEP}\ }\textbf {\bibinfo {volume} {2018}}\bibinfo
  {number} { (6)},\ \bibinfo {pages} {122}}\BibitemShut {NoStop}%
\bibitem [{\citenamefont {Jonay}\ \emph {et~al.}(2018)\citenamefont {Jonay},
  \citenamefont {Huse},\ and\ \citenamefont {Nahum}}]{Jonay2018}%
  \BibitemOpen
\bibfield  {number} {  }\bibfield  {author} {\bibinfo {author} {\bibfnamefont
  {C.}~\bibnamefont {Jonay}}, \bibinfo {author} {\bibfnamefont {D.~A.}\
  \bibnamefont {Huse}},\ and\ \bibinfo {author} {\bibfnamefont
  {A.}~\bibnamefont {Nahum}},\ }\bibfield  {title} {\bibinfo {title}
  {{Coarse-grained dynamics of operator and state entanglement}},\ }\href
  {https://arxiv.org/pdf/1803.00089.pdf http://arxiv.org/abs/1803.00089}
  {\bibfield  {journal} {\bibinfo  {journal} {arXiv:1803.00089}\ } (\bibinfo
  {year} {2018})}\BibitemShut {NoStop}%
\bibitem [{\citenamefont {Mezei}(2018)}]{Mezei2018}%
  \BibitemOpen
  \bibfield  {author} {\bibinfo {author} {\bibfnamefont {M.}~\bibnamefont
  {Mezei}},\ }\bibfield  {title} {\bibinfo {title} {{Membrane theory of
  entanglement dynamics from holography}},\ }\href
  {https://arxiv.org/abs/1803.10244 http://arxiv.org/abs/1803.10244
  http://dx.doi.org/10.1103/PhysRevD.98.106025
  https://link.aps.org/doi/10.1103/PhysRevD.98.106025} {\bibfield  {journal}
  {\bibinfo  {journal} {Phys. Rev. D}\ }\textbf {\bibinfo {volume} {98}},\
  \bibinfo {pages} {106025} (\bibinfo {year} {2018})}\BibitemShut {NoStop}%
\bibitem [{\citenamefont {You}\ and\ \citenamefont {Gu}(2018)}]{You2018}%
  \BibitemOpen
  \bibfield  {author} {\bibinfo {author} {\bibfnamefont {Y.-Z.}\ \bibnamefont
  {You}}\ and\ \bibinfo {author} {\bibfnamefont {Y.}~\bibnamefont {Gu}},\
  }\bibfield  {title} {\bibinfo {title} {{Entanglement features of random
  Hamiltonian dynamics}},\ }\href {https://arxiv.org/pdf/1803.10425.pdf
  http://arxiv.org/abs/1803.10425
  https://link.aps.org/doi/10.1103/PhysRevB.98.014309} {\bibfield  {journal}
  {\bibinfo  {journal} {Phys. Rev. B}\ }\textbf {\bibinfo {volume} {98}},\
  \bibinfo {pages} {014309} (\bibinfo {year} {2018})}\BibitemShut {NoStop}%
\bibitem [{\citenamefont {Chen}\ and\ \citenamefont {Zhou}(2018)}]{Chen2018}%
  \BibitemOpen
  \bibfield  {author} {\bibinfo {author} {\bibfnamefont {X.}~\bibnamefont
  {Chen}}\ and\ \bibinfo {author} {\bibfnamefont {T.}~\bibnamefont {Zhou}},\
  }\bibfield  {title} {\bibinfo {title} {{Operator scrambling and quantum
  chaos}},\ }\href {http://arxiv.org/abs/1804.08655} {\bibfield  {journal}
  {\bibinfo  {journal} {arXiv:1804.08655}\ } (\bibinfo {year}
  {2018})}\BibitemShut {NoStop}%
\bibitem [{\citenamefont {Swingle}\ \emph {et~al.}(2016)\citenamefont
  {Swingle}, \citenamefont {Bentsen}, \citenamefont {Schleier-Smith},\ and\
  \citenamefont {Hayden}}]{swingle2016}%
  \BibitemOpen
  \bibfield  {author} {\bibinfo {author} {\bibfnamefont {B.}~\bibnamefont
  {Swingle}}, \bibinfo {author} {\bibfnamefont {G.}~\bibnamefont {Bentsen}},
  \bibinfo {author} {\bibfnamefont {M.}~\bibnamefont {Schleier-Smith}},\ and\
  \bibinfo {author} {\bibfnamefont {P.}~\bibnamefont {Hayden}},\ }\bibfield
  {title} {\bibinfo {title} {{Measuring the scrambling of quantum
  information}},\ }\href {https://arxiv.org/pdf/1602.06271.pdf
  https://link.aps.org/doi/10.1103/PhysRevA.94.040302} {\bibfield  {journal}
  {\bibinfo  {journal} {Phys. Rev. A}\ }\textbf {\bibinfo {volume} {94}},\
  \bibinfo {pages} {040302} (\bibinfo {year} {2016})}\BibitemShut {NoStop}%
\bibitem [{\citenamefont {Zhu}\ \emph {et~al.}(2016)\citenamefont {Zhu},
  \citenamefont {Hafezi},\ and\ \citenamefont {Grover}}]{Zhu2016}%
  \BibitemOpen
  \bibfield  {author} {\bibinfo {author} {\bibfnamefont {G.}~\bibnamefont
  {Zhu}}, \bibinfo {author} {\bibfnamefont {M.}~\bibnamefont {Hafezi}},\ and\
  \bibinfo {author} {\bibfnamefont {T.}~\bibnamefont {Grover}},\ }\bibfield
  {title} {\bibinfo {title} {{Measurement of many-body chaos using a quantum
  clock}},\ }\href {https://link.aps.org/doi/10.1103/PhysRevA.94.062329}
  {\bibfield  {journal} {\bibinfo  {journal} {Phys. Rev. A}\ }\textbf {\bibinfo
  {volume} {94}},\ \bibinfo {pages} {062329} (\bibinfo {year}
  {2016})}\BibitemShut {NoStop}%
\bibitem [{\citenamefont {Yao}\ \emph {et~al.}(2016)\citenamefont {Yao},
  \citenamefont {Grusdt}, \citenamefont {Swingle}, \citenamefont {Lukin},
  \citenamefont {Stamper-Kurn}, \citenamefont {Moore},\ and\ \citenamefont
  {Demler}}]{Yao2016a}%
  \BibitemOpen
  \bibfield  {author} {\bibinfo {author} {\bibfnamefont {N.~Y.}\ \bibnamefont
  {Yao}}, \bibinfo {author} {\bibfnamefont {F.}~\bibnamefont {Grusdt}},
  \bibinfo {author} {\bibfnamefont {B.}~\bibnamefont {Swingle}}, \bibinfo
  {author} {\bibfnamefont {M.~D.}\ \bibnamefont {Lukin}}, \bibinfo {author}
  {\bibfnamefont {D.~M.}\ \bibnamefont {Stamper-Kurn}}, \bibinfo {author}
  {\bibfnamefont {J.~E.}\ \bibnamefont {Moore}},\ and\ \bibinfo {author}
  {\bibfnamefont {E.~A.}\ \bibnamefont {Demler}},\ }\bibfield  {title}
  {\bibinfo {title} {{Interferometric Approach to Probing Fast Scrambling}},\
  }\href {http://arxiv.org/abs/1607.01801} {\bibfield  {journal} {\bibinfo
  {journal} {arXiv:1607.01801}\ } (\bibinfo {year} {2016})}\BibitemShut
  {NoStop}%
\bibitem [{\citenamefont {{Yunger Halpern}}(2017)}]{Halpern2016}%
  \BibitemOpen
  \bibfield  {author} {\bibinfo {author} {\bibfnamefont {N.}~\bibnamefont
  {{Yunger Halpern}}},\ }\bibfield  {title} {\bibinfo {title} {{Jarzynski-like
  equality for the out-of-time-ordered correlator}},\ }\href
  {http://arxiv.org/abs/1609.00015 http://dx.doi.org/10.1103/PhysRevA.95.012120
  https://link.aps.org/doi/10.1103/PhysRevA.95.012120} {\bibfield  {journal}
  {\bibinfo  {journal} {Phys. Rev. A}\ }\textbf {\bibinfo {volume} {95}},\
  \bibinfo {pages} {012120} (\bibinfo {year} {2017})}\BibitemShut {NoStop}%
\bibitem [{\citenamefont {{Yunger Halpern}}\ \emph {et~al.}(2018)\citenamefont
  {{Yunger Halpern}}, \citenamefont {Swingle},\ and\ \citenamefont
  {Dressel}}]{Halpern2017}%
  \BibitemOpen
  \bibfield  {author} {\bibinfo {author} {\bibfnamefont {N.}~\bibnamefont
  {{Yunger Halpern}}}, \bibinfo {author} {\bibfnamefont {B.}~\bibnamefont
  {Swingle}},\ and\ \bibinfo {author} {\bibfnamefont {J.}~\bibnamefont
  {Dressel}},\ }\bibfield  {title} {\bibinfo {title} {{Quasiprobability behind
  the out-of-time-ordered correlator}},\ }\href
  {http://arxiv.org/abs/1704.01971
  https://link.aps.org/doi/10.1103/PhysRevA.97.042105} {\bibfield  {journal}
  {\bibinfo  {journal} {Phys. Rev. A}\ }\textbf {\bibinfo {volume} {97}},\
  \bibinfo {pages} {042105} (\bibinfo {year} {2018})}\BibitemShut {NoStop}%
\bibitem [{\citenamefont {Campisi}\ and\ \citenamefont
  {Goold}(2017)}]{Campisi2017}%
  \BibitemOpen
  \bibfield  {author} {\bibinfo {author} {\bibfnamefont {M.}~\bibnamefont
  {Campisi}}\ and\ \bibinfo {author} {\bibfnamefont {J.}~\bibnamefont
  {Goold}},\ }\bibfield  {title} {\bibinfo {title} {{Thermodynamics of quantum
  information scrambling}},\ }\href {http://arxiv.org/abs/1609.05848
  http://dx.doi.org/10.1103/PhysRevE.95.062127
  http://link.aps.org/doi/10.1103/PhysRevE.95.062127} {\bibfield  {journal}
  {\bibinfo  {journal} {Phys. Rev. E}\ }\textbf {\bibinfo {volume} {95}},\
  \bibinfo {pages} {062127} (\bibinfo {year} {2017})}\BibitemShut {NoStop}%
\bibitem [{\citenamefont {Yoshida}\ and\ \citenamefont
  {Kitaev}(2017)}]{Yoshida2017}%
  \BibitemOpen
  \bibfield  {author} {\bibinfo {author} {\bibfnamefont {B.}~\bibnamefont
  {Yoshida}}\ and\ \bibinfo {author} {\bibfnamefont {A.}~\bibnamefont
  {Kitaev}},\ }\bibfield  {title} {\bibinfo {title} {{Efficient decoding for
  the Hayden-Preskill protocol}},\ }\href {http://arxiv.org/abs/1710.03363}
  {\bibfield  {journal} {\bibinfo  {journal} {arXiv:1710.03363}\ } (\bibinfo
  {year} {2017})}\BibitemShut {NoStop}%
\bibitem [{\citenamefont {Garttner}\ \emph {et~al.}(2017)\citenamefont
  {Garttner}, \citenamefont {Bohnet}, \citenamefont {Safavi-Naini},
  \citenamefont {Wall}, \citenamefont {Bollinger},\ and\ \citenamefont
  {Rey}}]{Garttner2016}%
  \BibitemOpen
  \bibfield  {author} {\bibinfo {author} {\bibfnamefont {M.}~\bibnamefont
  {Garttner}}, \bibinfo {author} {\bibfnamefont {J.~G.}\ \bibnamefont
  {Bohnet}}, \bibinfo {author} {\bibfnamefont {A.}~\bibnamefont
  {Safavi-Naini}}, \bibinfo {author} {\bibfnamefont {M.~L.}\ \bibnamefont
  {Wall}}, \bibinfo {author} {\bibfnamefont {J.~J.}\ \bibnamefont
  {Bollinger}},\ and\ \bibinfo {author} {\bibfnamefont {A.~M.}\ \bibnamefont
  {Rey}},\ }\bibfield  {title} {\bibinfo {title} {{Measuring out-of-time-order
  correlations and multiple quantum spectra in a trapped-ion quantum magnet}},\
  }\href {http://arxiv.org/abs/1608.08938 http://dx.doi.org/10.1038/nphys4119}
  {\bibfield  {journal} {\bibinfo  {journal} {Nat. Phys.}\ }\textbf {\bibinfo
  {volume} {13}},\ \bibinfo {pages} {781} (\bibinfo {year} {2017})}\BibitemShut
  {NoStop}%
\bibitem [{\citenamefont {Wei}\ \emph {et~al.}(2018)\citenamefont {Wei},
  \citenamefont {Ramanathan},\ and\ \citenamefont {Cappellaro}}]{Wei2016}%
  \BibitemOpen
  \bibfield  {author} {\bibinfo {author} {\bibfnamefont {K.~X.}\ \bibnamefont
  {Wei}}, \bibinfo {author} {\bibfnamefont {C.}~\bibnamefont {Ramanathan}},\
  and\ \bibinfo {author} {\bibfnamefont {P.}~\bibnamefont {Cappellaro}},\
  }\bibfield  {title} {\bibinfo {title} {{Exploring Localization in Nuclear
  Spin Chains}},\ }\href {http://arxiv.org/abs/1612.05249
  https://link.aps.org/doi/10.1103/PhysRevLett.120.070501} {\bibfield
  {journal} {\bibinfo  {journal} {Phys. Rev. Lett.}\ }\textbf {\bibinfo
  {volume} {120}},\ \bibinfo {pages} {070501} (\bibinfo {year}
  {2018})}\BibitemShut {NoStop}%
\bibitem [{\citenamefont {Li}\ \emph {et~al.}(2017)\citenamefont {Li},
  \citenamefont {Fan}, \citenamefont {Wang}, \citenamefont {Ye}, \citenamefont
  {Zeng}, \citenamefont {Zhai}, \citenamefont {Peng},\ and\ \citenamefont
  {Du}}]{Li2017a}%
  \BibitemOpen
  \bibfield  {author} {\bibinfo {author} {\bibfnamefont {J.}~\bibnamefont
  {Li}}, \bibinfo {author} {\bibfnamefont {R.}~\bibnamefont {Fan}}, \bibinfo
  {author} {\bibfnamefont {H.}~\bibnamefont {Wang}}, \bibinfo {author}
  {\bibfnamefont {B.}~\bibnamefont {Ye}}, \bibinfo {author} {\bibfnamefont
  {B.}~\bibnamefont {Zeng}}, \bibinfo {author} {\bibfnamefont {H.}~\bibnamefont
  {Zhai}}, \bibinfo {author} {\bibfnamefont {X.}~\bibnamefont {Peng}},\ and\
  \bibinfo {author} {\bibfnamefont {J.}~\bibnamefont {Du}},\ }\bibfield
  {title} {\bibinfo {title} {{Measuring out-of-time-order correlators on a
  nuclear magnetic resonance quantum simulator}},\ }\href
  {http://link.aps.org/doi/10.1103/PhysRevX.7.031011} {\bibfield  {journal}
  {\bibinfo  {journal} {Phys. Rev. X}\ }\textbf {\bibinfo {volume} {7}},\
  \bibinfo {pages} {031011} (\bibinfo {year} {2017})}\BibitemShut {NoStop}%
\bibitem [{\citenamefont {Meier}\ \emph {et~al.}(2019)\citenamefont {Meier},
  \citenamefont {Ang'ong'a}, \citenamefont {An},\ and\ \citenamefont
  {Gadway}}]{Meier2017}%
  \BibitemOpen
  \bibfield  {author} {\bibinfo {author} {\bibfnamefont {E.~J.}\ \bibnamefont
  {Meier}}, \bibinfo {author} {\bibfnamefont {J.}~\bibnamefont {Ang'ong'a}},
  \bibinfo {author} {\bibfnamefont {F.~A.}\ \bibnamefont {An}},\ and\ \bibinfo
  {author} {\bibfnamefont {B.}~\bibnamefont {Gadway}},\ }\bibfield  {title}
  {\bibinfo {title} {{Exploring quantum signatures of chaos on a Floquet
  synthetic lattice}},\ }\href {http://arxiv.org/abs/1705.06714
  https://link.aps.org/doi/10.1103/PhysRevA.100.013623} {\bibfield  {journal}
  {\bibinfo  {journal} {Phys. Rev. A}\ }\textbf {\bibinfo {volume} {100}},\
  \bibinfo {pages} {013623} (\bibinfo {year} {2019})}\BibitemShut {NoStop}%
\bibitem [{\citenamefont {Chowdhury}\ and\ \citenamefont
  {Swingle}(2017)}]{Chowdhury2017}%
  \BibitemOpen
  \bibfield  {author} {\bibinfo {author} {\bibfnamefont {D.}~\bibnamefont
  {Chowdhury}}\ and\ \bibinfo {author} {\bibfnamefont {B.}~\bibnamefont
  {Swingle}},\ }\bibfield  {title} {\bibinfo {title} {{Onset of many-body chaos
  in the $O(N)$ model}},\ }\href {http://arxiv.org/abs/1703.02545
  http://dx.doi.org/10.1103/PhysRevD.96.065005
  https://link.aps.org/doi/10.1103/PhysRevD.96.065005} {\bibfield  {journal}
  {\bibinfo  {journal} {Phys. Rev. D}\ }\textbf {\bibinfo {volume} {96}},\
  \bibinfo {pages} {065005} (\bibinfo {year} {2017})}\BibitemShut {NoStop}%
\bibitem [{\citenamefont {Patel}\ \emph {et~al.}(2017)\citenamefont {Patel},
  \citenamefont {Chowdhury}, \citenamefont {Sachdev},\ and\ \citenamefont
  {Swingle}}]{Patel2017}%
  \BibitemOpen
  \bibfield  {author} {\bibinfo {author} {\bibfnamefont {A.~A.}\ \bibnamefont
  {Patel}}, \bibinfo {author} {\bibfnamefont {D.}~\bibnamefont {Chowdhury}},
  \bibinfo {author} {\bibfnamefont {S.}~\bibnamefont {Sachdev}},\ and\ \bibinfo
  {author} {\bibfnamefont {B.}~\bibnamefont {Swingle}},\ }\bibfield  {title}
  {\bibinfo {title} {{Quantum Butterfly Effect in Weakly Interacting Diffusive
  Metals}},\ }\href {https://journals.aps.org/prx/pdf/10.1103/PhysRevX.7.031047
  https://link.aps.org/doi/10.1103/PhysRevX.7.031047} {\bibfield  {journal}
  {\bibinfo  {journal} {Phys. Rev. X}\ }\textbf {\bibinfo {volume} {7}},\
  \bibinfo {pages} {031047} (\bibinfo {year} {2017})}\BibitemShut {NoStop}%
\bibitem [{\citenamefont {Sachdev}\ and\ \citenamefont
  {Ye}(1993)}]{Sachdev1993}%
  \BibitemOpen
  \bibfield  {author} {\bibinfo {author} {\bibfnamefont {S.}~\bibnamefont
  {Sachdev}}\ and\ \bibinfo {author} {\bibfnamefont {J.}~\bibnamefont {Ye}},\
  }\bibfield  {title} {\bibinfo {title} {{Gapless spin-fluid ground state in a
  random quantum Heisenberg magnet}},\ }\href
  {https://link.aps.org/doi/10.1103/PhysRevLett.70.3339} {\bibfield  {journal}
  {\bibinfo  {journal} {Phys. Rev. Lett.}\ }\textbf {\bibinfo {volume} {70}},\
  \bibinfo {pages} {3339} (\bibinfo {year} {1993})}\BibitemShut {NoStop}%
\bibitem [{\citenamefont {Kitaev}(2015)}]{kitaev2015}%
  \BibitemOpen
  \bibfield  {author} {\bibinfo {author} {\bibfnamefont {A.}~\bibnamefont
  {Kitaev}},\ }\bibfield  {title} {\bibinfo {title} {{A simple model of quantum
  holography}},\ }in\ \href@noop {} {\emph {\bibinfo {booktitle} {KITP Progr.
  Entanglement Strongly-Correlated Quantum Matter}}}\ (\bibinfo {year}
  {2015})\BibitemShut {NoStop}%
\bibitem [{\citenamefont {Gu}\ \emph {et~al.}(2017)\citenamefont {Gu},
  \citenamefont {Qi},\ and\ \citenamefont {Stanford}}]{Gu2017}%
  \BibitemOpen
  \bibfield  {author} {\bibinfo {author} {\bibfnamefont {Y.}~\bibnamefont
  {Gu}}, \bibinfo {author} {\bibfnamefont {X.-L.}\ \bibnamefont {Qi}},\ and\
  \bibinfo {author} {\bibfnamefont {D.}~\bibnamefont {Stanford}},\ }\bibfield
  {title} {\bibinfo {title} {{Local criticality, diffusion and chaos in
  generalized Sachdev-Ye-Kitaev models}},\ }\href
  {https://link.springer.com/content/pdf/10.1007%2FJHEP05%282017%29125.pdf
  http://link.springer.com/10.1007/JHEP05(2017)125} {\bibfield  {journal}
  {\bibinfo  {journal} {JHEP}\ }\textbf {\bibinfo {volume} {2017}}\bibinfo
  {number} { (5)},\ \bibinfo {pages} {125}}\BibitemShut {NoStop}%
\bibitem [{\citenamefont {Luitz}\ and\ \citenamefont {{Bar
  Lev}}(2017)}]{Luitz2017}%
  \BibitemOpen
\bibfield  {number} {  }\bibfield  {author} {\bibinfo {author} {\bibfnamefont
  {D.~J.}\ \bibnamefont {Luitz}}\ and\ \bibinfo {author} {\bibfnamefont
  {Y.}~\bibnamefont {{Bar Lev}}},\ }\bibfield  {title} {\bibinfo {title}
  {{Information propagation in isolated quantum systems}},\ }\href
  {http://link.aps.org/doi/10.1103/PhysRevB.96.020406} {\bibfield  {journal}
  {\bibinfo  {journal} {Phys. Rev. B}\ }\textbf {\bibinfo {volume} {96}},\
  \bibinfo {pages} {020406} (\bibinfo {year} {2017})}\BibitemShut {NoStop}%
\bibitem [{\citenamefont {Bohrdt}\ \emph {et~al.}(2016)\citenamefont {Bohrdt},
  \citenamefont {Mendl}, \citenamefont {Endres},\ and\ \citenamefont
  {Knap}}]{Bohrdt2017a}%
  \BibitemOpen
  \bibfield  {author} {\bibinfo {author} {\bibfnamefont {A.}~\bibnamefont
  {Bohrdt}}, \bibinfo {author} {\bibfnamefont {C.~B.}\ \bibnamefont {Mendl}},
  \bibinfo {author} {\bibfnamefont {M.}~\bibnamefont {Endres}},\ and\ \bibinfo
  {author} {\bibfnamefont {M.}~\bibnamefont {Knap}},\ }\bibfield  {title}
  {\bibinfo {title} {{Scrambling and thermalization in a diffusive quantum
  many-body system}},\ }\href
  {http://stacks.iop.org/1367-2630/19/i=6/a=063001?key=crossref.13dc6c66bffaebd5c787acc0f3465b17
  http://arxiv.org/abs/1612.02434 http://dx.doi.org/10.1088/1367-2630/aa719b}
  {\bibfield  {journal} {\bibinfo  {journal} {New J. Phys.}\ }\textbf {\bibinfo
  {volume} {19}},\ \bibinfo {pages} {063001} (\bibinfo {year}
  {2016})}\BibitemShut {NoStop}%
\bibitem [{\citenamefont {Heyl}\ \emph {et~al.}(2018)\citenamefont {Heyl},
  \citenamefont {Pollmann},\ and\ \citenamefont {D{\'{o}}ra}}]{Heyl2018}%
  \BibitemOpen
  \bibfield  {author} {\bibinfo {author} {\bibfnamefont {M.}~\bibnamefont
  {Heyl}}, \bibinfo {author} {\bibfnamefont {F.}~\bibnamefont {Pollmann}},\
  and\ \bibinfo {author} {\bibfnamefont {B.}~\bibnamefont {D{\'{o}}ra}},\
  }\bibfield  {title} {\bibinfo {title} {{Detecting Equilibrium and Dynamical
  Quantum Phase Transitions in Ising Chains via Out-of-Time-Ordered
  Correlators}},\ }\href {http://arxiv.org/abs/1801.01684
  https://link.aps.org/doi/10.1103/PhysRevLett.121.016801} {\bibfield
  {journal} {\bibinfo  {journal} {Phys. Rev. Lett.}\ }\textbf {\bibinfo
  {volume} {121}},\ \bibinfo {pages} {016801} (\bibinfo {year}
  {2018})}\BibitemShut {NoStop}%
\bibitem [{\citenamefont {Lin}\ and\ \citenamefont
  {Motrunich}(2018)}]{Lin2018}%
  \BibitemOpen
  \bibfield  {author} {\bibinfo {author} {\bibfnamefont {C.-J.}\ \bibnamefont
  {Lin}}\ and\ \bibinfo {author} {\bibfnamefont {O.~I.}\ \bibnamefont
  {Motrunich}},\ }\bibfield  {title} {\bibinfo {title} {{Out-of-time-ordered
  correlators in a quantum Ising chain}},\ }\href
  {http://arxiv.org/abs/1801.01636 http://dx.doi.org/10.1103/PhysRevB.97.144304
  https://link.aps.org/doi/10.1103/PhysRevB.97.144304} {\bibfield  {journal}
  {\bibinfo  {journal} {Phys. Rev. B}\ }\textbf {\bibinfo {volume} {97}},\
  \bibinfo {pages} {144304} (\bibinfo {year} {2018})}\BibitemShut {NoStop}%
\bibitem [{\citenamefont {Nahum}\ \emph {et~al.}(2017)\citenamefont {Nahum},
  \citenamefont {Ruhman}, \citenamefont {Vijay},\ and\ \citenamefont
  {Haah}}]{Nahum2017}%
  \BibitemOpen
  \bibfield  {author} {\bibinfo {author} {\bibfnamefont {A.}~\bibnamefont
  {Nahum}}, \bibinfo {author} {\bibfnamefont {J.}~\bibnamefont {Ruhman}},
  \bibinfo {author} {\bibfnamefont {S.}~\bibnamefont {Vijay}},\ and\ \bibinfo
  {author} {\bibfnamefont {J.}~\bibnamefont {Haah}},\ }\bibfield  {title}
  {\bibinfo {title} {{Quantum Entanglement Growth under Random Unitary
  Dynamics}},\ }\href
  {https://journals.aps.org/prx/pdf/10.1103/PhysRevX.7.031016
  http://link.aps.org/doi/10.1103/PhysRevX.7.031016} {\bibfield  {journal}
  {\bibinfo  {journal} {Phys. Rev. X}\ }\textbf {\bibinfo {volume} {7}},\
  \bibinfo {pages} {031016} (\bibinfo {year} {2017})}\BibitemShut {NoStop}%
\bibitem [{\citenamefont {Rakovszky}\ \emph {et~al.}(2018)\citenamefont
  {Rakovszky}, \citenamefont {Pollmann},\ and\ \citenamefont {von
  Keyserlingk}}]{Rakovszky2017}%
  \BibitemOpen
  \bibfield  {author} {\bibinfo {author} {\bibfnamefont {T.}~\bibnamefont
  {Rakovszky}}, \bibinfo {author} {\bibfnamefont {F.}~\bibnamefont
  {Pollmann}},\ and\ \bibinfo {author} {\bibfnamefont {C.~W.}\ \bibnamefont
  {von Keyserlingk}},\ }\bibfield  {title} {\bibinfo {title} {{Diffusive
  Hydrodynamics of Out-of-Time-Ordered Correlators with Charge Conservation}},\
  }\href {https://arxiv.org/pdf/1710.09827.pdf http://arxiv.org/abs/1710.09827
  http://dx.doi.org/10.1103/PhysRevX.8.031058
  https://link.aps.org/doi/10.1103/PhysRevX.8.031058} {\bibfield  {journal}
  {\bibinfo  {journal} {Phys. Rev. X}\ }\textbf {\bibinfo {volume} {8}},\
  \bibinfo {pages} {031058} (\bibinfo {year} {2018})}\BibitemShut {NoStop}%
\bibitem [{\citenamefont {Khemani}\ \emph
  {et~al.}(2018{\natexlab{a}})\citenamefont {Khemani}, \citenamefont
  {Vishwanath},\ and\ \citenamefont {Huse}}]{Khemani2017}%
  \BibitemOpen
  \bibfield  {author} {\bibinfo {author} {\bibfnamefont {V.}~\bibnamefont
  {Khemani}}, \bibinfo {author} {\bibfnamefont {A.}~\bibnamefont
  {Vishwanath}},\ and\ \bibinfo {author} {\bibfnamefont {D.~A.}\ \bibnamefont
  {Huse}},\ }\bibfield  {title} {\bibinfo {title} {{Operator Spreading and the
  Emergence of Dissipative Hydrodynamics under Unitary Evolution with
  Conservation Laws}},\ }\href {https://arxiv.org/pdf/1710.09835.pdf
  http://arxiv.org/abs/1710.09835
  https://link.aps.org/doi/10.1103/PhysRevX.8.031057} {\bibfield  {journal}
  {\bibinfo  {journal} {Phys. Rev. X}\ }\textbf {\bibinfo {volume} {8}},\
  \bibinfo {pages} {031057} (\bibinfo {year} {2018}{\natexlab{a}})}\BibitemShut
  {NoStop}%
\bibitem [{\citenamefont {Swingle}\ and\ \citenamefont
  {Chowdhury}(2017)}]{Swingle2016a}%
  \BibitemOpen
  \bibfield  {author} {\bibinfo {author} {\bibfnamefont {B.}~\bibnamefont
  {Swingle}}\ and\ \bibinfo {author} {\bibfnamefont {D.}~\bibnamefont
  {Chowdhury}},\ }\bibfield  {title} {\bibinfo {title} {{Slow scrambling in
  disordered quantum systems}},\ }\href {https://arxiv.org/pdf/1608.03280.pdf
  http://arxiv.org/abs/1608.03280 http://dx.doi.org/10.1103/PhysRevB.95.060201
  https://link.aps.org/doi/10.1103/PhysRevB.95.060201} {\bibfield  {journal}
  {\bibinfo  {journal} {Phys. Rev. B}\ }\textbf {\bibinfo {volume} {95}},\
  \bibinfo {pages} {060201} (\bibinfo {year} {2017})}\BibitemShut {NoStop}%
\bibitem [{\citenamefont {Chen}(2016)}]{Chen2016a}%
  \BibitemOpen
  \bibfield  {author} {\bibinfo {author} {\bibfnamefont {Y.}~\bibnamefont
  {Chen}},\ }\bibfield  {title} {\bibinfo {title} {{Universal Logarithmic
  Scrambling in Many Body Localization}},\ }\href
  {https://arxiv.org/pdf/1608.02765.pdf http://arxiv.org/abs/1608.02765}
  {\bibfield  {journal} {\bibinfo  {journal} {arXiv:1608.02765}\ } (\bibinfo
  {year} {2016})}\BibitemShut {NoStop}%
\bibitem [{\citenamefont {Fan}\ \emph {et~al.}(2017)\citenamefont {Fan},
  \citenamefont {Zhang}, \citenamefont {Shen},\ and\ \citenamefont
  {Zhai}}]{Fan2017}%
  \BibitemOpen
  \bibfield  {author} {\bibinfo {author} {\bibfnamefont {R.}~\bibnamefont
  {Fan}}, \bibinfo {author} {\bibfnamefont {P.}~\bibnamefont {Zhang}}, \bibinfo
  {author} {\bibfnamefont {H.}~\bibnamefont {Shen}},\ and\ \bibinfo {author}
  {\bibfnamefont {H.}~\bibnamefont {Zhai}},\ }\bibfield  {title} {\bibinfo
  {title} {{Out-of-time-order correlation for many-body localization}},\ }\href
  {https://arxiv.org/pdf/1608.01914.pdf http://arxiv.org/abs/1608.01914
  http://dx.doi.org/10.1016/j.scib.2017.04.011
  http://linkinghub.elsevier.com/retrieve/pii/S2095927317301925} {\bibfield
  {journal} {\bibinfo  {journal} {Sci. Bull.}\ }\textbf {\bibinfo {volume}
  {62}},\ \bibinfo {pages} {707} (\bibinfo {year} {2017})}\BibitemShut
  {NoStop}%
\bibitem [{\citenamefont {Huang}\ \emph {et~al.}(2017)\citenamefont {Huang},
  \citenamefont {Zhang},\ and\ \citenamefont {Chen}}]{Huang2017}%
  \BibitemOpen
  \bibfield  {author} {\bibinfo {author} {\bibfnamefont {Y.}~\bibnamefont
  {Huang}}, \bibinfo {author} {\bibfnamefont {Y.-L.}\ \bibnamefont {Zhang}},\
  and\ \bibinfo {author} {\bibfnamefont {X.}~\bibnamefont {Chen}},\ }\bibfield
  {title} {\bibinfo {title} {{Out-of-time-ordered correlators in many-body
  localized systems}},\ }\href {http://arxiv.org/abs/1608.01091
  http://dx.doi.org/10.1002/andp.201600318
  http://doi.wiley.com/10.1002/andp.201600318} {\bibfield  {journal} {\bibinfo
  {journal} {Ann. Phys.}\ }\textbf {\bibinfo {volume} {529}},\ \bibinfo {pages}
  {1600318} (\bibinfo {year} {2017})}\BibitemShut {NoStop}%
\bibitem [{\citenamefont {Khemani}\ \emph
  {et~al.}(2018{\natexlab{b}})\citenamefont {Khemani}, \citenamefont {Huse},\
  and\ \citenamefont {Nahum}}]{Khemani2018}%
  \BibitemOpen
  \bibfield  {author} {\bibinfo {author} {\bibfnamefont {V.}~\bibnamefont
  {Khemani}}, \bibinfo {author} {\bibfnamefont {D.~A.}\ \bibnamefont {Huse}},\
  and\ \bibinfo {author} {\bibfnamefont {A.}~\bibnamefont {Nahum}},\ }\bibfield
   {title} {\bibinfo {title} {{Velocity-dependent Lyapunov exponents in
  many-body quantum, semiclassical, and classical chaos}},\ }\href
  {https://arxiv.org/pdf/1803.05902.pdf http://arxiv.org/abs/1803.05902
  https://link.aps.org/doi/10.1103/PhysRevB.98.144304} {\bibfield  {journal}
  {\bibinfo  {journal} {Phys. Rev. B}\ }\textbf {\bibinfo {volume} {98}},\
  \bibinfo {pages} {144304} (\bibinfo {year} {2018}{\natexlab{b}})}\BibitemShut
  {NoStop}%
\bibitem [{\citenamefont {Leviatan}\ \emph {et~al.}(2017)\citenamefont
  {Leviatan}, \citenamefont {Pollmann}, \citenamefont {Bardarson},
  \citenamefont {Huse},\ and\ \citenamefont {Altman}}]{Leviatan2017}%
  \BibitemOpen
  \bibfield  {author} {\bibinfo {author} {\bibfnamefont {E.}~\bibnamefont
  {Leviatan}}, \bibinfo {author} {\bibfnamefont {F.}~\bibnamefont {Pollmann}},
  \bibinfo {author} {\bibfnamefont {J.~H.}\ \bibnamefont {Bardarson}}, \bibinfo
  {author} {\bibfnamefont {D.~A.}\ \bibnamefont {Huse}},\ and\ \bibinfo
  {author} {\bibfnamefont {E.}~\bibnamefont {Altman}},\ }\bibfield  {title}
  {\bibinfo {title} {{Quantum thermalization dynamics with Matrix-Product
  States}},\ }\href {http://arxiv.org/abs/1702.08894} {\bibfield  {journal}
  {\bibinfo  {journal} {arXiv:1702.08894}\ } (\bibinfo {year}
  {2017})}\BibitemShut {NoStop}%
\bibitem [{\citenamefont {Lashkari}\ \emph {et~al.}(2013)\citenamefont
  {Lashkari}, \citenamefont {Stanford}, \citenamefont {Hastings}, \citenamefont
  {Osborne},\ and\ \citenamefont {Hayden}}]{Lashkari2012}%
  \BibitemOpen
  \bibfield  {author} {\bibinfo {author} {\bibfnamefont {N.}~\bibnamefont
  {Lashkari}}, \bibinfo {author} {\bibfnamefont {D.}~\bibnamefont {Stanford}},
  \bibinfo {author} {\bibfnamefont {M.}~\bibnamefont {Hastings}}, \bibinfo
  {author} {\bibfnamefont {T.}~\bibnamefont {Osborne}},\ and\ \bibinfo {author}
  {\bibfnamefont {P.}~\bibnamefont {Hayden}},\ }\bibfield  {title} {\bibinfo
  {title} {{Towards the fast scrambling conjecture}},\ }\href
  {https://arxiv.org/pdf/1111.6580.pdf
  http://link.springer.com/10.1007/JHEP04(2013)022} {\bibfield  {journal}
  {\bibinfo  {journal} {JHEP}\ }\textbf {\bibinfo {volume} {2013}}\bibinfo
  {number} { (4)},\ \bibinfo {pages} {22}}\BibitemShut {NoStop}%
\bibitem [{\citenamefont {Shenker}\ and\ \citenamefont
  {Stanford}(2015)}]{Shenker2014a}%
  \BibitemOpen
\bibfield  {number} {  }\bibfield  {author} {\bibinfo {author} {\bibfnamefont
  {S.~H.}\ \bibnamefont {Shenker}}\ and\ \bibinfo {author} {\bibfnamefont
  {D.}~\bibnamefont {Stanford}},\ }\bibfield  {title} {\bibinfo {title}
  {{Stringy effects in scrambling}},\ }\href {http://arxiv.org/abs/1412.6087
  http://link.springer.com/10.1007/JHEP05(2015)132} {\bibfield  {journal}
  {\bibinfo  {journal} {JHEP}\ }\textbf {\bibinfo {volume} {2015}}\bibinfo
  {number} { (5)},\ \bibinfo {pages} {132}}\BibitemShut {NoStop}%
\bibitem [{\citenamefont {Fisher}(1937)}]{Fisher1937}%
  \BibitemOpen
\bibfield  {number} {  }\bibfield  {author} {\bibinfo {author} {\bibfnamefont
  {R.~A.}\ \bibnamefont {Fisher}},\ }\bibfield  {title} {\bibinfo {title} {{The
  wave of advance of advantageous genes}},\ }\href
  {https://doi.org/10.1111/j.1469-1809.1937.tb02153.x} {\bibfield  {journal}
  {\bibinfo  {journal} {Ann. Eugen.}\ }\textbf {\bibinfo {volume} {7}},\
  \bibinfo {pages} {355} (\bibinfo {year} {1937})}\BibitemShut {NoStop}%
\bibitem [{\citenamefont {Kolmogorov}\ \emph {et~al.}(1937)\citenamefont
  {Kolmogorov}, \citenamefont {Petrovsky},\ and\ \citenamefont
  {Piscounov}}]{Kolmogorov1937}%
  \BibitemOpen
  \bibfield  {author} {\bibinfo {author} {\bibfnamefont {A.}~\bibnamefont
  {Kolmogorov}}, \bibinfo {author} {\bibfnamefont {I.}~\bibnamefont
  {Petrovsky}},\ and\ \bibinfo {author} {\bibfnamefont {N.}~\bibnamefont
  {Piscounov}},\ }\bibfield  {title} {\bibinfo {title} {{A study of the
  diffusion equation with increase in the amount of substance}},\ }\href@noop
  {} {\bibfield  {journal} {\bibinfo  {journal} {Bull. Univ. {\'{E}}tat
  Moscou}\ }\textbf {\bibinfo {volume} {1}},\ \bibinfo {pages} {1} (\bibinfo
  {year} {1937})}\BibitemShut {NoStop}%
\bibitem [{\citenamefont {Aleiner}\ \emph {et~al.}(2016)\citenamefont
  {Aleiner}, \citenamefont {Faoro},\ and\ \citenamefont {Ioffe}}]{Aleiner2016}%
  \BibitemOpen
  \bibfield  {author} {\bibinfo {author} {\bibfnamefont {I.~L.}\ \bibnamefont
  {Aleiner}}, \bibinfo {author} {\bibfnamefont {L.}~\bibnamefont {Faoro}},\
  and\ \bibinfo {author} {\bibfnamefont {L.~B.}\ \bibnamefont {Ioffe}},\
  }\bibfield  {title} {\bibinfo {title} {{Microscopic model of quantum
  butterfly effect: Out-of-time-order correlators and traveling combustion
  waves}},\ }\href {https://arxiv.org/pdf/1609.01251.pdf} {\bibfield  {journal}
  {\bibinfo  {journal} {Ann. Phys. (N. Y).}\ }\textbf {\bibinfo {volume}
  {375}},\ \bibinfo {pages} {378} (\bibinfo {year} {2016})}\BibitemShut
  {NoStop}%
\bibitem [{\citenamefont {Grozdanov}\ \emph {et~al.}(2019)\citenamefont
  {Grozdanov}, \citenamefont {Schalm},\ and\ \citenamefont
  {Scopelliti}}]{Grozdanov2018}%
  \BibitemOpen
  \bibfield  {author} {\bibinfo {author} {\bibfnamefont {S.}~\bibnamefont
  {Grozdanov}}, \bibinfo {author} {\bibfnamefont {K.}~\bibnamefont {Schalm}},\
  and\ \bibinfo {author} {\bibfnamefont {V.}~\bibnamefont {Scopelliti}},\
  }\bibfield  {title} {\bibinfo {title} {{Kinetic theory for classical and
  quantum many-body chaos}},\ }\href {https://arxiv.org/pdf/1804.09182.pdf
  http://arxiv.org/abs/1804.09182 http://dx.doi.org/10.1103/PhysRevE.99.012206
  https://link.aps.org/doi/10.1103/PhysRevE.99.012206} {\bibfield  {journal}
  {\bibinfo  {journal} {Phys. Rev. E}\ }\textbf {\bibinfo {volume} {99}},\
  \bibinfo {pages} {012206} (\bibinfo {year} {2019})}\BibitemShut {NoStop}%
\bibitem [{\citenamefont {Brunet}\ and\ \citenamefont
  {Derrida}(1997)}]{Brunet1997}%
  \BibitemOpen
  \bibfield  {author} {\bibinfo {author} {\bibfnamefont {E.}~\bibnamefont
  {Brunet}}\ and\ \bibinfo {author} {\bibfnamefont {B.}~\bibnamefont
  {Derrida}},\ }\bibfield  {title} {\bibinfo {title} {{Shift in the velocity of
  a front due to a cutoff}},\ }\href {https://doi.org/10.1103/PhysRevE.56.2597}
  {\bibfield  {journal} {\bibinfo  {journal} {Phys. Rev. E}\ }\textbf {\bibinfo
  {volume} {56}},\ \bibinfo {pages} {2597} (\bibinfo {year}
  {1997})}\BibitemShut {NoStop}%
\bibitem [{\citenamefont {Brunet}\ \emph {et~al.}(2006)\citenamefont {Brunet},
  \citenamefont {Derrida}, \citenamefont {Mueller},\ and\ \citenamefont
  {Munier}}]{Brunet2006}%
  \BibitemOpen
  \bibfield  {author} {\bibinfo {author} {\bibfnamefont {E.}~\bibnamefont
  {Brunet}}, \bibinfo {author} {\bibfnamefont {B.}~\bibnamefont {Derrida}},
  \bibinfo {author} {\bibfnamefont {A.~H.}\ \bibnamefont {Mueller}},\ and\
  \bibinfo {author} {\bibfnamefont {S.}~\bibnamefont {Munier}},\ }\bibfield
  {title} {\bibinfo {title} {{Phenomenological theory giving the full
  statistics of the position of fluctuating pulled fronts}},\ }\href
  {https://doi.org/10.1103/PhysRevE.73.056126} {\bibfield  {journal} {\bibinfo
  {journal} {Phys. Rev. E}\ }\textbf {\bibinfo {volume} {73}},\ \bibinfo
  {pages} {056126} (\bibinfo {year} {2006})}\BibitemShut {NoStop}%
\bibitem [{\citenamefont {Vidal}(2003)}]{Vidal2003}%
  \BibitemOpen
  \bibfield  {author} {\bibinfo {author} {\bibfnamefont {G.}~\bibnamefont
  {Vidal}},\ }\bibfield  {title} {\bibinfo {title} {{Efficient classical
  simulation of slightly entangled quantum computations}},\ }\href
  {https://doi.org/10.1103/PhysRevLett.91.147902} {\bibfield  {journal}
  {\bibinfo  {journal} {Phys. Rev. Lett.}\ }\textbf {\bibinfo {volume} {91}},\
  \bibinfo {pages} {147902} (\bibinfo {year} {2003})}\BibitemShut {NoStop}%
\bibitem [{\citenamefont {Vidal}(2004)}]{Vidal2004}%
  \BibitemOpen
  \bibfield  {author} {\bibinfo {author} {\bibfnamefont {G.}~\bibnamefont
  {Vidal}},\ }\bibfield  {title} {\bibinfo {title} {{Efficient simulation of
  one-dimensional quantum many-body systems}},\ }\href
  {https://doi.org/10.1103/PhysRevLett.93.040502} {\bibfield  {journal}
  {\bibinfo  {journal} {Phys. Rev. Lett.}\ }\textbf {\bibinfo {volume} {93}},\
  \bibinfo {pages} {040502} (\bibinfo {year} {2004})}\BibitemShut {NoStop}%
\bibitem [{\citenamefont {Johnson}\ \emph {et~al.}(2010)\citenamefont
  {Johnson}, \citenamefont {Clark},\ and\ \citenamefont
  {Jaksch}}]{Johnson2010}%
  \BibitemOpen
  \bibfield  {author} {\bibinfo {author} {\bibfnamefont {T.~H.}\ \bibnamefont
  {Johnson}}, \bibinfo {author} {\bibfnamefont {S.~R.}\ \bibnamefont {Clark}},\
  and\ \bibinfo {author} {\bibfnamefont {D.}~\bibnamefont {Jaksch}},\
  }\bibfield  {title} {\bibinfo {title} {{Dynamical simulations of classical
  stochastic systems using matrix product states}},\ }\href
  {https://doi.org/10.1103/PhysRevE.82.036702} {\bibfield  {journal} {\bibinfo
  {journal} {Phys. Rev. E}\ }\textbf {\bibinfo {volume} {82}},\ \bibinfo
  {pages} {036702} (\bibinfo {year} {2010})}\BibitemShut {NoStop}%
\bibitem [{\citenamefont {White}\ \emph {et~al.}(2018)\citenamefont {White},
  \citenamefont {Zaletel}, \citenamefont {Mong},\ and\ \citenamefont
  {Refael}}]{White2017}%
  \BibitemOpen
  \bibfield  {author} {\bibinfo {author} {\bibfnamefont {C.~D.}\ \bibnamefont
  {White}}, \bibinfo {author} {\bibfnamefont {M.}~\bibnamefont {Zaletel}},
  \bibinfo {author} {\bibfnamefont {R.~S.~K.}\ \bibnamefont {Mong}},\ and\
  \bibinfo {author} {\bibfnamefont {G.}~\bibnamefont {Refael}},\ }\bibfield
  {title} {\bibinfo {title} {{Quantum dynamics of thermalizing systems}},\
  }\href {https://arxiv.org/pdf/1707.01506.pdf
  http://dx.doi.org/10.1103/PhysRevB.97.035127
  https://link.aps.org/doi/10.1103/PhysRevB.97.035127} {\bibfield  {journal}
  {\bibinfo  {journal} {Phys. Rev. B}\ }\textbf {\bibinfo {volume} {97}},\
  \bibinfo {pages} {035127} (\bibinfo {year} {2018})}\BibitemShut {NoStop}%
\bibitem [{\citenamefont {Maldacena}\ \emph {et~al.}(2016)\citenamefont
  {Maldacena}, \citenamefont {Shenker},\ and\ \citenamefont
  {Stanford}}]{Maldacena2016}%
  \BibitemOpen
  \bibfield  {author} {\bibinfo {author} {\bibfnamefont {J.}~\bibnamefont
  {Maldacena}}, \bibinfo {author} {\bibfnamefont {S.~H.}\ \bibnamefont
  {Shenker}},\ and\ \bibinfo {author} {\bibfnamefont {D.}~\bibnamefont
  {Stanford}},\ }\bibfield  {title} {\bibinfo {title} {{A bound on chaos}},\
  }\href {http://link.springer.com/10.1007/JHEP08(2016)106} {\bibfield
  {journal} {\bibinfo  {journal} {JHEP}\ }\textbf {\bibinfo {volume}
  {2016}}\bibinfo  {number} { (8)},\ \bibinfo {pages} {106}}\BibitemShut
  {NoStop}%
\bibitem [{\citenamefont {Bramson}(1983)}]{Bramson1983}%
  \BibitemOpen
\bibfield  {number} {  }\bibfield  {author} {\bibinfo {author} {\bibfnamefont
  {M.}~\bibnamefont {Bramson}},\ }\bibfield  {title} {\bibinfo {title}
  {{Convergence of solutions of the Kolmogorov equation to travelling waves}},\
  }\href {https://doi.org/10.1090/memo/0285} {\bibfield  {journal} {\bibinfo
  {journal} {Mem. Am. Math. Soc.}\ }\textbf {\bibinfo {volume} {44}},\ \bibinfo
  {pages} {0} (\bibinfo {year} {1983})}\BibitemShut {NoStop}%
\bibitem [{\citenamefont {Berestycki}\ \emph {et~al.}(2017)\citenamefont
  {Berestycki}, \citenamefont {Brunet}, \citenamefont {Harris},\ and\
  \citenamefont {Roberts}}]{Berestycki2017}%
  \BibitemOpen
  \bibfield  {author} {\bibinfo {author} {\bibfnamefont {J.}~\bibnamefont
  {Berestycki}}, \bibinfo {author} {\bibfnamefont {{\'{E}}.}~\bibnamefont
  {Brunet}}, \bibinfo {author} {\bibfnamefont {S.~C.}\ \bibnamefont {Harris}},\
  and\ \bibinfo {author} {\bibfnamefont {M.}~\bibnamefont {Roberts}},\
  }\bibfield  {title} {\bibinfo {title} {{Vanishing Corrections for the
  Position in a Linear Model of FKPP Fronts}},\ }\href
  {https://doi.org/10.1007/s00220-016-2790-9} {\bibfield  {journal} {\bibinfo
  {journal} {Commun. Math. Phys.}\ }\textbf {\bibinfo {volume} {349}},\
  \bibinfo {pages} {857} (\bibinfo {year} {2017})}\BibitemShut {NoStop}%
\end{thebibliography}%

\appendix
\begin{appendices}
\section{Derivation of the master equation of the Brownian coupled cluster}
\label{appsec:master_eq}

In this appendix, the derivation of master equation governing the operator dynamics of the Brownian coupled cluster model is presented. Consider a chain of $L$ clusters with open boundary conditions where each cluster contains $N$ qubits. At every time-step, all the qubits within a cluster interact and all the qubits between neighboring clusters interact. The time evolution operator is stochastic and obeys
\bea
    &U(t+dt)-U(t) \\
    &=  - \frac{N( L + g^2(L-1))}{2} U(t) dt \nonumber \\
    & - i A \sum_{r=1}^L \sum_{b>a=1}^N \sum_{\alpha,\beta} \sigma_{r,a}^\alpha \sigma_{r,b}^\beta U(t) dB_{r,a,b,\alpha,\beta} \nonumber \\
    & -i g \sqrt{\frac{N-1}{2N}}A \sum_{r=1}^{L-1} \sum_{a,b=1}^N \sum_{\alpha,\beta} \sigma_{r,a}^\alpha \sigma_{r+1,b}^\beta U(t) d\tilde{B}_{r,a,b,\alpha,\beta}.
\eea
Here $r$ labels the cluster, $a,b$ are labels within a cluster, and $\alpha,\beta$ label Pauli matrices. The coupling constant $g$ can be used to dial the relative strength of the within-cluster and between-cluster interactions. Some of the other factors are chosen for convenience, and the sum over $a,b$ is unconstrained in the between-cluster term. The coefficient $A$ is determined by demanding that $U U^\dagger = 1$ on average, leading to
\be
A = \sqrt{\frac{1}{8(N-1)}}.
\ee

Given an operator $W$, the Heisenberg operator is $W(t)=U W U^\dagger$. We may expand $W(t)$ in a complete basis of operators,
\be
W(t) = \sum_{\mathcal{S}} c(\mathcal{S}) \mathcal{S},
\ee
where $\mathcal{S}$ is a product of cluster operators $\mathcal{S}_r$ with each $\mathcal{S}_r$ a product of Pauli operators within the cluster. The coefficients $c(\mathcal{S})$ can be determined from
\be
c(\mathcal{S}) = \frac{1}{2^{NL}} \tr\left(\mathcal{S} W(t) \right).
\ee
We will study the average operator probabilities,
\be
h(\mathcal{S}) = \overline{c(\mathcal{S})^2}.
\ee

To determine the equation of motion of $h(\mathcal{S})$ we must compute $dh(\mathcal{S})$. There are two kinds of terms, depending on whether $dB$ or $d\tilde{B}$ appear in the same trace or not. When they appear in the same trace, we find
\bea
dh_1 \\
= & -2N(L +g^2(L-1)) h dt + 2 A^2 \left(\sum_{r,a<b,\alpha,\beta} q_{r,a,b}^{\alpha,\beta}(\mathcal{S})\right) h dt  \nonumber \\
& + 2 g^2 \frac{N-1}{2N} A^2 \left(\sum_{r,a<b,\alpha,\beta} \tilde{q}_{r,a,b}^{\alpha,\beta}(\mathcal{S})\right) h dt,
\eea
where $q_{r,a,b}^{\alpha,\beta} =\pm 1$ depending on whether $\sigma_{r,a}^\alpha \sigma_{r,b}^\beta$ commutes or anti-commutes with $\mathcal{S}$ and $\tilde{q}_{r,a,b}^{\alpha,\beta} =\pm 1$ depending on whether $\sigma_{r,a}^\alpha \sigma_{r+1,b}^\beta$ commutes or anti-commutes with $\mathcal{S}$.

Let $w_r$ denote the total weight of $\mathcal{S}$ on site $r$, $w_r=0,\cdots,N$. The sums above can be written in terms of the $w_r$. The first sum is
\be
\sum_{r,a<b,\alpha,\beta}  q_{r,a,b}^{\alpha,\beta} = 16 \sum_{r=1}^L \frac{(N-w_r)(N-w_r-1)}{2}.
\ee
This is because, if $\mathcal{S}$ has a non-zero Pauli on spin $r,a$ or $r,b$, then $\sum_{\alpha,\beta} q^{\alpha,\beta}_{r,a,b}=0$, otherwise it is $16$. Thus we need to count the number of pairs $a,b$ which are both the identity operator. This number is $\frac{(N-w_r)(N-w_r-1)}{2}$. Similarly, the second sum is
\be
\sum_{r,a<b,\alpha,\beta} \tilde{q}_{r,a,b}^{\alpha,\beta} =  16 \sum_{r=1}^{L-1}(N- w_r)(N- w_{r+1}).
\ee
Putting everything together gives
\bea
dh_1 = & - 2 N(L+g^2(L-1)) h dt \\
&+ \frac{2}{N-1} \sum_r (N-w_r)(N-w_r-1) h dt  \nonumber \\
& + \frac{2 g^2}{N} \sum_r (N-w_r)(N-w_{r+1}) h dt.
\eea

When the $dB$ or $d\tilde{B}$ factors appear in different traces, then we get terms connecting $h(\mathcal{S})$ to $h(\sigma \sigma \mathcal{S})$. These terms are
\bea
d&h_2 =\\
 & - \frac{2}{8(N-1)}\sum_{r=1}^L \sum_{a<b} \sum_{\alpha,\beta} q_{r,a,b}^{\alpha,\beta} (1 - q_{r,a,b}^{\alpha,\beta}) h(\sigma_{r,a}^\alpha \sigma^\beta_{r,b}\mathcal{S})dt  \\
& - \frac{2 g^2}{16N} \sum_{r=1}^{L-1} \sum_{a,b} \sum_{\alpha,\beta} \tilde{q}_{r,a,b}^{\alpha,\beta} (1 - \tilde{q}_{r,a,b}^{\alpha,\beta}) h(\sigma_{r,a}^\alpha \sigma^\beta_{r+1,b}\mathcal{S})dt.
\eea

To proceed further, let us assume that $h$ depends only on the total weight $w_r$ on site $r$ and not on the particular operator $\mathcal{S}$. The $dh_1$ terms already manifestly depend just on the total weight. The $dh_2$ terms connect probabilities for different weights. The $dh_1$ term can be written

\bea
\label{eq:dh1}
dh_1 =& \sum_{r=1}^{L} \frac{-2\left[(2N-1) w_r - w_r^2\right]}{N-1} h dt \\
 -&\frac{g^2}{N}\sum\limits_{\braket{rr'}}(2Nw_r-w_rw_{r'}) hdt+r\longleftrightarrow r'
\eea

To compute the $dh_2$ terms, we can consider a particular operator of the desired weight. Consider first the onsite terms. Suppose $\mathcal{S}_r = \sigma^x_{r,1} \cdots \sigma^x_{r,w_r} I_{r,w_r+1} \cdots I_{r,N}$. Now consider all $\sigma^\alpha_{r,a} \sigma^\beta_{r,b}$. If the $\sigma \sigma$ commutes with $\mathcal{S}$, then the $dh_2$ contribution vanishes. If $\sigma \sigma$ anticommutes with $\mathcal{S}$, then the $q(1-q)$ factor is $-2$. If $\mathcal{S}_r$ is the identity on one of $a$ or $b$, say $b$, then the anticommuting terms are $\sigma^y\sigma^\beta$ and $\sigma^z \sigma^\beta$. Of these eight, two keep the weight the same and six increase the weight by one. There are $w_r(N-w_r)$ choices of $a,b$ in this class. The contribution is thus
\be
-2 w_r (N-w_r)[ 2 h(\mathbf{w}) dt + 6 h(\mathbf{w}+\mathbf{e}_r)dt].
\ee
Here $\mathbf{e}_r = (0,\cdots,0,1_r,0,\cdots,0)$ is a unit vector that adds one to weight $w_r$.

If $\mathcal{S}_r$ is non-identity on both $a$ and $b$, then the anticommuting terms are $\text{($y$ or $z$)} \text{($I$ or $x$)}$ and $\text{($I$ or $x$)} \text{($y$ or $z$)}$. Of these eight, four keep the weight the same and four decrease the weight by one. There are $w_r (w_r-1)/2$ such choices of $a,b$. The total contribution from these terms is thus
\be
-2 \frac{w_r(w_r-1)}{2} [ 4 h(\mathbf{w}) dt + 4 h(\mathbf{w}-\mathbf{e}_r)dt].
\ee

The total onsite contributions are thus
\bea
\label{eq:dh2onsite}
&dh_{2,\text{onsite}} \\
= & \frac{1}{2(N-1)} \sum_{r=1}^L  w_r (N-w_r)[ 2 h(\mathbf{w}) dt + 6 h(\mathbf{w}+\mathbf{e}_r)dt] \nonumber \\
& + \frac{1}{2(N-1)} \sum_{r=1}^L \frac{w_r(w_r-1)}{2} [ 4 h(\mathbf{w}) dt + 4 h(\mathbf{w}-\mathbf{e}_r)dt].
\eea

For the non-onsite contributions, we choose $\mathcal{S}$ to be an operator with $w_r$ $\sigma^x$s on cluster $r$ and $w_{r+1}$ $\sigma^x$s on cluster $r+1$. If $\mathcal{S}$ is the identity on $r,a$, then the anti-commuting operators are $\sigma^\alpha_{r,a} \sigma^y_{r+1,b} $ and $\sigma^\alpha_{r,a} \sigma^z_{r+1,b}$. Of these eight, two keep $w_r$ the same and two increase $w_r$ by one. There are $(N-w_r)w_{r+1}$ such $a,b$. The contribution is
\be
-2 (N-w_r) w_{r+1} [ 2 h(\mathbf{w})dt + 6 h(\mathbf{w}+\mathbf{e}_r)dt].
\ee
Similarly, if $\mathcal{S}$ is the identity on $r+1,b$, then by the same logic we find a contribution of
\be
-2  w_r (N-w_{r+1}) [ 2 h(\mathbf{w})dt + 6 h(\mathbf{w}+\mathbf{e}_{r+1})dt].
\ee
Finally, suppose $\mathcal{S}$ is not the identity on both $r,a$ and $r+1,b$. Then the anti-commuting operators are $\text{($y$ or $z$)} \text{($I$ or $x$)}$ and $\text{($I$ or $x$)} \text{($y$ or $z$)}$. Of these eight, four leave both weights unchanged, two decrease weight $w_r$ by one, and two decrease weight $w_{r+1}$ by one. There are $w_r w_{r+1}$ such $a,b$. The contribution is
\be
-2 w_r w_{r+1} [ 4 h(\mathbf{w})dt + 2 h(\mathbf{w}-\mathbf{e}_{r+1})dt + 2 h(\mathbf{w}-\mathbf{e}_{r})dt].
\ee
The total non-onsite contribution is thus
\bea
 \label{eq:dh2nononsite}
&dh_{2,\text{non-onsite}} \\
&=\frac{g^2}{4N}\sum\limits_{\braket{r,r'}}Nw_{r'}(2h(\mathbf{w})+6h(\mathbf{w}+\mathbf{e}_r))-\\
&\ \ \ \ \ \ \ \  w_r w_{r'} (6h(\mathbf{w}+\mathbf{e}_r)-2h(\mathbf{w}-\mathbf{e}_r))+r\longleftrightarrow r'\\
\eea

By combining Eqs.~\eqref{eq:dh1}, \eqref{eq:dh2onsite}, and \eqref{eq:dh2nononsite} together, we obtain the complete equation of motion for $h(\mathbf{w})$ in the case where $h(\mathcal{S})$ depends only on the weights $w_r$ of $P$ at the different clusters. It is, however, convenient to take one more step and include degeneracy factors.

The number of operators with weights $w_r$ are
\be
D(\mathbf{w}) = \prod_r \binom{N}{w_r} 3^{w_r}.
\ee
While $h(\mathbf{w})$ is the probability of a single operator with weights $w_r$, the object $\tilde{h}(\mathbf{w})$, defined by
\be
\tilde{h}(\mathbf{w}) = D(\mathbf{w}) h(\mathbf{w}),
\ee
is the total probability of all operators with weight $w_r$, i.e., the probability of weights $w_r$.

The equation of motion for $\tilde{h}$ can be obtained from that of $h$. The $dh_1$ terms immediately translate to $d\tilde{h}_1$ terms since the weights are the same on both sides of the equation. However, the $dh_2$ must by modified. We replace each $h(\mathbf{w})$ with a $\tilde{h}(\mathbf{w})/D(\mathbf{w})$. Then the various rates are modified by ratios of $D(\mathbf{w})$ to $D(\mathbf{w}\pm \mathbf{e}_r)$. These ratios are
\be
\frac{D(\mathbf{w})}{D(\mathbf{w}+\mathbf{e}_r)} = \frac{w_r+1}{3(N-w_r)}
\ee
and
\be
\frac{D(\mathbf{w})}{D(\mathbf{w}-\mathbf{e}_r)} = \frac{3(N-w_r+1)}{w_r}.
\ee

Thus we have
\bea
\label{eq:dDh1}
\frac{d\tilde{h}_1}{dt} =& \sum_{r=1}^{L} \frac{-2\left[(2N-1) w_r - w_r^2\right]}{N-1} \tilde{h}  \\
&  -\frac{ g^2}{N} \sum\limits_{\braket{r,r'}} (2Nw_{r'}-w_rw_{r'})\tilde{h}+r\longleftrightarrow r'
\eea
\bea
\label{eq:dDh2onsite}
&\frac{d\tilde{h}_{2,\text{onsite}}}{dt} \\
& =\sum_{r=1}^L  \frac{w_r}{(N-1)}\left[  \tilde{h} + \frac{w_r+1}{N-w_r}\tilde{h}(\mathbf{w}+\mathbf{e}_r)\right] \nonumber \\
& +  \sum_{r=1}^L \frac{w_r(w_r-1)}{(N-1)} \left[ \tilde{h} + \frac{3(N-w_r+1)}{w_r} \tilde{h}(\mathbf{w}-\mathbf{e}_r)\right],
\eea
and
\bea
 \label{eq:dDh2nononsite}
&\frac{d\tilde{h}_{2,\text{non-onsite}}}{dt} \\
=&\frac{g^2}{2N}\sum\limits_{\braket{rr'}}3(N-w_r+1)w_{r'}\tilde{h}(\w-\e_r)\\
&\ \ \ \ \ \ \ \ \ \ +Nw_{r'}\tilde{h}+w_{r'}(w_r+1)\tilde{h}(\w+\e_r)+r\longleftrightarrow r'
\eea
Combining every thing together, we obtain the master equation of $\tilde{h}$ presented in Eq.~\eqref{eq:master} in the main text.

\section{More details on the Brownian coupled cluster in the infinite-$N$ limit}
\label{appsec:inf_N}
In this appendix, some additional details on the infinite $N$ noiseless limit of the BCC model are presented. The continuum limit of Eq.~\eqref{eq:langevin} resembles a FKPP-type partial differential equation, Eq.~\eqref{eq:fisher}.

\begin{figure}
\includegraphics[height=0.49\columnwidth, width=0.49\columnwidth]
{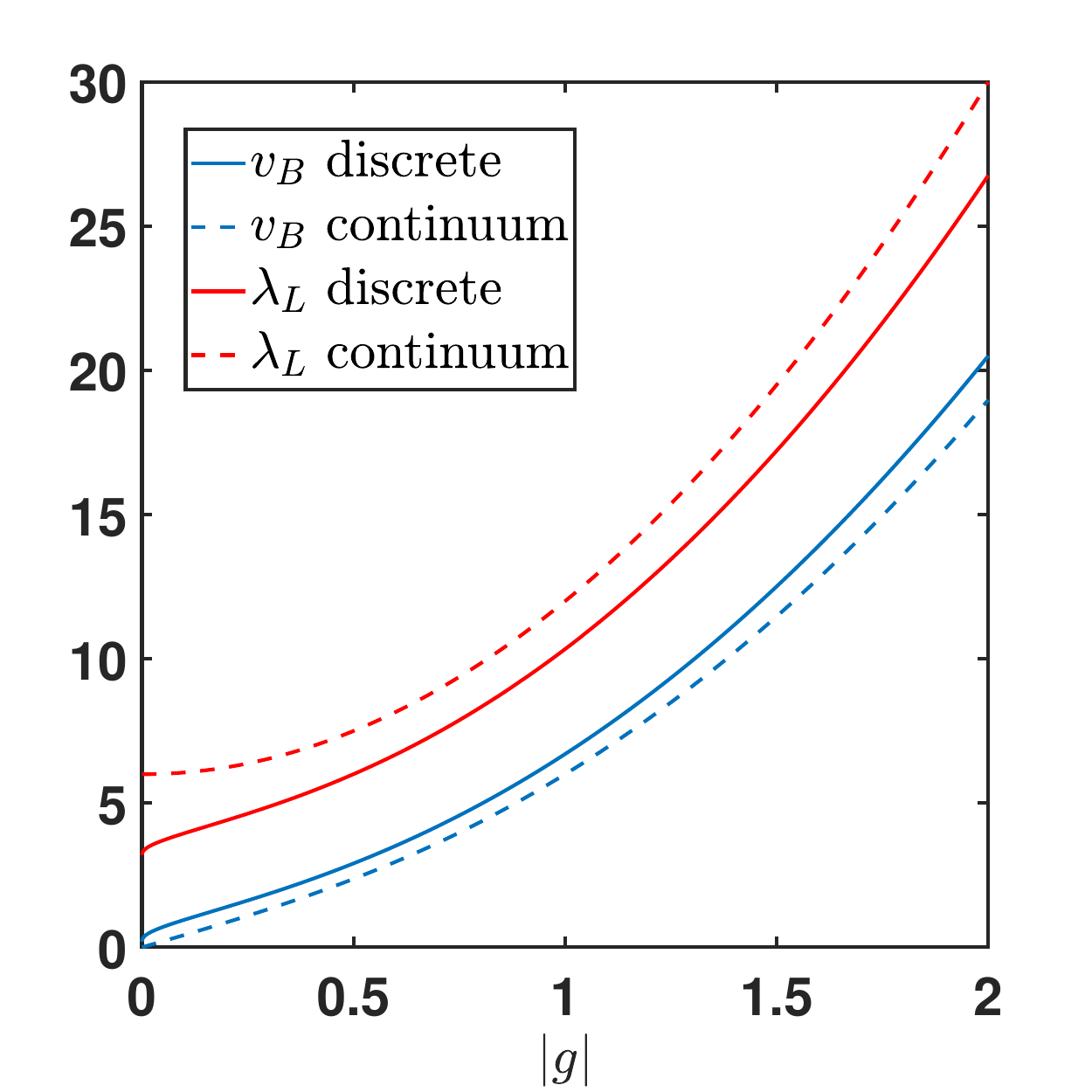}
\includegraphics[height=0.49\columnwidth, width=0.49\columnwidth]
{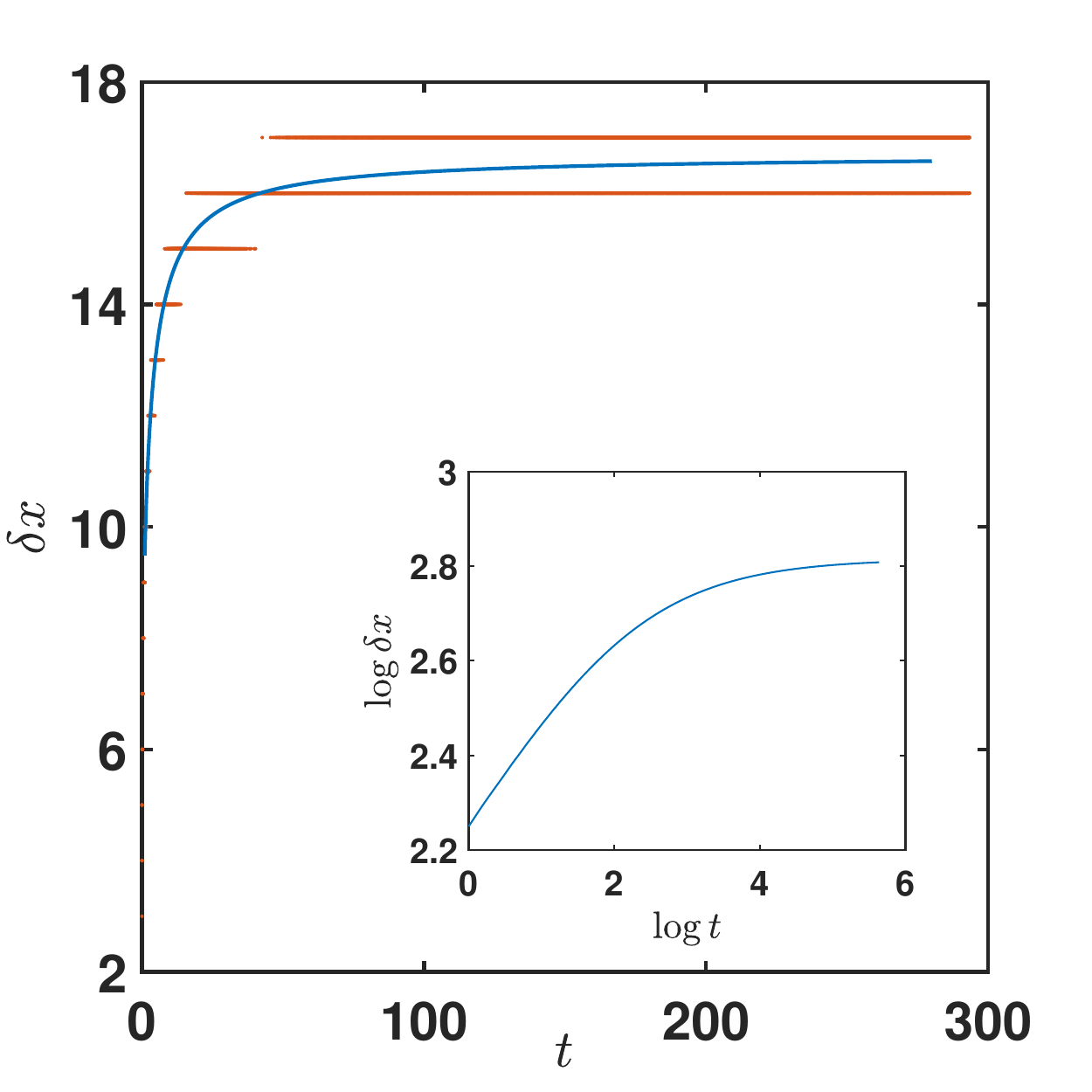}
\caption{(a) The difference of $v_B$ and $\lambda_L$ between the discrete infinite-$N$ BCC and the continuum approximation. (b) Direct simulation of the discrete model by numerically solving the differential equation Eq.~\eqref{eq:inf_N_discrete}. The spatial distance between two contours of $\log C(r,t)$ saturates in the late time, showing that the front is sharp. On a log-log plot, the slope of the curve decreases to zero. This is in the sharp contrast of the local Hamiltonian systems, where the slope increases to $1/2$ and $1/3$ for the chaotic case and non-interacting case respectively as shown in Fig.~\ref{fig:chaotic_broadening} and Fig.~\ref{fig:free_broadening}.  }
\label{fig:inf_N}
\end{figure}

To justify the continuum approximation, here we directly study the original discrete ordinary differential equation on the lattice. Starting with
\be
\partial_t \phi(r,t)=3\left(\phi(r,t)+\frac{g^2}{2}\left (\phi(r-1,t)+\phi(r+1,t)\right )\right),
\label{eq:inf_N_discrete}
\ee
we look for the traveling wave solution $\exp(\lambda (t-r/v)$ in the small $\phi$ limit, so that the nonlinearity can be safely ignored. The relation between the velocity $v$ and the growth rate is
\be
\lambda=3+3g^2\cosh \frac{\lambda}{v}.
\label{eq:inf_N_discrete_v}
\ee
The minimum velocity for positive growth rate is the butterfly velocity $v_B$ and the corresponding growth rate is the Lyapunov exponent. In Fig.~\ref{fig:inf_N}(a), we plot $v_B$ and $\lambda_L$ as a function of $g$ and compare them with the analytical result from the continuum limit. Overall, the result from the continuum approximation tracks that from the discrete model. The main difference occurs in the limit that $g\ll 1$: Analyzing Eq.~\eqref{eq:inf_N_discrete_v} shows that
\bea
v_B&\sim -\frac{3}{2\log |g|+1}\\
\lambda_L &\sim 3 -\frac{3}{2\log |g|+1},
\eea
while the continuum approximation predicts that $v_B\sim 3\sqrt{2} |g|$ and $\lambda_L \sim 6(1+g^2)$.

Furthermore, we directly simulate the discrete nonlinear ordinary differential equation. The result is shown in Fig.~\ref{fig:inf_N}(b). As time increases, the spatial difference between two contours of $\log C(r,t)$ saturates. This explicitly verifies the sharp wavefront and the exponential growth of the squared commutator.

\section{The butterfly velocity and the diffusion constant from the noisy FKPP equation}
\label{appsec:noisy_FKPP}
In this appendix, we discuss how to obtain Eq.~\eqref{eq:large_N_v_d} for large but finite $N$ BCC by analyzing Eq.~\eqref{eq:fisher} with the hard-cutoff approximation and the noise term Eq.~\eqref{eq:noise_fisher}. This material is essentially a review of the analysis of Brunet et. al. \cite{Brunet1997,Brunet2006}.
To simplify the notation, we rescale $\phi$, $r$ and $t$ as following,
\bea
\phi &\rightarrow \frac{4}{3}\phi\\
r &\rightarrow \sqrt{2(1+g^2)/g^2}\\
t &\rightarrow 3(1+g^2) t
\eea
After the rescaling, Eq. \eqref{eq:fisher} becomes
\be
\partial_t \phi =(1-\phi)(\partial_r^2 \phi+\phi)
\label{eq:fisher_rescale}
\ee
and the noise term becomes
\bea
f_{noise}=\frac{(2+2g^2)^{1/4}}{(3g)^{1/2}}\sqrt{\frac{1}{2N}(2-\phi)(\partial_r^2 \phi+\phi)}\eta(r,t)
\label{eq:noise_fisher_rescale}
\eea
In this section, we will mainly focus on Eq. \eqref{eq:fisher_rescale} and Eq. \eqref{eq:noise_fisher_rescale}.

\subsection{The noiseless case}
We first discuss the case without noise, corresponding to the infinite-$N$ limit of the BCC model.
Eq.~\eqref{eq:fisher_rescale} is similar to the Fisher-KPP equation,
\be
\partial_t \phi =\partial_r^2 \phi+\phi(1-\phi)
\ee
describing a growth-saturation process occurring in a wide class of systems including population dynamics, combustion, and reaction-diffusion systems. One of the interesting features of the Fisher-KPP equation is that it admits traveling wave solutions $\phi(r,t)=w(r-vt)$ that the initial configurations converge to \cite{Bramson1983}.
We expect that our equation Eq.~\eqref{eq:fisher_rescale} obtained from unitary dynamics also falls into the FKPP universality class, because the linearized version of Eq.~\eqref{eq:fisher_rescale} is the linearized Fisher-KPP equation, and because Eq.~\eqref{eq:fisher_rescale} also has saturation physics so that $\phi=1$ is a stable solution.

For the Fisher-KPP equation, given a initial condition $\phi(r,0)$ that is sufficiently well localized, it asymptotically approaches the traveling wave solution,
\bea
&\lim \limits_{t\rightarrow \infty}\phi(z+m(t),t)=w_v(z)\\
&\lim \limits_{t\rightarrow \infty} \frac{m(t)}{t}=v,
\eea
where $m(t)$ is the position of the wave front defined by the equation $\phi(m(t),t) = \text{constant}$.

\textit{Traveling waves.}---The first question to answer is the shape of the traveling wave solution $w_v(z)$ for different velocities. It can be determined from the following equation,
\be
-v \partial_z w(z)=\left(\partial_z^2 w(z)+ w(z)\right ) \left (1-w(z)\right )
\ee
 with the boundary conditions $w(\infty)\rightarrow 0$ and $w(-\infty)\rightarrow 1$.  In the region that $z \gg 0$ and $w(z)\ll 1$, the shape can be understood from the linearized equation. For each velocity, there are two modes $e^{-\gamma z}$ and $e^{-z/\gamma}$, where $\gamma+1/\gamma=v$. At the critical velocity $v_c=2$, $\gamma=1/\gamma =1$ and the two modes are $z e^{-z}$ and $z e^{-z}$. On the level of the linear equation, for a given velocity, any combination of the two modes is valid.
The effect of the non-linearity can be regarded as setting a boundary condition for the linearized equation, say at $z=0$,
\be
w(0)=\alpha(v),  \ \ \ \ w'(0)=\beta(v)
\ee
where $\alpha(v)$ and $\beta(v)$ are tied to each other based on the solution to the full nonlinear equation. The above boundary condition forces both decay modes to appear, with the slower decaying mode dominating the behavior of $w(z)$ in the large $z$ limit. In the case that $v<2$, $\gamma$ becomes complex , and both modes can appear as long as the combination is real.  Therefore we have
\bea
w(z)\sim \begin{cases}
e^{-\gamma z}, \ \ \ \ \gamma <1 \ \ \ \ &\text{if $v>2$},\\
z e^{-\gamma z}\ \ \ \ &\text{if $v=2$}, \\
a e^{-\gamma z}+a^* e^{-\gamma^* z}  \ \ \ \ &\text{if $v<2$}.
\end{cases}
\label{eq:tail}
\eea
In the current context, $\phi$ is interpreted as the operator weight and is always positive. Therefore, the last case where $w(z)$ oscillates is physically irrelevant, setting the minimal physical velocity to $v_c=2$. However the last case becomes important for the noisy case discussed below.

\textit{Approaching to the traveling waves.}---Great efforts have been made to understand the relationship between the initial configuration and the final traveling wave it asymptotes to. It is found that for an initial perturbation which is sufficiently localized, $\phi(r,t)$ approaches the critical traveling wave in the long time limit.
This can be understood from the linearized equation,
\be
\partial_t \phi=\partial_r^2 \phi+\phi.
\label{eq:linear}
\ee

Using Green's function, we can write down the general solution as
\be
\phi(r,t)=\int dr' \frac{e^{t-\frac{(r'-r)^2}{4 t}}}{2 \sqrt{\pi } \sqrt{t}}\phi(r',0)
\ee
where $\phi(r',0)$ is given by the initial condition. Starting with $\phi(r,0)=e^{-\lambda |r|}$, we obtain that
\be
\lim \limits_{t\rightarrow \infty} \phi(vt+z,t)\sim
\begin{cases}
\frac{1}{\sqrt{t}} e^{t \left(1-\frac{v^2}{4}\right)-\frac{v z}{2}} \ \ \ \text{if $v<2\lambda$} \\
e^{(1-\lambda v+\lambda^2)t-\lambda z} \ \ \ \text{if $v\geq2\lambda$}.
\end{cases}
\ee
The velocity of the wavefront is determined by choosing $v$ to cancel the $t$ dependence in the exponent so that $\phi$ approaches a constant in the traveling frame. As a result,
\be
v_B=
\begin{cases}
2, \ \ \  &\text{if $\lambda\geq 1 $} \\
\lambda+\frac{1}{\lambda} \ \ \ &\text{if $\lambda<1$}
\end{cases}
\ee
This demonstrates that, an initial sufficiently-localized configuration travels with the minimal velocity $v_c=2$, i.e., the leading term of the wavefront position $m(t)$ is $2t$. The asymptotic form is,
\be
\phi(v_c t+z, t)\sim \frac{1}{\sqrt{t}}e^{-z}.
\label{eq:linear_tail}
\ee
From this, one can also obtain the sub-leading term of $m(t)$ by canceling the $1/\sqrt{t}$ prefactor, which gives rise to $m(t)\sim 2t-\frac{1}{2} \log(t) $. Then $\phi(m(t)+z,t)\rightarrow e^{-z} $.

The linearized equation gives the right velocity. However, the sub-leading term in $m(t)$ is not correct. The nonlinearity forces the waveform at the critical velocity to decay like $z e^{-z}$, slower than the $e^{-z}$ form obtained above. To take into account the nonlinear effects, we note that $\partial_r \phi(r,t) $ is also a solution to the linearized equation, and we can combine $\partial_r \phi(r,t)$ and $\phi(r,t)$ to obtain a new solution $\tilde{\phi}$ that minimizes the effect of the nonlinear term $\phi(r,t)^2$. By expanding Eq.~\eqref{eq:linear_tail} to the next order, one finds that
\be
\tilde{\phi}(v_c t+z,t)=(\phi+\partial_r \phi) \rightarrow \frac{z}{t^{3/2}}e^{-z},
\ee
which indeed has the corrected asymptotic behavior as a function of $z$. We can again obtain the sub-leading term in $m(t)$ by canceling the time dependence,
\be
 m(t)\sim 2t-\frac{3}{2} \log(t).
\ee
The second term was found by Bramson \cite{Bramson1983} and turns out to be independent of the specific shape of the initial condition as long as it is sufficiently localized.

To incorporate a simple saturation mechanism into the linearized equations and enforce the asymptotic shape of the traveling wave solution, Berestycki, et al. \cite{Berestycki2017} recently solved the linearized equation with the following moving boundary condition,
\be
\phi(m(t),t)=\alpha, \ \ \ \partial_r \phi(m(t),t)=\beta,
\label{eq:moving_boundary}
\ee
in order to obtain the vanishing correction of $m(t)$. They found that
\be
m(t)\sim 2t -\frac{3}{2}\log (t)-\frac{3\sqrt{\pi}}{\sqrt{t}}+\cdots.
\ee
The same correction has been identified in other models and is therefore expected to be universal, applying to the original Fisher-KPP equation and Eq. \ref{eq:fisher_rescale} as well.

\subsection{The noisy case}
The above picture applies to the infinite-$N$ limit of the BCC model. As stated in the main text, $1/N$ expansion away from the limit has two main effects. First, we need cutoff $\phi(r,t)$ to $0$ whenever it is below $1/N$; second, we need include the noise term Eq.~\eqref{eq:noise_fisher_rescale} in the Eq.~\eqref{eq:fisher_rescale}.

\textit{Cutoff-velocity.}---We first discuss the effect of the cutoff without considering the noise. As discussed above, due to the positivity of $\phi$, the velocity of the traveling wave can never go below $v_c=2$. However, with the cutoff, the part of $\phi$ below $1/N$ is set to zero by hand, and a smaller velocity becomes possible.
The scaling behavior of the velocity as a function $N$ can be obtained following Ref.~\citep{Brunet1997}. When $v<2$, the tail of the traveling acquires a oscillating part with a long wavelength in addition to the decay,
\bea
w(z)\sim \sin (\gamma_I z) e^{-\gamma_R z}
\label{eq: sine_wave}
\eea
with $\gamma_I$ and $\gamma_R$ the real and imaginary part of $\gamma$ respectively. Even with the cut-off, $w(z)$ should still remain positive until it decays to $1/N$. This imposes a constraint on the wave-length of the oscillation part which is determined by $\gamma_I$, requiring $\gamma_I < \frac{\pi}{\log N}$. In consequence, $v>v_c-\frac{\pi^2}{\log^2 N}$. Therefore, the velocity correction scales as $1/\log^2 N$, which is consistent with the numerical result presented in the inset of Fig.~\ref{fig:brownian_vd}(a).

To quantify the above picture, and especially to understand how different initial conditions approach the asymptotic traveling wave with the cut-off, Brunet et al. \cite{Brunet1997} introduced a third boundary condition $\phi(m(t)+L,t)=0$ with $L\sim \log N$ to the linearized equation in addition to Eq.~\eqref{eq:moving_boundary}. This new boundary condition is to account for the hard-cutoff occurring at the leading edge of the traveling wave. They also set $\alpha=0$ for simplicity, which seems unnatural but does not affect the shape of the traveling wave in the large $z$ limit.
In the following, we repeat their analysis in some detail. With this setup, it is convenient to go to the traveling frame by performing the substitution $\phi(r,t)=w(r-m(t),t)$. Then the boundary conditions are
\bea
&\partial_t w=\dot{m}(t)\partial_z w+\partial_z^2 w+w \\
&w(0, t)=0, \ \  w(L, t)=0  \\
&\partial_z w(0, t )=1.
 \eea
We first approximate $\dot{m}(t)$ as its asymptotic value $v$, to be determined. By setting the time derivative of $\dot{m}$ to zero, the boundary conditions uniquely determines the form of the asymptotic traveling wave,
\be
w(z)=\frac{L}{\pi} \sin \frac{n\pi }{L}z \exp(-\frac{v}{2}z),
\ee
with the velocity $v=2\sqrt{1-\frac{\pi^2}{L^2}}$.

Understanding the asymptotic form, we restore $\dot{m}(t)$ in the equation and study the full dynamics of $w(z,t)$. To the leading order of $L$, $w(z,t)$ can be written as a superposition of eigenmodes,
\bea
&w(z,t )\\
&=\sum a_n \frac{L}{\pi}\sin \frac{n\pi}{L}z \exp\left (-z+\frac{\pi^2}{L^2}(1-n^2)t +v t-m(t) \right )
\eea
where $a_n$ is obtained from Fourier expanding the initial condition, and $m(t)$ is tuned so that $\partial_z w(0,t)=0$ for all time. All modes with $n>1$ decay exponentially.  In the long-time limit, since $m(t)\rightarrow v t$, we obtain
\bea
w(t\rightarrow \infty,z)= a_1 \frac{L}{\pi}\sin \frac{\pi z}{L}\exp \left(-z+v t-m(t)\right)\\
a_1=\frac{2\pi}{L^2}\int _0^L w(0,z) \exp(z)\sin \frac{\pi z}{L} dz.
\eea
Therefore,
\be
m(t)-vt=\log a_1,
\label{eq:shift}
\ee
in order to match the boundary condition. In other words, the relaxation to the asymptotic traveling wave causes an additional shift to the wavefront position.

\textit{Noise-induced diffusive motion.}---Now let us analyze the role of the noise term on top of the hard cut-off. Based on Eq.~\eqref{eq:noise_fisher_rescale}, the noise scales as $\sqrt{\phi/N}$ and therefore is most prominent at the leading edge the traveling wave where $\phi\sim \frac{1}{N}$.
We will argue that the wavefront obeys
\bea
m(t)\sim vt+\delta v t+ X,
\eea
where $X$ is a random diffusive process with $\braket{X}=0$ and $\braket{X^2}=2Dt$.

Different from the deterministic model studied above, in the actual noisy equation, it is possible that $\phi(r, t)$ is on the order of $1/N$ even when $z>L$. Let $L+\delta$ denote the maximal distance that $w(r,t)\neq 0$, where $\delta$ is a random variable. Then in the region that $L<z<L+\delta$, the noise term scales as $e^{\delta }/N$, significantly larger than $w(z,t)$ itself which scales as $1/N$. The wavefront shift caused by the noise in a unit time is
\bea
\Delta z (\delta) &\sim \log \left  (1+\frac{2\pi}{L^2}\int \limits_{L-\delta} ^L e^{  \delta +z-L} \sin \frac{\pi z}{L} dz \right ) \\
& \sim \log \left ( 1+ \frac{e^{\delta}}{L^3} \right ).
\eea

Based on a phenomenological reaction-diffusion model, Brunet et al. \citep{Brunet2006} obtained the probability for a large $\delta$ to appear in a unit time as
\be
p(\delta)\sim e^{- \delta},
\ee
decaying with the natural decay constant in the system. Then the additional velocity correction and the diffusion constant can be calculated straightforwardly as
\bea
 \delta v\sim \int \Delta z(\delta) p(\delta) d \delta \sim \frac{1}{L^3}\\
 D \sim \int (\Delta z(\delta) )^2p(\delta) d \delta \sim \frac{1}{L^3}.
\eea
This suggests that the system approaches its infinite-$N$ limit extremely slowly.

\section {S-MPS simulation of stochastic processes}
\label{appsec:SMPS}
In this appendix, we present more details on using matrix product state techniques to simulate the master equation Eq.~\eqref{eq:master}.

\subsection{Matrix product state for simulating a stochastic process (S-MPS)}
The key idea of S-MPS is to represent a probability distribution in a matrix product form and update the MPS based on the master equation for the probability distribution. A probability distribution $\rho$ (viewed as a diagonal density matrix) and a quantum state $\psi$ have similar structures, both containing $L$ indices with dimension $d$, where $L$ is the number of sites and $d$ is the number of physical states per site. But the normalization is different. The normalization of a quantum state is $\langle \psi |\psi \rangle =1 $, a 2-norm condition, while the normalization of a probability distribution is $\text{Tr} (\rho)=1$, a 1-norm condition. Furthermore, each element of $\rho$ must be positive. Most conventional MPS techniques, such as the canonical form, are built around the 2-norm structure. They can still be applied to a probability distribution $\rho$ for small system sizes, but numerical stability issues are encountered in larger systems. One reason is that the 1-norm is much larger than one if the 2-norm is kept to one. Therefore, conventional MPS techniques need modification for simulating large scale stochastic processes.

\textit{Decomposition.}--- The first goal is to decompose a probability into a matrix product form that facilitates the calculation of local observables, similar to the infamous canonical form of matrix product states. To realize this goal, consider a probability distribution $\rho ^{\alpha\beta}$, where $\alpha$ represents the state in the first site and $\beta$ represents the rest. The system is assumed to consist of $L$ sites with open boundary conditions. In the following, we will use superscripts for physical degrees of freedom and subscripts for auxiliary indices.

\textbf{Step 1:} Break $\rho$ into two pieces $\rho^\alpha_{(l),m} $ and $\rho^\beta_{(r),m}$, so that $\sum\limits_m \rho^\alpha_{(l),m} \rho^\beta_{(r),m}=\rho^{\alpha \beta}$. Here $(l)$ and $(r)$ stand for left and right, as we have in mind a sweeping procedure. This can be achieved using the usual Schmidt decomposition or an LQ decomposition.

\textbf{Step 2:} Do a local basis change on the auxiliary dimension so that $\sum \limits_\alpha \rho ^\alpha _{(l),m}  =\delta _{m,1}$. This can be achieved by performing an LQ decomposition on $\sum\limits_\alpha \rho^\alpha_{(l),m}$,
\bea
\sum_\alpha \rho^\alpha_{(l),m}=\sum \limits_{m'}\lambda \delta_{m',1} Q_{m'm}
\eea
where $Q$ is unitary and $\lambda$ is a number.
Then  $\rho^\alpha_{(l),m}$ and $\rho^\beta_{(r),m}$ are updated as follows,
\bea
\rho^\alpha_{(l),m}\rightarrow\frac{1}{\lambda}\sum\limits_{m'}\rho^\alpha_{(l),m'}Q^{-1}_{m'm}\\
\rho^\beta_{(r),m}\rightarrow \lambda \sum\limits_{m'} Q_{m m'} \rho^\beta_{(r),m'}
\label{appeq:tensor_update}
\eea
 Now we factor out the first tensor $\rho(1) ^\alpha_{m}=\rho^\alpha
 _{(l),m}$ to yield the first matrix in the matrix product form.

\textbf{Step 3:} The goal is to factor out the next tensor from $\rho^\beta_{(r),m}$. First rewrite it as $\rho^{\alpha \beta'}_{m_1}$ by dropping the $(r)$ label and breaking $\beta$ into $\alpha$ and $\beta'$ with $\alpha$ representing the state on site $2$ and $\beta'$ the states on sites $3$ to $L$. The label $m_1$ indicates that this auxiliary index is associated with the first link in the MPS. Because of the previous  steps, $\rho^{\alpha \beta'}_{m_1=1}$ is the reduced probability distribution for sites $2$ to $L$. We again decompose it into $\sum\limits_m\rho^\alpha_{(l),m_1, m} \rho^\beta_{(r),m}$ by, say, an SVD. Keep in mind that the auxiliary bond $m$ now has dimension $d^2$ since the combined index $\alpha m_1$ has dimension $d^2$.

\textbf{Step 4 (optional):} Perform a local basis change on the auxiliary space $m$ so that $\rho ^{\alpha}_{(l),m_1=1, m}=0$ for $m>d$. This can be achieved by performing an LQ on $\rho^{\alpha}_{(l),m_1=1,m}$ and using the unitary matrix $Q$ to update the tensors as follows,
\bea
\rho^{\alpha}_{(l),m_1=1, m}&=\sum\limits_{m'}\rho^{\alpha}_{(l),m_1=1,m'}Q_{m'm}\\
\rho^{\alpha}_{(l),m_1>1,m}&\rightarrow \sum\limits_{m'}\rho^\alpha_{(l),m_1>1,m'}Q^{-1}_{m'm} \\
\rho^{\beta'}_{(r),m1,m}&\rightarrow \sum\limits_{m'} Q_{m m'}\rho^{\beta'}_{(r),m'}. \\
\eea
This step is not necessary to bring the probability distribution into the S-MPS canonical form, but is important for the purpose of truncation as discussed later.

\textbf{Step 5:} Similar to \textbf{Step 2}, we perform another local basis change to make sure that $\sum \limits_\alpha \rho^\alpha_{(l), m_1=1,m}=\delta_{m,1} $ by doing a LQ decomposition on $\sum \limits_\alpha \rho^\alpha_{(l), m_1=1,m}$ and updating the tensors $\rho ^{\alpha}_{(l),m_1,m}$ and $\rho^\beta_{(r),m}$ as in Eq.~\eqref{appeq:tensor_update}.

\textbf{Step 6:} Iterate \textbf{Step 3} through \textbf{Step 5} until site $L$ is reached. The whole procedure is called a right-sweep and it produces 1-norm right canonical form of S-MPS. One could equally well start with site $L$ and perform a left-sweep by changing the LQ decomposition to a QR decomposition to obtain the left canonical form.

\begin{figure}
\includegraphics[height=0.6\columnwidth, width=0.7\columnwidth]
{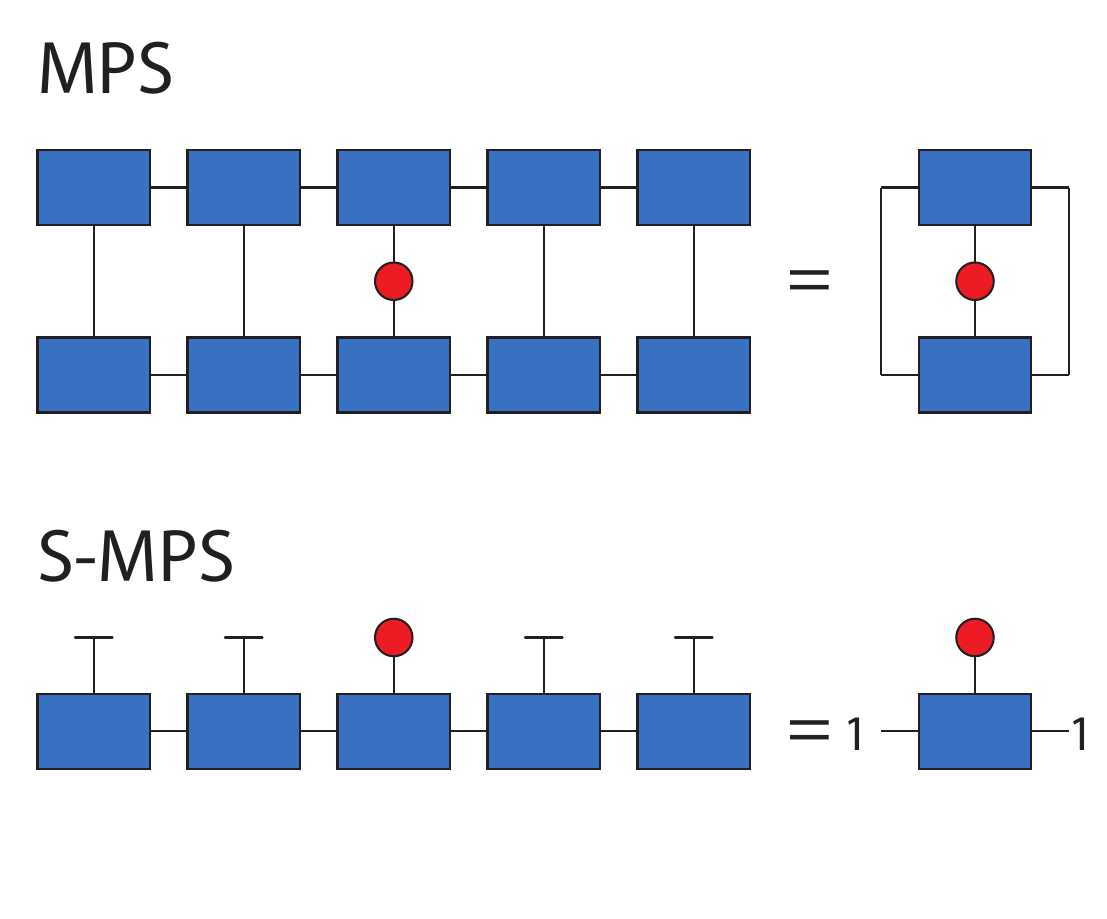}
\caption{Procedures for measuring local observables in an MPS in canonical form (top) and an S-MPS in canonical form (bottom). }
\label{fig:SMPS_local}
\end{figure}

\textbf{Local observables:} To measure a local observable at site $i$, one can first right-sweep from site $1$ to $i-1$ and then left-sweep from site $L$ to site $i+1$. Then the probability distribution is in the so-called mixed canonical form,
\be
\rho^{\alpha \gamma \beta} =\sum \limits_{mm'}\rho^{\alpha}_m \rho^\gamma_{m m'} \rho^{\beta}_{m'}
\ee
where $\gamma$ is the index for the states at site $i$, $\alpha$ is for the states to the left of $i$ and $\beta$ is for the states to the right of $i$.  By construction, $\rho^\gamma_{1,1}$ is the reduced probability distribution for site $i$, $\sum\limits_{\alpha \beta} \rho ^{\alpha\gamma\beta}$, from which the calculation of local observables is straightforward. Thus we have realized the goal sketched in Fig.~\ref{fig:SMPS_local}, where calculation of local observables is reduced to manipulating local data.

\textit{Simulating master equation and truncation.}---To simulate a local master equation such as Eq.~\eqref{eq:master}, we view the generator of the stochastic evolution as a non-hermitian Hamiltonian which explicitly conserves 1-norm. Then we can roughly follow the usual time evolving block decimation (TEBD) steps to update the S-MPS after a short time step, but with the sweeping procedures replaced by those described in the last section.

Similar to TEBD, the bond dimension of the S-MPS typically grows rapidly with time and truncation is always necessary. Consider the situation after updating the tensors at site $i$ and site $i+1$. The probability distribution is in the following form,
\bea
\rho^{\alpha \gamma_1\gamma_2\beta}=\sum \limits _{m_l,m_r}\rho^{\alpha}_{m_l} \rho^{\gamma_1\gamma_2}_{m_l m_r} \rho^{\beta}_{m_r}
\label{appeq:trnctn1}
\eea
where $\gamma_1$ and $\gamma_2$ represent the states on site $i$ and $i+1$ respectively, and by construction $\rho^{\gamma_1 \gamma_2}_{11}$ is the reduced probability distribution of site $i$ and site $i+1$. Assuming that the stochastic simulation starts with an S-MPS with bond dimension $\chi$, the dimension of the middle matrix $\rho^{\gamma_1\gamma_2}_{m_l m_r} $ is $\chi d $, i.e., the bond dimension at the bond linking sites $i$ and $i+1$ is $\chi d$. The goal is to reduce it back to $\chi$ by breaking $\rho^{\gamma_1\gamma_2}_{m_l m_r} $ into two pieces, $\rho^{\gamma_1}_{m_l, m}$ and $\rho^{\gamma_2}_{m, m_r}$, where the dimension of $m$ is $\chi$. Then one can continue to right-sweep or left-sweep to update the next bond.

There are two comparable methods to achieve this. The most straightforward method is to perform an SVD on $\rho^{\gamma_1\gamma_2}_{m_l m_r} $ by regarding it as a matrix with dimension $\chi d$. After keeping only the leading $\chi$ singular values, the bond dimension is reduced back to $\chi$.

The second method is one developed by White et. al in a recent paper studying density matrix dynamics. They first break $\rho_{m_l m_r}^{\gamma_1\gamma_2}$ into two pieces and perform a single right/left-sweeping step (\textbf{Step 3} to \textbf{Step 5}) on the left/right piece, so that $\rho_{m_l m_r}^{\gamma_1\gamma_2}$ is expressed as,
\be
\rho^{\gamma_1\gamma_2}_{m_l m_r}=\sum\limits_{mm'}\rho^{\gamma_1}_{m_l ,m} Q_{m m'} \rho^{\gamma_2}_{m', m_r},
\label{appeq:trnctn2}
\ee
with the properties that $\rho^{\gamma_1}_{1,m>d}=0$ (due to \textbf{Step 4} in the last section), $\sum \limits_{\gamma_1}\rho^{\gamma_1}_{1,m}=\delta_{m,1}$, and similarly for $\rho^{\gamma_2}_{m',1}$. In general, the rank of the matrix $Q$ is $\chi d$.

As they pointed out, given Eq. \eqref{appeq:trnctn1} and Eq. \eqref{appeq:trnctn2}, the right lower $(\chi-1)d \times (\chi-1)d$ section of $Q$ does not affect the reduced probability distribution of site $1$ to site $i$, and site $i+1$ to site $L$, precisely due to \textbf{Step 4} in the sweeping procedure. This gives some freedom to manipulate that section of the matrix in order to reduce the rank of $Q$ without affecting local observables (they were concerned with developing a truncation scheme that respects local observables). For example, if one performed an SVD on that section and kept the leading $\chi-2d$ singular values, then the resulting $Q$ matrix would have rank $\chi$ as required. One could also follow White et. al and reduce the rank while minimizing the error occurring in the correlation functions between the left part ($1\sim i$) and the right part ($i+1 \sim L$) of the system.

Comparing the two methods, the second one is more appealing since a truncation step at a particular bond does not change the reduced probability distribution of the left and right parts of the system. Furthermore, the truncation scheme automatically preserves conserved quantities, like the 1-norm or the total amount of some conserved charge, for all time regardless of the bond dimension. However, the accuracy of the time-dependence of local quantities still depends on the bond dimension. After one sweeps through the system and performs the truncation on each bond, only local observables with support on up to two neighboring sites remain unaffected. Furthermore, as time increases in the simulation, the errors made in observables with larger support could feedback to the dynamics of the two-site observables.

In practice, for our goal of calculating the squared commutator from the master equation Eq.~\eqref{eq:master}, we find that the results obtained from these two truncation schemes are similar, and the first scheme is $2$ to $3$ times faster since it requires fewer steps of SVD.

\subsection{Applying S-MPS to the master equation of the Brownian couple cluster}

\begin{figure}
\includegraphics[height=0.49\columnwidth, width=0.49\columnwidth]
{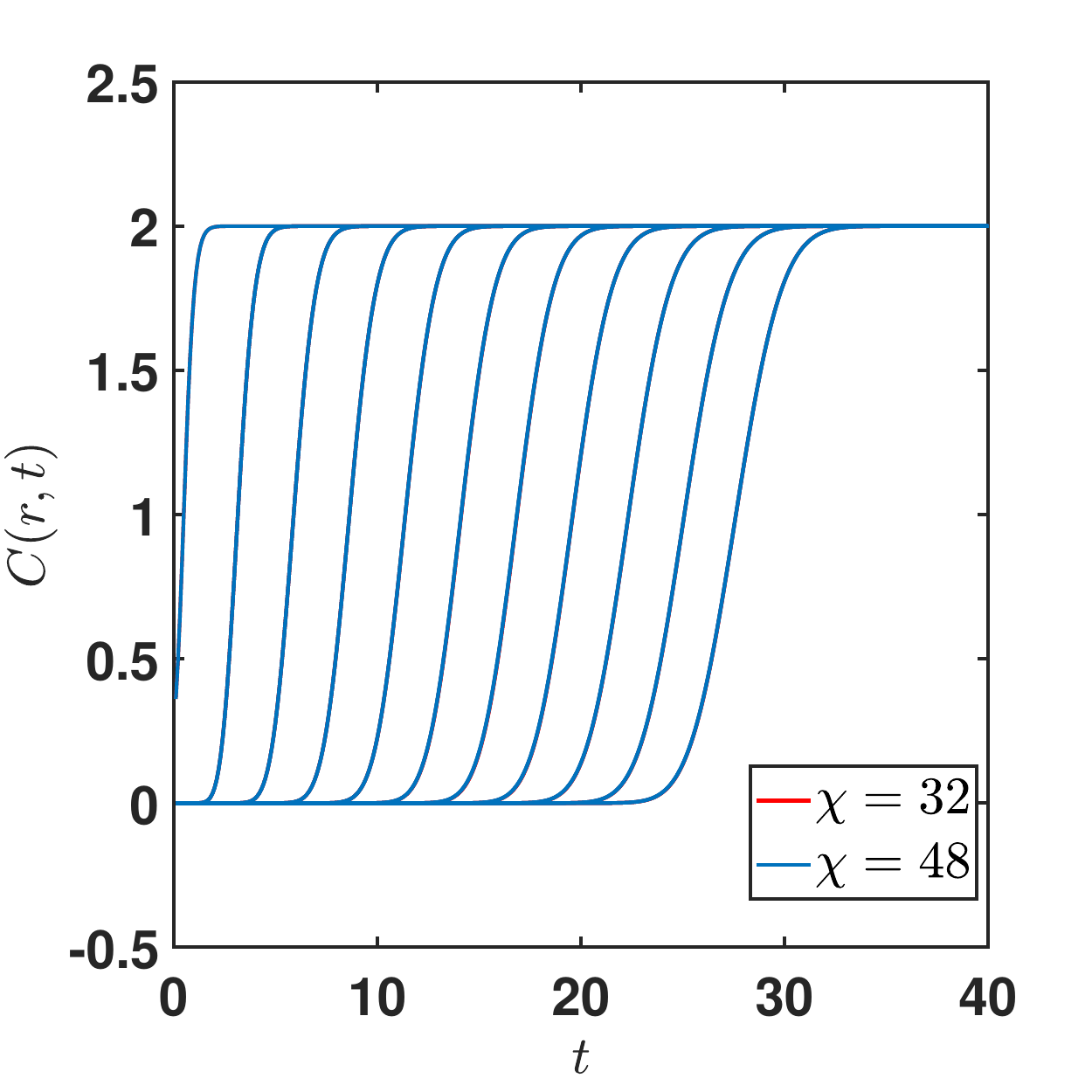}
\includegraphics[height=0.49\columnwidth, width=0.49\columnwidth]
{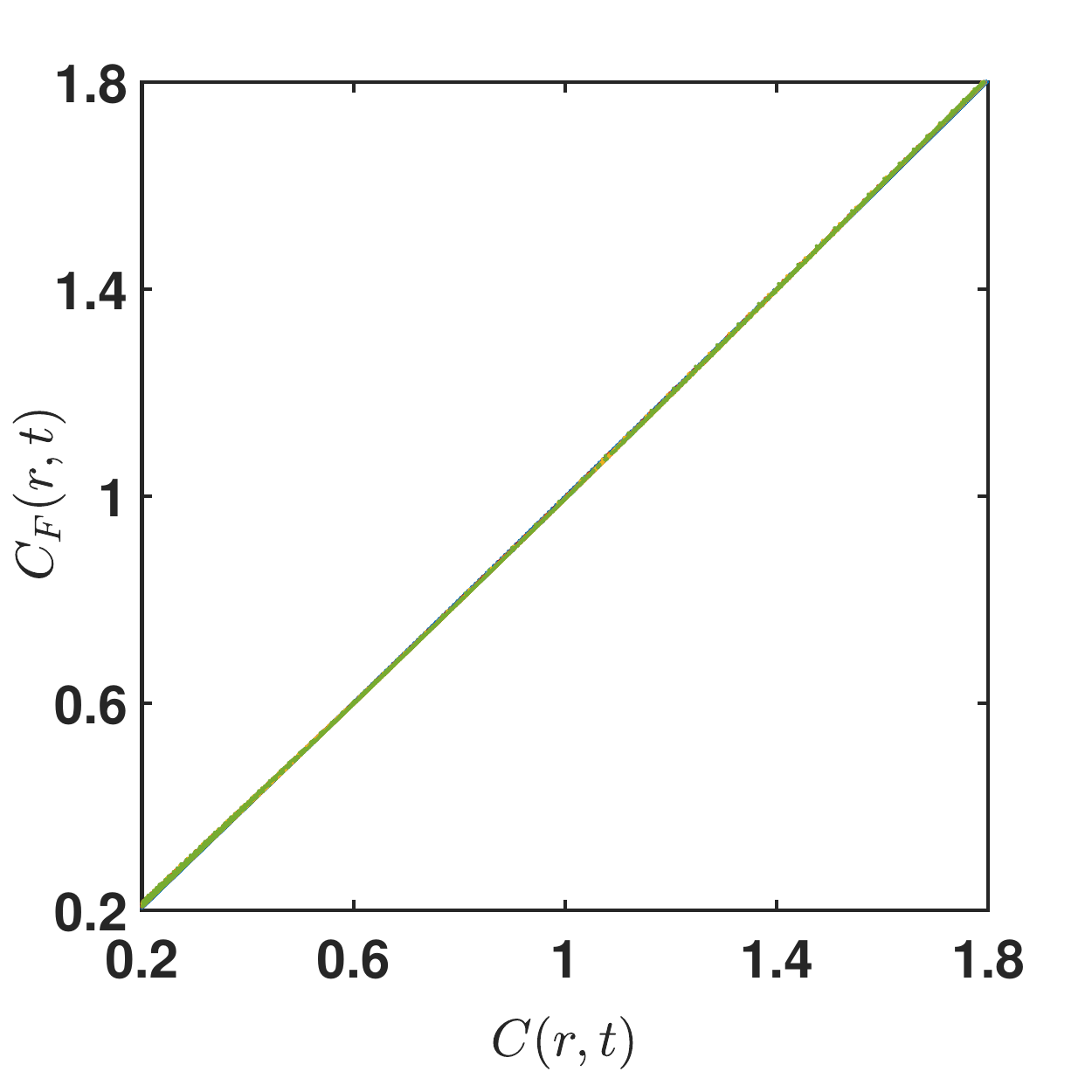}
\caption{(a) Up $N=10$ spins in each cluster and $L=200$ clusters, the numerical data for the squared commutator converges excellently with the bond dimension of the S-MPS. Bond dimension $\chi=32$ and bond dimension $\chi=48$ give rise to almost identical results for all time scales. This allows us to access the entire scrambling behavior from early growth to late-time saturation. (b) Near the wavefront, $C(r,t)$ perfectly agrees with the fitting function $C_{\text{fit}}(r,t)=1+\erf((v_B t-r-r_0)/\sqrt{4Dt})$ for all $N$. }
\label{fig:SMPS}
\end{figure}

We apply the technique described above to simulate the master equation Eq.~\eqref{eq:master}. We first check the convergence of the resulting squared commutator with the bond dimension of the S-MPS. As shown in Fig.~\ref{fig:SMPS}, an S-MPS with bond dimension $\chi=32$ already produces converged results for all time scales for $N=10$ spins within a single cluster and a total of $L=200$ clusters. This allows us to access both early growth and late-time saturation of scrambling in the BCC. We then fit $C(r,t)$ in the region near the wavefront with an error function $1+\erf \left(\frac{r-v_Bt-r_0}{\sqrt{4Dt}}\right)$ to extract the butterfly velocity $v_B$ and the diffusion constant $D$. The fitting quality is shown in Fig.~\ref{fig:SMPS}(b).

\end{appendices}

\end{document}